%% file: main.tex
\documentclass[nonblindrev,11pt]{informs3} 
\usepackage{endnotes}

\newcommand{\theMANUSCRIPTNO}{}

 \usepackage{setspace}
\usepackage[hyperfootnotes=true]{hyperref}
\usepackage[english]{babel}
\usepackage[utf8]{inputenc} 
\usepackage{booktabs} 
\usepackage{threeparttable}
\usepackage[table,xcdraw]{xcolor} 
\usepackage{array} 
\usepackage{graphicx}
\usepackage[colorinlistoftodos]{todonotes}
\usepackage[toc,page]{appendix}
\usepackage{geometry}
\usepackage{CJKutf8}
\usepackage{bbding}
\usepackage{csquotes}
\usepackage{algorithm} 
\usepackage{lscape}
\usepackage{tabularx}
\usepackage{algpseudocode} 
\usepackage[T1]{fontenc}
\usepackage[style=authoryear, backend=biber, natbib=true, maxbibnames=50, maxcitenames=2, uniquename=false, uniquelist=false]{biblatex}
\usepackage{colortbl} 
\usepackage{amsmath}
\usepackage{multirow}
\usepackage{subfiles}

\usepackage{titlesec}

\titleformat{\paragraph}
{\normalfont\normalsize\bfseries}{\theparagraph}{1em}{}
\titlespacing*{\paragraph}
{0pt}{3.25ex plus 1ex minus .2ex}{1.5ex plus .2ex}

\renewcommand{\theARTICLETOP}{}

\begin{document}

\renewcommand{\thefootnote}{\fnsymbol{footnote}} 
 
	\RUNTITLE{LLMs for Traffic and Transportation Research}

	\TITLE{\Large Large Language Models for Traffic and Transportation Research: \\Methodologies, State of the Art, and Future Opportunities\thanks{This manuscript has been authored in part by UT-Battelle, LLC, under contract DE-AC05-00OR22725 with the US Department of Energy (DOE). The publisher acknowledges the US government license to provide public access under the DOE Public Access Plan (\url{https://www.energy.gov/doe-public-access-plan}).}}
	\ARTICLEAUTHORS{%
\AUTHOR{Yimo Yan$^{a,b}$, Yejia Liao$^c$, Guanhao Xu$^d$, Ruili Yao$^e$, Huiying Fan$^f$, Jingran Sun$^g$, \\Xia Wang$^h$, Jonathan Sprinkle$^h$, Ziyan An$^h$, Meiyi Ma$^h$, Xi Cheng$^i$, Tong Liu$^j$, \\Zemian Ke$^k$, Bo Zou$^b$, Matthew Barth$^l$, Yong-Hong Kuo$^a$}
\AFF{\raggedright $^a$Department of Data and Systems Engineering, the University of Hong Kong, Hong Kong SAR, China\\
$^b$Department of Civil, Materials, and Environmental Engineering, University of Illinois Chicago, IL, USA\\
$^c$Department of Electrical Engineering, University of California, Riverside, CA, USA\\
$^d$Buildings and Transportation Science Division, Oak Ridge National Laboratory, TN, USA\\
$^e$University of California, Riverside, CA, USA\\
$^f$School of Civil and Environmental Engineering, Georgia Institute of Technology, GA, USA\\
$^g$University of Texas at Austin, TX, USA\\
$^h$Department of Computer Science, Vanderbilt University, TN, USA\\
$^i$Cornell University, NY, USA\\
$^j$Department of Civil and Environmental Engineering, University of Illinois Urbana-Champaign, IL, USA\\
$^k$Google Inc., CA, USA\\
$^l$Department of Electrical Engineering, University of California, Riverside, CA, USA\\
}
\vspace{-0.2in}
}

	\maketitle
\vspace{-0.2in}
\setcounter{page}{1}

\textbf{Abstract}
The rapid rise of Large Language Models (LLMs) is transforming traffic and transportation research, with significant advancements emerging between the years 2023 and 2025 -- a period marked by the inception and swift growth of adopting and adapting LLMs for various traffic and transportation applications. However, despite these significant advancements, a systematic review and synthesis of the existing studies remain lacking. To address this gap, this paper provides a comprehensive review of the methodologies and applications of LLMs in traffic and transportation, highlighting their ability to process unstructured textual data to advance transportation research. We explore key applications, including autonomous driving, travel behavior prediction, and general transportation-related queries, alongside methodologies such as zero- or few-shot learning, prompt engineering, and fine-tuning. Our analysis identifies critical research gaps. From the methodological perspective, many research gaps can be addressed by integrating LLMs with existing tools and refining LLM architectures. From the application perspective, we identify numerous opportunities for LLMs to tackle a variety of traffic and transportation challenges, building upon existing research. By synthesizing these findings, this review not only clarifies the current state of LLM adoption and adaptation in traffic and transportation but also proposes future research directions, paving the way for smarter and more sustainable transportation systems.

\input{Introduction}
\input{Method/Methodmain.tex}

\input{Application/Applications}

\input{FutureDirections/FutureDirections}

\input{Conclusion}

%\ACKNOWLEDGMENT{.}

\addcontentsline{toc}{section}{References}

\begin{CJK}{UTF8}{min}
\newpage
\printbibliography
\input{Appendix}
%\subfile{Appendix3}
\end{CJK}
\newpage

\end{document}

%% file: Introduction.tex
\section{Introduction}
\label{section:Introduction}

\quad Traffic and transportation have been pivotal in shaping human civilization throughout history. From the rise and fall of empires driven by maritime trade routes to the development of intricate road networks facilitating urban expansion, the movement of people and goods has always been a cornerstone of societal advancement since the 20th century Before Christ \autocite{gianpaolo2013introduction}. Efficient transportation systems have enabled economic growth, cultural exchange, and technological progress, while also presenting challenges related to congestion, safety, and environmental impact.

In the 20th century, the advent of computer technologies revolutionized traffic and transportation research. The introduction of optimization algorithms and prediction models has allowed for more systematic and efficient planning of transportation networks. These advancements have enabled better traffic management, route optimization, and forecasting of transportation demands, significantly improving the functionality of transportation systems. However, despite these technological strides, several persistent issues remain unresolved. Modern transportation systems generate vast amounts of heterogeneous data, encompassing numerical metrics, videos, images, and unstructured textual information from diverse sources such as traffic reports, social media, and sensor logs. Traditional optimization and prediction algorithms, while powerful, often struggle to integrate and interpret this multifaceted data effectively.

Recent developments in artificial intelligence, particularly Large Language Models (LLMs), have the potential to address these challenges. LLMs, such as generative pre-trained transformer(GPT)-4, bidirectional encoder representations from transformers (BERT), and their derivatives are advanced artificial intelligence (AI) systems trained on extensive datasets to understand, generate, and manipulate human language with high proficiency. These models leverage Transformer architectures \citep{vaswani2017attention}, enabling them to capture complex linguistic patterns and contextual relationships. Beyond natural language processing (NLP), LLMs exhibit capabilities in reasoning, data integration, and multimodal understanding, making them well-suited for applications in traffic and transportation research.

LLMs can undertake a variety of tasks critical to enhancing transportation systems. They can automate the extraction and summarization of information from unstructured data sources, improve the accuracy of traffic forecasts by integrating textual and numerical data, assist in scenario generation for planning and emergency response, and facilitate better decision-making through sophisticated data analysis and interpretation. These capabilities not only enhance the efficiency and safety of transportation systems but also contribute to sustainability by optimizing resource allocation and reducing emissions.

The purpose of this paper is to provide a comprehensive review of recent methodologies and applications of LLMs in traffic and transportation research. Our goal is to present and highlight the state of the art and the potential of LLMs within the traffic and transportation research community, thereby outlining promising directions for future research. The specific research questions to be addressed include:
\begin{itemize}
    \item In which areas of traffic and transportation research are LLMs more promising for adoption?    
    \item Which LLM methods are more appropriate to tackle specific traffic and transportation problems?
    \item What are the challenges and future opportunities for LLMs in traffic and transportation research?
\end{itemize}

Our paper is organized as follows. In Section \ref{sec:Methodology}, we introduce the background and the core methodologies in LLMs. In Section \ref{section:Application}, applications are classified into two broad categories, namely \textit{traffic} and \textit{transportation}. In Section \ref{section:FD}, we present statistics of current research trends and future directions. Finally, we conclude this paper in Section \ref{section:Conclusion}. The abbreviations used in the paper are presented in Table \ref{Table:abbr}.

\begin{table}[htbp]
\footnotesize
\label{Table:abbr}
\begin{center}
\caption{Table of Abbreviations}
\begin{tabular}{ p{2cm} p{5.8cm} p{2cm} p{5.8cm} }
\hline
AD & Autonomous Driving & AI & Artificial Intelligence \\ 
ALPACA & Instruction-Fine-Tuned LLaMA & ASR & Attack Success Rate \\
ASR & Automatic Speech Recognition & ATC & Air Traffic Control \\
ATFM & Air Traffic Flow Management & AVs & Autonomous Vehicles \\
BERT & Bidirectional Encoder Representations from Transformers & BLIP & Bootstrapped Language Image Pretraining\\ 
C-CLUE & Chinese Corpus for Language Understanding Evaluation & CLIP & Contrastive Language–Image Pretraining \\
CMA & Cascaded Multi-Scale Attention & COPT & Cardinal Optimizer \\
CoT & Chain-of-Thought & DL & Deep Learning \\
DPO & Direct Preference Optimization & DRL & Deep Reinforcement Learning \\
FL & Federated Learning & FLAN & Finetuned Language-Action Network \\
FLRT & Fluent Student-Teacher Redteaming & FSL & Few-Shot Learning  \\
FSM & Flight Schedule Manager & FT & Fine-Tuning\\
GCG & Greedy Coordinate Gradient & GDP & Ground Delay Program \\
GRU & Gated Recurrent Unit & GPT & Generative Pre-trained Transformer \\
GUI & Graphical User Interface & ICL & In-Context Learning \\
IR & Information Retrieval & KD & Knowledge Distillation \\ 
KG & Knowledge Graph & LLaMA & Large Language Model Meta AI \\
LLMs & Large Language Models & LVLM & Large Vision Language Model \\
LSTM & Long Short-Term Memory & LoRA & Low-Rank Adaptation \\
MC & Model Construction / Multimodal Content & MFD & Multi-Modal Fusion Discriminator \\
ML & Machine Learning & MLLM & Multi-modal Large Language Model \\ MMLMs & Multimodal Large Language Models & MMBench & Multimodal Model Benchmark \\
MT & Machine Translation & MT-Bench & Multitask Benchmark \\ NER & Named Entity Recognition & NLG & Natural Language Generation \\
NLP & Natural Language Processing & NLU & Natural Language Understanding \\ 
NAS & National Airspace System & NN & Neural Network \\
PaLM & Pathways Language Model& PEFT & Parameter-Efficient Fine-Tuning \\ 
PPO & Proximal Policy Optimization & POS & Part-of-Speech (Tagging) \\ 
QA & Question Answering & QLoRA & Quantized Low-Rank Adaptation \\ 
RAG & Retrieval-Augmented Generation & RLAIF & Reinforcement Learning from AI Feedback \\ 
RLHF & Reinforcement Learning from Human Feedback & RL & Reinforcement Learning \\ 
RNN & Recurrent Neural Network & SA & Sentiment Analysis \\
T0 & T5-like Multitask Text-to-Text Transfer
Transformer & T5 & Text-to-Text Transfer Transformer \\ TTS & Text-to-Speech & TC & Text Classification 
\\TL & Transfer Learning & TSC & Traffic Signal Control \\
UAV & Uncrewed Aerial Vehicle & ULM & Unsupervised Language Model \\
V2I & Vehicle-to-Infrastructure & V2V & Vehicle-to-Vehicle \\ V2X & Vehicle-to-Everything & ViT & Vision Transformer \\
ZSL & Zero-Shot Learning  \\\hline \end{tabular} 
\end{center} 
\end{table}

%% file: Method/Methodmain.tex
\section{Background of LLMs}
% Place holder for methodology
\label{sec:Methodology}
\input{Method/Method1.tex}

%temporarily removed for loading efficiency
% Place holder for subsections

%% file: Method/Method1.tex
\quad Between six and eleven months, a child typically starts to learn its language from the surrounding environment \autocite{stanford-age-milestones}. A newborn is exposed to an overwhelming amount of linguistic input -- parents talking, sibling chattering, TV sounds, and even books they see. Initially, these exposures are noise. But gradually, through consistent exposure and the pattern recognition repertoire of infants' brains, the child begins to make sense of the sea of information \autocite{jurafsky-martin-2025-slp3}.
%\textcolor{red}{[YH: is there any reference for this statement (especially for ``between six and nine months)?]} 

This remarkable process mirrors how LLMs learn, beginning with their data foundation. Just as children absorb massive amounts of language input during their formative years, LLMs begin their development with enormous text datasets. The parallel extends to the processing mechanism: just as human sensory organs (eyes and ears) perform initial pre-processing of linguistic input before neural transmission, LLMs employ sophisticated pre-processing techniques to transform raw text into processable tokens. The neural networks underlying LLMs mirror our brain's architectural principles, though in a simplified and artificial form.

Research shows that children learn approximately seven to ten words daily through reading and exposure, accumulating thousands of words by adulthood \autocite{jurafsky-martin-2025-slp3}. Similarly, LLMs process vast amounts of text during pretraining, building their foundational knowledge. This learning process is guided by the ``distributional hypothesis,'' which suggests that both children and LLMs can learn meaning through observing how words appear together in context, forming intricate patterns of understanding through repeated exposure and contextual learning.

%\textcolor{red}{[YH: we may also need a reference here]}

Knowledge acquisition follows a similar trajectory in both systems. Children progress from basic vocabulary to complex language understanding, much like how LLMs develop from basic pattern recognition to sophisticated language processing. After acquiring basic language ability, children begin to learn domain-specific knowledge, very much similar to that of post-training procedures such as fine-tuning of LLMs. This specialization phase allows both human learners and artificial systems to develop expertise in specific areas while building upon their foundational language understanding.

This acquired knowledge then transforms into practical application. Children eventually use their language knowledge for various purposes -- asking questions, telling stories, expressing emotions, and engaging in complex dialogues. Similarly, LLMs can apply their learned patterns to diverse tasks ranging from summarization and question answering to text completion and translation under appropriate interaction techniques. The versatility of both systems demonstrates how fundamental language understanding can adapt to serve numerous practical purposes.

\begin{figure}[h]
\centering
\includegraphics[trim=2.3cm 3.5cm 2.3cm 3.5cm, clip, width=0.98\textwidth]{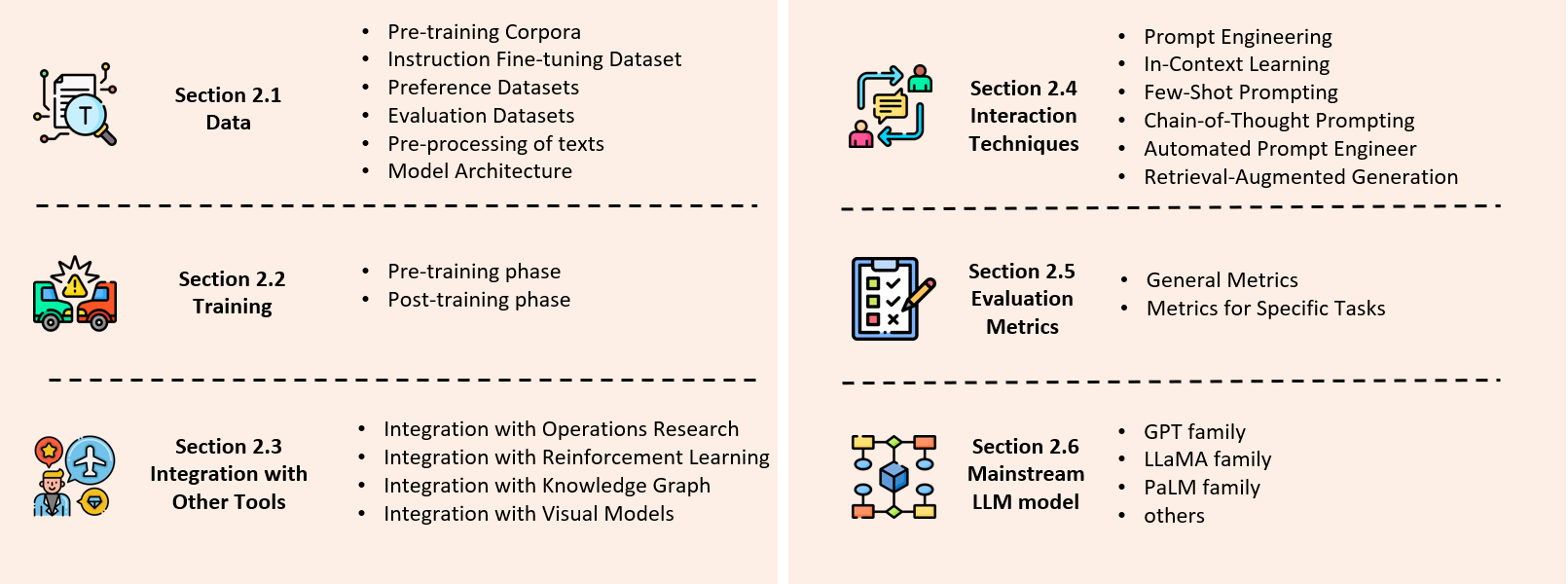}
\caption{Key aspects of LLMs}
\label{fig:mindmap}
\end{figure}

This section establishes the foundation for understanding LLMs in traffic and transportation applications. It encompasses the essential components: data, training, integration with other tools, interaction techniques, evaluation metrics, and mainstream LLMs. Figure \ref{fig:mindmap} shows some key aspects of LLMs. %\textcolor{blue}{[Zou: Strictly speaking, subsection 2.6 is about different LLM models but not about how it contributes to the overall capability of LLMs]}.

\subsection{Data} %summer
\quad Just as a baby learns to understand and communicate by being exposed to varied and abundant stimuli -- listening to conversations, observing facial expressions, and interacting with its surroundings -- LLMs also rely on their ``environment'' to develop their extraordinary capabilities. For LLMs, this environment is the vast array of datasets that serve as the foundation for training, fine-tuning, and evaluation. Without high-quality datasets akin to the rich experiences of a child, an LLM cannot achieve its full potential in understanding and generating language.

In this section, we explore the ``lifecycles'' of datasets that shape LLMs, from the foundational pre-training corpora that teach basic language understanding to the instruction fine-tuning datasets that help LLMs follow human commands. We also examine preference datasets that align models with human expectations and evaluation datasets that measure their performance. These datasets, much like the milestones in a child’s learning journey, are crucial for the development of an LLM’s capabilities. 

\subsubsection{Pre-training Corpora}

The foundation of an LLM’s ``early learning'' lies in its pre-training corpora, akin to the vast sensory input a baby receives during its formative years. These extensive collections of text data enable LLMs to acquire fundamental language skills, such as grammar, semantics, and contextual understanding. Pre-training corpora are the largest and most diverse datasets in the LLM lifecycle, encompassing both general knowledge and specialized domain expertise.

\paragraph*{General Pre-Training Corpora} 
\quad General pre-training corpora, like a child’s exposure to everyday conversations and activities, provide LLMs with broad and diverse language knowledge from mixed domains. These corpora include a few major categories:

\textbf{Webpages}: Crawled texts from the internet, characterized by their massive scale, dynamic updates, multilingual content, and diverse themes~\autocite{xue2020mt5, penedo2023refinedweb, raffel2020exploring}.

\textbf{Code}: Datasets containing programming languages like Python, Java, and C++, enabling LLMs to perform tasks such as code comprehension and generation. Examples include The Stack~\autocite{kocetkov2022stack} and BIGQUERY~\autocite{nijkamp2022codegen}.

\textbf{Parallel Corpus}: Text pairs in different languages, vital for machine translation and cross-lingual tasks~\autocite{banon2020paracrawl, ziemski2016united}.

\textbf{Encyclopedia}: Authoritative resources like Wikipedia and Baidu Baike, offering structured knowledge edited by experts.

\paragraph*{Domain-Specific Pre-Training Corpora}
\quad As a child matures, it gains a deeper understanding in specific areas, like recognizing the nuances of a parent’s profession or learning specialized vocabulary. Similarly, domain-specific pre-training corpora are tailored to particular fields, such as transportation, machine learning, or environmental protection. These datasets enhance LLMs' performance in specialized domains by building on the foundational knowledge acquired from general corpora.

For example, in the transportation field, the TransGPT-pt corpus~\footnote{\hyperlink{https://github.com/DUOMO/TransGPT}{https://github.com/DUOMO/TransGPT}} provides data on transportation literature, engineering, and management, supporting model development in traffic-related applications.
\subsubsection{Instruction Fine-tuning Datasets}

Instruction fine-tuning datasets act like a child learning to follow specific instructions or commands, such as “tie your shoes” or “say thank you.” These datasets teach LLMs to follow human instructions across tasks like classification, summarization, and code generation, improving their ability to align with human expectations.
\paragraph*{General Instruction Fine-tuning Datasets}
\quad General instruction datasets are designed to help LLMs follow commands across a wide range of tasks. These datasets are constructed using four main methods:

\textbf{Human Generation (HG)}: Manually created by annotators, offering high quality but limited scalability due to cost and time constraints (e.g., Databricks-dolly-15K~\autocite{conover2023free}, OASST1~\autocite{wang2023openchat}).  

\textbf{Model Construction (MC)}: Generated by LLMs, providing abundant and cost-effective data (e.g., Alpaca~\autocite{taori2023instruction}, UltraChat~\autocite{ding2023enhancing}).

\textbf{Collection and Improvement of Existing Datasets (CI)}: Integrating and refining open-source datasets for diversity and scale (e.g., Flan 2022~\autocite{longpre2023flan}, InstructDial~\autocite{gupta2022instructdial}).

\textbf{Combination Method}: Combining approaches for optimal results (e.g., Firefly~\footnote{\hyperlink{https://github.com/yangjianxin1/Firefly}{https://github.com/yangjianxin1/Firefly}}, COIG~\autocite{zhang2023chinese}).

\paragraph*{Domain-specific Instruction Fine-tuning Datasets}
\quad Domain-specific datasets, like a child practicing a specific skill (e.g., playing an instrument), fine-tune LLMs for specialized tasks. For instance, TransGPT-sft~\footnote{\hyperlink{https://github.com/DUOMO/TransGPT}{https://github.com/DUOMO/TransGPT}} is a transportation LLM fine-tuned based on domain-specific documents and can generate traffic-related dialogues.

\subsubsection{Preference Datasets}

Preference datasets are akin to a parent guiding a child’s behavior by providing feedback on what actions are good or bad. These datasets evaluate multiple responses to a single instruction, capturing human or model preferences through voting, sorting, or scoring \autocite{zhao2023survey}.

\textbf{Vote}: Selecting the better option between two or more answers, offering simplicity but limited granularity (e.g., Chatbot-arena-conversations~\autocite{zheng2023judging}, PKU-SafeRLHF~\autocite{ji2024beavertails}, CValues~\autocite{xu2023cvalues}, MT-Bench-human-judgments~\autocite{zheng2023judging}).

\textbf{Sort}: Ranking responses based on predefined criteria, providing detailed insights but requiring significant effort (e.g., OASST1~\autocite{wang2023openchat}).

\textbf{Score}: Assigning numerical values to responses for nuanced reflections of preferences (e.g., WebGPT~\autocite{nakano2021webgpt}, Alpaca\_comparison\_data~\autocite{peng2023instruction}).

\textbf{Other}: Alternative approaches, such as preference modeling (e.g., Medical-rlhf~\footnote{\hyperlink{https://github.com/shibing624/MedicalGPT}{https://github.com/shibing624/MedicalGPT}}).

A balanced approach, combining human and model feedback, ensures alignment with human expectations while mitigating bias.

\subsubsection{Evaluation Datasets}
Evaluation datasets represent the ``student report cards'' for LLMs, testing their performance across various domains and tasks. These datasets assess an LLM's knowledge, capabilities, and alignment with human values. The corresponding evaluation methods could be divided into three categories: code evaluation, human evaluation, and model evaluation.

\textbf{General}: Measures versatility across domains and ability to follow complex instructions (e.g., AlpacaEval~\autocite{dubois2024alpacafarm}, MT-Bench~\autocite{zheng2023judging}).

\textbf{Reasoning}: Tests logical reasoning and inference skills (e.g., Chain-of-Thought (CoT) Hub~\autocite{fu2023chain}, TabMWP~\autocite{lu2022dynamic}). 

\textbf{Code}: Evaluates programming proficiency, including code generation and debugging (e.g., HumanEval~\autocite{chen2021evaluating}, CodeXGLUE~\autocite{lu2021codexglue}).

\textbf{Others}: Covers specialized areas like safety, multilingualism, and academic tasks (e.g., SafetyBench~\autocite{zhang2023safetybench}, C-CLUE~\footnote{\hyperlink{https://github.com/jizijing/C-CLUE}{https://github.com/jizijing/C-CLUE}}).

Much like the various stages of a child’s development, the datasets used in LLMs play unique, critical roles in shaping their capabilities. From foundational learning in pre-training corpora to task-specific instruction fine-tuning, alignment with preferences, and rigorous evaluation, datasets are the driving force behind the success of LLMs. 

\subsubsection{Pre-processing of Texts}
Pre-processing transforms raw text into a standardized format for LLM analysis, critical for handling domain-specific data like transportation terminology, traffic reports, or infrastructure records. Key steps include data cleaning, normalization, and tokenization, which ensure consistency and computational interpretability.

\paragraph*{Data Cleaning and Normalization}
\quad Cleaning removes irrelevant elements (e.g., HTML tags, sensor noise in traffic datasets) and standardizes domain-specific content, such as unifying road naming conventions (e.g., ``St,'' ``Street'') or vehicle classifications. Normalization enforces uniformity by converting text to lowercase, expanding contractions, and harmonizing units (e.g., converting ``mph'' and ``km/h'' to a single metric). This step mirrors standardizing traffic signal terminology (e.g., ``red phase'' vs. ``stop interval'') to reduce ambiguity.

\paragraph*{Tokenization}
\quad Tokenization splits text into subwords or characters, enabling LLMs to process complex domain vocabulary (e.g., decomposing ``deceleration'' into [``de,'' ``celer'', ``ation''] or handling technical terms like ``LiDAR''). Subword tokenization balances vocabulary size and out-of-vocabulary resilience, which is crucial for evolving domains like autonomous vehicles (AVs). Special tokens (e.g., classification tokens and separator tokens) structure inputs for tasks such as classifying traffic incident reports or segmenting infrastructure descriptions. This approach aligns with scaling laws \autocite{kaplan2020scaling}, optimizing model performance on sparse, domain-specific datasets.

\subsubsection{Model Architecture}
\quad Modern LLMs rely on advanced neural architectures to process tokenized text and generate human-like responses. The Transformer architecture, introduced by Vaswani et al.~\autocite{vaswani2017attention}, revolutionized NLP by overcoming the limitations of earlier sequential models, such as RNNs and LSTMs. Its ability to handle long-range dependencies and enable parallel processing makes it the foundation of modern LLMs.

At the core of the Transformer are three key components: attention mechanism, which dynamically focuses on relevant input parts; positional encoding, which incorporates token order; and the encoder-decoder framework, which processes inputs and outputs. These components enable Transformers to excel in both understanding and generating text.

Transformers have evolved into three main architectural variants tailored to different tasks: decoder-only, encoder-only, and encoder-decoder models. Table \ref{table:comparisonoftransformer} summarizes their primary characteristics and applications. Decoder-only models (e.g., GPT) specialize in text generation; encoder-only models (e.g., BERT) are optimized for understanding tasks; and encoder-decoder models (e.g., T5) excel in sequence-to-sequence tasks. This section provides a concise overview of these architectures.

For further details on the implementation of Transformers, readers can refer to more comprehensive resources~\autocite{hadi2023survey, kalyan2021ammus, huang2023advancing}.

\paragraph*{Attention Mechanism}
\quad The attention mechanism, particularly self-attention, is central to the Transformer’s success~\autocite{vaswani2017attention}. It evaluates relationships between tokens, enabling parallel processing and efficient modeling of long-range dependencies. Multi-head attention extends this by capturing diverse relationships within sequences. Variants like multi-query attention~\autocite{shazeer2019fast} and grouped-query attention~\autocite{ainslie2023gqa} further improve scalability.

\paragraph*{Positional Encoding}
\quad Transformers lack inherent sequential structure, so positional encoding incorporates token order into the model. Absolute positional encoding~\autocite{vaswani2017attention} assigns unique embeddings to each position, while relative positional encoding focuses on token distances, improving performance in tasks requiring contextual understanding.

\paragraph*{Overview of Transformer Variants}
\quad Decoder-only models process text left-to-right for autoregressive generation tasks like text completion (e.g., GPT~\autocite{radford2018improving,brown2020language}). Encoder-only models, such as BERT~\autocite{devlin2018bert}, use bidirectional self-attention for understanding tasks like classification and question answering. Encoder-decoder models (e.g., T5~\autocite{raffel2020exploring}) balance generation and understanding tasks, excelling in translation and summarization.

\begin{table}[htbp]
\footnotesize
\centering
\caption{Comparison of Transformer Architectures}
\begin{threeparttable}
\label{table:comparisonoftransformer}
\begin{tabular}{>{\raggedright}p{2.5cm} >{\raggedright}p{3.5cm} >{\raggedright}p{3cm} >{\raggedright}p{3cm} >{\raggedright\arraybackslash}p{3.5cm}}
\toprule
\rowcolor{gray!20}
\textbf{Architecture} & \textbf{Components} & \textbf{Attention} & \textbf{Training Objective} & \textbf{Primary Use Cases} \\
\midrule
\textbf{Decoder-Only} & Stacked Decoders & Masked Self-Attention (Unidirectional) & Autoregressive Next-Token Prediction & Text Generation (e.g., GPT) \\
\textbf{Encoder-Only} & Stacked Encoders & Self-Attention (Bidirectional) & Masked Language Modeling (MLM) & Understanding Tasks (e.g., BERT) \\
\textbf{Encoder-Decoder} & Stacked Encoders \& Decoders & Self-Attention (Encoder) \& Cross-Attention (Decoder) & Sequence-to-Sequence Loss & Translation, Summarization (e.g., T5, BART) \\
\bottomrule
\end{tabular}
\end{threeparttable}
\end{table}

\subsection{Training}

\quad Building upon pre-processed data, LLMs undergo a two-stage training process akin to how children acquire language skills. This consists of pre-training for general language understanding and post-training for specialization. During pre-training, LLMs learn fundamental language patterns through tasks like next-token prediction \autocite{radford2018improving} and Masked Language Modeling (MLM) \autocite{devlin2018bert}. In post-training, LLMs are fine-tuned to follow instructions, align with human preferences, or improve specific skills like coding and reasoning. For domain-specific tasks, fine-tuning on related datasets improves performance, enabling LLMs to better respond to domain-specific contexts.

\subsubsection{Pre-training Phase}

\quad Pre-training is unsupervised, involving large, diverse text corpora (e.g., books, websites). Two common pre-training methods are:

\paragraph*{Next-token Prediction} 
\quad This method \autocite{radford2018improving} involves predicting the next word in a sequence, such as completing the sentence ``The sun is shining in the...'' with ``sky.'' The objective is to minimize the autoregressive loss as shown in Equation (\ref{eq:nexttoken_pred}):
\begin{equation}
\mathcal{L}_{\text{AR}}(\theta) = -\mathbb{E}_{(x_1, \dots, x_T) \sim \mathcal{D}} \left[ \sum_{t=1}^{T} \log P_\theta(x_t \mid x_{1:t-1}) \right]
\label{eq:nexttoken_pred}
\end{equation}

\paragraph*{Masked Language Modeling (MLM)}
\quad MLM \autocite{devlin2018bert} masks tokens in a sentence, requiring the model to predict them using bidirectional context. For example, in ``The cat is [MASK] on the mat,'' the model predicts ``sitting.'' The MLM loss is as shown in Equation (\ref{eq:MLM}):
\begin{equation}
\mathcal{L}_{\text{MLM}}(\theta) = -\mathbb{E}_{(x_1, \dots, x_T) \sim \mathcal{D}} \left[ \sum_{t \in M} \log P_\theta(x_t \mid x_{\backslash t}) \right]
\label{eq:MLM}
\end{equation}
Where $\backslash t$ denotes the sequence without the token at position $t$.

\subsubsection{Post-training Phase}

Post-training adapts pre-trained models to specific tasks and user needs.

\paragraph*{Instruction Tuning} 
\quad Fine-tuning refines models using datasets containing instructions, such as ``Summarize this text,'' to improve their ability to handle diverse user requests. The associated loss function is based on next-token prediction and is applied to task-specific data, as shown in Equation (\ref{eq:instruction}):
\begin{equation}
\mathcal{L}_{\text{FT}}(\theta) = -\mathbb{E}_{(x_1, \dots, x_T) \sim \mathcal{D}_{\text{task}}} \left[ \sum_{t=1}^{T} \log P_\theta(x_t \mid x_{1:t-1}) \right]
\label{eq:instruction}
\end{equation}

\paragraph*{Alignment with Human Preferences} 
\quad After fine-tuning for task-specific skills, aligning model behavior with human values involves methods like Reinforcement Learning from Human Feedback (RLHF) \autocite{ouyang2022training}. Human evaluators rank outputs (e.g., preferring $x_1$ over $x_2$), and a reward model $R_\phi(x)$ is trained using pairwise ranking loss as shown in Equation (\ref{eq:alignment}):
\begin{equation}
\mathcal{L}_{\text{reward}}(\phi) = -\mathbb{E}_{(x_1, x_2) \sim \mathcal{D}_{\text{human}}} \left[ \log \frac{e^{R_\phi(x_1)}}{e^{R_\phi(x_1)} + e^{R_\phi(x_2)}} \right]
\label{eq:alignment}
\end{equation}

The LLM is fine-tuned using reinforcement learning to maximize expected rewards, with algorithms like Proximal Policy Optimization (PPO) as shown in Equation (\ref{eq:PPO}):
\begin{equation}
\mathcal{L}_{\text{PPO}}(\theta) = \mathbb{E}_{(s, a) \sim \pi_{\text{old}}} \left[ \min\left( r(\theta) A(s, a), \text{clip}\big(r(\theta), 1 - \epsilon, 1 + \epsilon\big) \cdot A(s, a) \right) \right]
\label{eq:PPO}
\end{equation}
where $r(\theta)$ is the probability ratio between policies,
$A(s,a)$ is the advantage function.

\paragraph*{Alignment with Human Preferences} 
\quad Fine-tuning trains models on domain-specific datasets, improving their performance in fields like coding, law, or medicine. For example, fine-tuning for code generation uses the following loss as shown in Equation (\ref{eq:finetuning}):
\begin{equation}
\mathcal{L}_{\text{code}}(\theta) = -\mathbb{E}_{(c_1, \dots, c_T) \sim \mathcal{D}_{\text{code}}} \left[ \sum_{t=1}^{T} \log P_\theta(c_t \mid c_{1:t-1}) \right]
\label{eq:finetuning}
\end{equation}

Efficient techniques like adapter-based fine-tuning \autocite{houlsby2019parameter} or parameter-efficient fine-tuning (PEFT) \autocite{hu2021lora} reduce computational costs by only updating small parts of the model.
\subsection{Integrating with Other Tools}
Very much like a child learns to use different tools to facilitate his/her life, LLMs can learn to understand and utilize tools in the post-training phase, such as optimization solvers, reinforcement learning (RL), and knowledge graphs (KGs) to further enhance their productivity.

\subsubsection{Integration with Operations Research}
\quad Operations research (OR) is widely utilized in traffic and transportation domains to optimize decision-making under complex constraints and objectives. We find two streams of research that integrate OR with LLMs. The first stream of literature addresses the challenge of a steep learning curve for beginners and limited adoption in small businesses. The second stream of literature improves prompts with discrete optimization methods.

The first stream of literature combines LLMs with OR techniques to make these tools more accessible for industrial applications and education. Recently, a model named ``Operations Research Language Model'' (ORLM) based on fine-tuning is introduced to automate the process of optimization modeling \autocite{huang2024orlm}. Users can describe an optimization problem in natural language, and the model processes the input to identify key components such as decision variables, constraints, and objectives. It then translates these into a standardized mathematical representation. Once the mathematical model is generated, ORLM further converts it into code compatible with optimization solvers, such as COPT or Gurobi, enabling direct execution without manual intervention.

\begin{figure}
    \centering
    \includegraphics[trim=6cm 2cm 6cm 2cm, clip, width=0.7\textwidth]{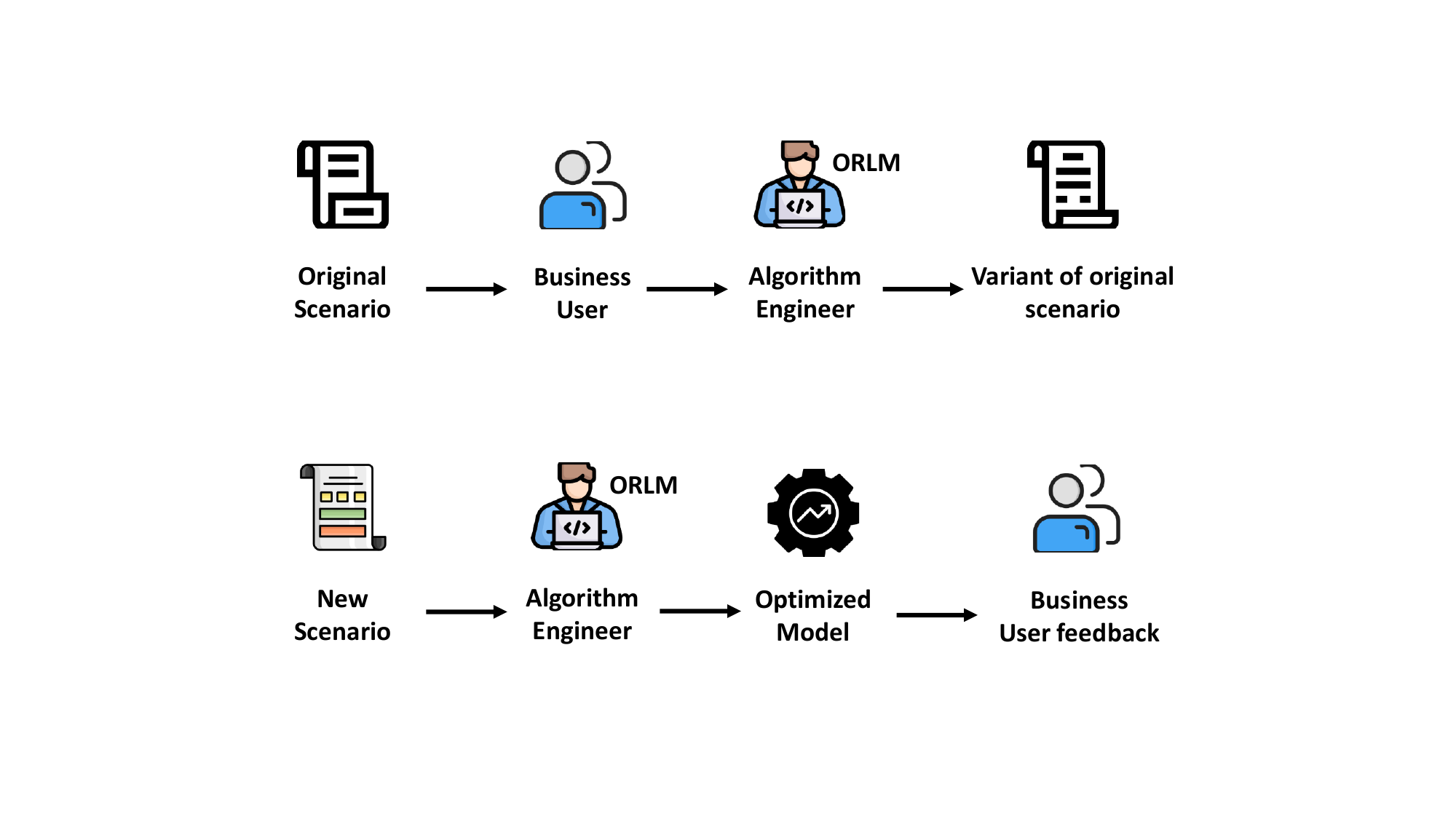}
    \caption{ORLM \autocite{huang2024orlm}}
    \label{fig:ORLM}
\end{figure}

In addition to generating initial solutions, ORLM offers flexibility for dynamic business requirements. For instance, if a constraint like ``airplanes must be used if ships are used'' is introduced, ORLM can seamlessly update both the mathematical model and the solver-compatible code to reflect the new condition. This adaptability reduces the time and effort required to adjust models to evolving scenarios. An illustration is shown in Figure \ref{fig:ORLM}. In our literature search, we find that similar methods have already been used before the introduction of ORLM to address supply chain management problems \autocite{li2023large}.

The second stream examines the application of operations research techniques to optimize prompts. However, relevant research has not been found in our literature search. Therefore, we direct readers to Subsection \ref{subsubsection:FDMethods} under future research directions section.

\subsubsection{Integration with Reinforcement Learning}

\begin{figure}[h]
\centering
\includegraphics[trim=3cm 2cm 3cm 2cm, clip,width=0.9\textwidth]{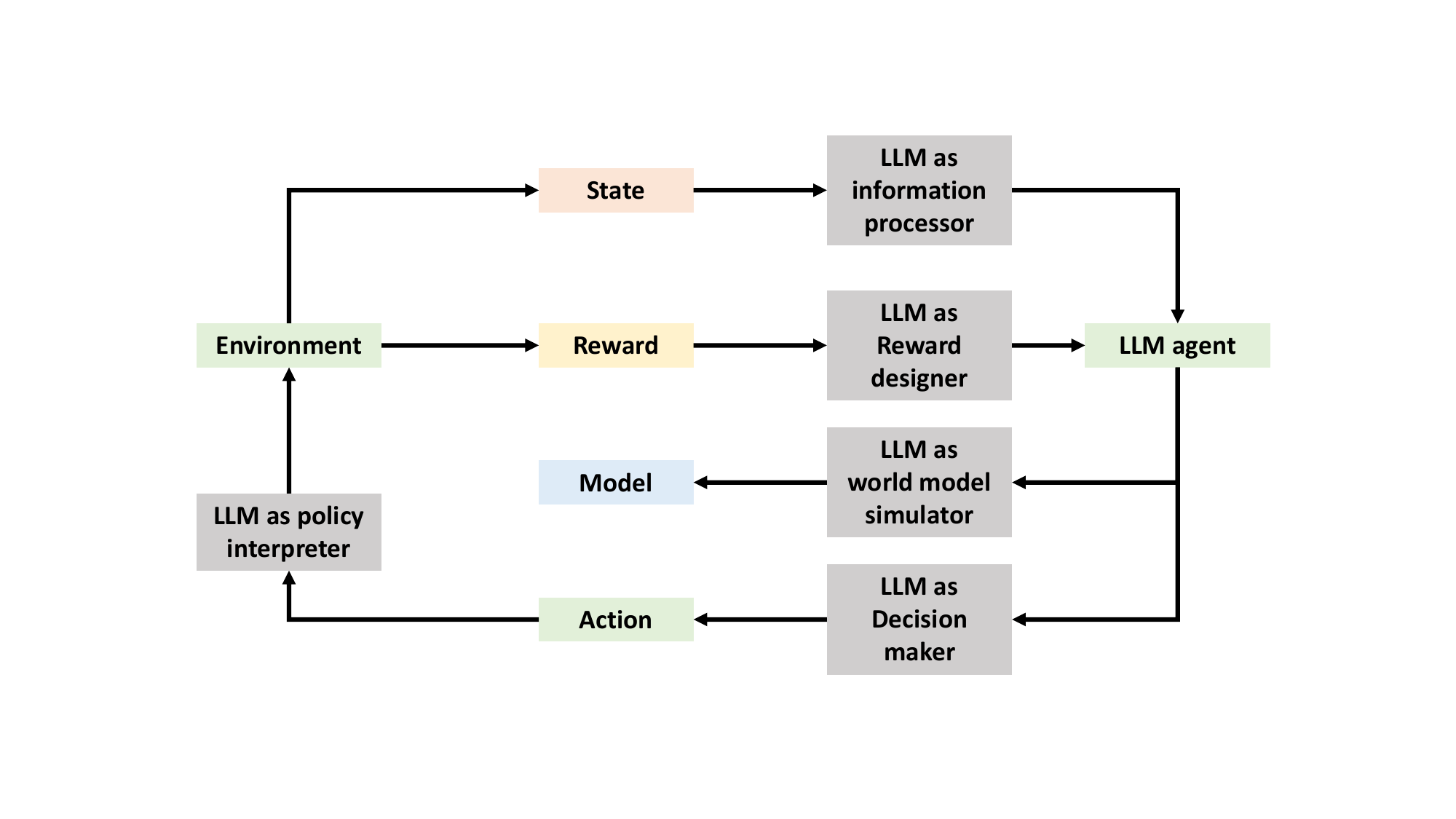}
\caption{Integration of LLMs with Reinforcement Learning \autocite{cao2024survey}}
\label{fig:RLguidedLLMs}
\end{figure}

LLMs are increasingly being used to improve RL. A comprehensive review~\autocite{cao2024survey} highlights how LLMs leverage their pretrained capabilities -- such as understanding, reasoning, and multimodal processing -- to enhance RL efficiency and generalization in complex environments. There are four important aspects of research, as shown in Figure \ref{fig:RLguidedLLMs}.

\textbf{Information Processing}: LLMs convert unstructured inputs like natural language instructions or visual data into structured representations usable by RL agents. This method has been used for traffic signal control \autocite{villarreal2023can_TS9, pang2024illm}, which reduces ambiguity and noise in state features.

\textbf{Reward Design}: In environments with sparse feedback, LLMs automate reward function creation through two methods: generating implicit reward values via semantic analysis (e.g., prioritizing stability over speed for delicate objects) and producing explicit reward code that encodes task logic \autocite{ma2023eureka}. This method has been used in the public transportation domain for bus holding control to address the problem of sparse reward \autocite{yu2024large}.

\textbf{Decision Guidance}: LLMs improve exploration efficiency by either directly generating actions (framing RL as sequence prediction) or constraining action spaces using semantic priors. This method has been used in traffic signal control \autocite{pang2024illm}, drone control \autocite{yang2024large} and bus holding control in public transportation \autocite{prabhod2023advanced}, where unreasonable actions are removed to enhance efficiency.

\textbf{Environment Generation}: LLMs synthesize plausible environment dynamics and policy explanations, addressing model-based RL's need for accurate simulators. For example, Minecraft, a sandbox game, subgoal sequences are hypothesized to (e.g., ``collect wood → build tools'') improve sample efficiency over pure RL methods \autocite{nottingham2023embodied}. %We have not identified relevant papers in our field.

\subsubsection{Integration with Knowledge Graph}
\quad LLMs are trained on vast amounts of unstructured textual data. Their knowledge is implicit, derived from statistical patterns in the data, rather than being explicitly organized or structured. While this allows them to generate coherent and contextually relevant responses, their outputs are not anchored in factual relationships or structured reasoning, resulting in hallucinations. To address this issue, researchers have explored integrating KGs with LLMs to enhance their reasoning capabilities and produce more reliable outputs.

There are two primary methods for combining KGs with LLMs: integration during pre-training and integration during inference (prompting). The latter is the most common and practical approach due to its flexibility.

A recent method introduces a plug-and-play, prompt-based approach to integrating KGs with LLMs. This approach involves using a KG as an input to the LLM during inference and designing prompts that guide the LLM to leverage the structured information from the KG for reasoning. For example, if a KG contains relationships about historical events, a prompt could explicitly reference this knowledge, enabling the LLM to generate a factually accurate and logically consistent response \autocite{li2024enhanced}.

The knowledge solver framework, introduced by \cite{feng2023knowledge}, transforms information retrieval into an interactive, multistep decision-making process. Instead of treating KGs as static sources of information, this framework enables the LLM to interact dynamically with the KGs, retrieving and reasoning over relevant knowledge in iterative steps. This enhances the LLM's ability to answer complex, multi-faceted questions as well as solving problems requiring connections between multiple facts or entities. For example, if a KG contains relationships about historical events, a prompt could explicitly reference this knowledge, enabling the LLM to generate a factually accurate and logically consistent response.

\cite{pan2024unifying} provides a comprehensive review of recent efforts to integrate LLMs with KGs. We direct readers to this review for more detailed methodologies. In our literature review, we have identified papers applying combined LLM and KG methods in supply chain management \autocite{almahri2024enhancing} and autonomous driving (AD)
\autocite{hussien2025rag}.

\subsubsection{Integration with Visual Model} 
\quad Large vision-language models (LVLMs) are sophisticated AI systems designed to process and reason about both visual (e.g., images and videos) and textual (natural language) data simultaneously. By combining computer vision and NLP, LVLMs can understand visual content, generate descriptions, and answer questions about images or videos in a coherent and context-aware manner.

Popular vision-language models include contrastive language-image pretraining (CLIP) by OpenAI, Flamingo by DeepMind, and bootstrapped language-image pretraining (BLIP) by Salesforce. These models are generally pre-trained on large datasets of image-caption pairs, enabling them to associate visual features with text representations. Their core methodologies involve leveraging contrastive learning (e.g., CLIP aligns visual and textual embeddings in a shared latent space to maximize the similarity between matching image-text pairs) and multimodal transformers (e.g., Flamingo and BLIP use cross-attention mechanisms to integrate vision and language modalities). These architectural designs allow LVLMs to effectively adapt to new data and excel at diverse multimodal tasks with minimal additional fine-tuning (FT).

A common way to utilize these models is visual question answering (VQA), where LVLMs analyze an image and respond to specific queries about its content using zero-shot learning (ZSL) or few-shot learning (FSL). For example, when asked, ``What does this sign represent?'' while providing an image of a road sign, an LVLM might respond: ``This sign indicates that all vehicles are only permitted to turn right or left'' \autocite{wang2024transgpt}. LVLMs also excel in object detection and scene understanding, where they identify objects in an image and reason about their spatial and contextual relationships. For instance, LVLMs can detect and describe the motion of vehicles to identify drivable paths \autocite{wang2024omnidrive, pan2024vlp, li2024automated, li2024UnstrPromptLargeLanguage}, detect ships in maritime imagery \autocite{zhang2024popeye}, or interpret environmental contexts for drone navigation \autocite{de2023semantic}.

In autonomous systems, such as self-driving vehicles or drones, LVLMs combine visual perception with language-based reasoning to perform critical tasks. These include identifying obstacles, interpreting traffic signs, or following complex textual instructions. LVLMs can even be integrated into end-to-end systems that directly generate vehicle trajectory plans based on visual and textual inputs \autocite{pan2024vlp}. Their ability to unify vision and language into a single reasoning framework makes them a promising tool for intelligent and adaptive decision-making in real-world settings.

\subsection{Interaction Techniques}
\quad Building upon the trained model, effective interaction techniques are crucial for extracting meaningful responses from LLMs. Just as children need appropriate prompting and access to reference materials to articulate their knowledge effectively, LLMs require carefully crafted prompts and supplementary information to generate accurate and contextually relevant outputs. Two primary interaction techniques have emerged: prompt engineering and retrieval augmented generation (RAG), each serving distinct yet complementary roles in enhancing LLM performance in transportation applications.

\subsubsection{Prompt Engineering} %Ziyan

Prompt engineering serves as the art of communication with LLMs, akin to how teachers carefully phrase questions to elicit specific responses from students. This technique involves crafting inputs that guide models toward desired outputs, especially critical in transportation contexts where precision and safety are paramount.

The practice of prompt engineering varies across application domains, with widely adopted approaches emerging. The simplest form is zero-shot prompting, where a task description is provided without examples or additional context. For instance, Prompt: ``Is the following statement true or false: `The first autonomous vehicle to complete a cross-country trip without human intervention was developed by Carnegie Mellon University in 1995.''' Output: ``False.'' Here, the LLM relies solely on its pre-existing knowledge. While sufficient for straightforward tasks, zero-shot prompting often fails for complex tasks like traffic flow prediction or route optimization.

For more advanced use cases, prompts can include a task description, the LLM's role, specific examples, and relevant context to improve responses. Just as teachers provide structured guidance with examples, these elements enhance LLM accuracy. Prompts can also vary structurally, using techniques such as few-shot prompting, CoT prompting~\autocite{wei2022chain}, tree-of-thought prompting~\autocite{yao2024tree}, and graph-of-thought prompting~\autocite{besta2024graph}. Some less common prompt techniques are listed in the Appendix.

\paragraph*{In-Context Learning}

In-context learning (ICL) involves providing demonstrations to LLMs as context~\autocite{dong2024surveyincontextlearning}, without requiring parameter updates. Effective ICL depends on selecting and organizing examples~\autocite{qin2023context,he2023icl}. Techniques like reformatting demonstrations with optimization methods~\autocite{dong2022survey} further enhance performance, making ICL a powerful tool for guiding LLM behavior.

\paragraph*{Few-Shot Prompting}

Few-shot prompting provides a few input-output examples to help LLMs recognize patterns and generalize. For instance, in transportation-related tasks, examples like ``The train from Boston to New York departs at 4:30 PM and arrives at 7:45 PM.'' paired with the question ``How long is the journey?'' and the answer ``The journey is 3 hours and 15 minutes.'' guide the model for further questions. However, as noted by~\cite{lu2021fantastically}, the order of examples can significantly influence performance, necessitating careful prompt design.

\paragraph*{Chain-of-Thought Prompting}

Few-shot prompting can fail to reveal the logic behind a task, especially for problems requiring complex reasoning. CoT prompting improves performance by encouraging step-by-step reasoning~\autocite{wei2022chain}. For example: ``Bill has 5 apples. He buys 2 buckets, each containing 6 apples. How many apples does he have?'' CoT: ``Bill starts with 5 apples and buys 2 buckets, each containing 6 apples. Total is $5 + 2 \times 6 = 17$. Output: 17.'' CoT prompting is particularly effective for tasks like mathematical reasoning and decision-making~\autocite{feng2024towards}.

\paragraph*{Automated Prompt Engineering}

Automated prompt engineering (APE) reduces reliance on manual trial-and-error by generating and optimizing prompts automatically. In retrieval and reasoning tasks, APE has been shown to outperform manual prompts (e.g., ``let's think step by step'')~\autocite{jin2024apeer,sahoo2024systematic}. Introduced by~\cite{zhou2022large}, APE uses LLMs to search over candidate prompts, framing the task as a black-box optimization problem. Each prompt is scored, and Monte Carlo search methods iteratively refine the best-performing candidates, significantly improving efficiency and accuracy.

\subsubsection{Retrieval-Augmented Generation} %Ziyan 
While prompt engineering focuses on how we ask questions, RAG enhances how LLMs access and utilize information, much like students consulting reference materials. RAG enables LLMs to complement their pre-trained knowledge with specific, up-to-date information, improving factual accuracy and reducing hallucinations~\autocite{lewis2020retrieval}.

RAG consists of two components: a retriever and a generator. The retriever fetches relevant information from an external knowledge source, while the generator combines this retrieved context with the original prompt to produce informed responses. For instance, in transportation tasks requiring domain-specific knowledge, such as traffic regulations or infrastructure updates, RAG retrieves relevant sections from a knowledge base and integrates them into the generation process.

A common implementation involves segmenting the knowledge base into chunks, encoding them into vector representations, and ranking them by similarity to the prompt (e.g., using cosine similarity). The top-ranked chunks are passed to the generator, enabling more accurate, contextually enriched responses.

Advanced variations, such as advanced RAG and modular RAG~\autocite{gao2023retrieval}, further optimize retrieval and generation. Modular RAG introduces additional tools, such as search modules, to integrate multiple data sources. For example,~\cite{wang2023knowledgpt} demonstrate improved accuracy by combining transportation databases to provide real-time updates on traffic conditions, route changes, and infrastructure.

\subsection{Evaluation Metrics of LLMs}  % zemian
\quad Just as educators assess children's language development through tests, LLMs require evaluation metrics to gauge their performance. This is especially important for transportation applications, where accuracy and reliability are critical. Evaluations encompass both general language capabilities and domain-specific competencies.

\subsubsection{General metrics}
LLMs are assessed using standard metrics that evaluate accuracy, fairness, robustness, and calibration across a variety of tasks \autocite{pmlr-v70-guo17a, wang2021adversarial, zhu2023promptbench}:

\paragraph*{Accuracy:} 
Accuracy measures how well model outputs match the ground truth. Key metrics include:
\begin{itemize}
    \item \textbf{Exact Match}: Evaluates whether the generated text matches the reference exactly.
    \item \textbf{F1 Score}: Combines precision and recall for classification tasks.
    \item \textbf{ROUGE Score}: Measures overlap between generated and reference texts, commonly used in summarization.
\end{itemize}
\paragraph*{Calibration:} 
Calibration assesses how well a model's confidence aligns with its correctness. Examples include:
\begin{itemize}
    \item \textbf{Expected Calibration Error (ECE)}: Measures the gap between confidence and accuracy. 
    \item \textbf{Area Under the Curve (AUC)}: Evaluates performance in coverage and accuracy for selective predictions.
\end{itemize}
\paragraph*{Fairness:} 
Fairness metrics ensure equal treatment across demographic groups. Examples include:
\begin{itemize}
    \item \textbf{Demographic Parity Difference}: Measures differences in positive outcomes across groups.
    \item \textbf{Equalized Odds Difference}: Evaluates whether outcomes are consistent across groups, controlling for actual results.
\end{itemize}
\paragraph*{Robustness:} 
Robustness measures a model's resilience to adversarial inputs or changes. Common metrics include:
\begin{itemize}
    \item \textbf{Attack Success Rate (ASR)}: Evaluates vulnerability to adversarial attacks.
    \item \textbf{Performance Drop Rate}: Quantifies performance degradation after adversarial modifications.
\end{itemize}

\subsubsection{Benchmarks for Assessing Specific Tasks}

In addition to general evaluation metrics, many customized metrics are developed to assess LLMs in specific domains or tasks. These metrics are tailored to the unique demands of particular applications and are used to evaluate capabilities that general metrics might not adequately capture.

\textbf{Multimodal Task Benchmarks:} For evaluating the performance of multimodal large language models (MLLMs), benchmarks like MMBench \autocite{liu2023mmbench} focus on assessing vision-language models by evaluating the model's capability to process and understand both visual and textual data. This includes performance across tasks such as perception and cognition.
    
\textbf{Reasoning Tasks:} The advanced reasoning benchmark \autocite{sawada2023arb} focuses on evaluating LLMs in complex reasoning tasks across multiple domains, pushing models to handle more sophisticated problem-solving. More recently, a few evaluation methods have been proposed to evaluate specific reasoning tasks, such as coding (e.g., SWE bench \autocite{jimenez2024swebench}) and machine learning engineering \autocite{chan2024mlebenchevaluatingmachinelearning}.

\textbf{Ethics and Bias in LLMs:} TRUSTGPT is a customized benchmark designed to test the ethical considerations of LLMs, including metrics related to toxicity, bias, and value alignment \autocite{huang2023trustgpt}.

\subsection{Comparison of Main-stream LLMs}
%(GPT-3, GPT-4, GPT-4o, Mistral, Vicuna-13B, Falcon, Davinci, Claude-V1, Alpaca-13B)
\quad LLMs are transformer-based pre-trained models (PLMs) with tens to hundreds of billions of parameters, demonstrating superior language understanding, generation, and emergent capabilities compared to smaller models. Notable LLM families include GPT, LLaMA, and PaLM~\autocite{zhao2023survey,minaee2024large}.

\paragraph*{GPT Family}

Developed by OpenAI, the GPT family includes some of the most widely used and influential models in NLP.
\begin{itemize}
    \item \textbf{GPT-3}~\autocite{brown2020language}: With 175B parameters, it introduced emergent capabilities like ICL and excels in diverse tasks.
    \item \textbf{ChatGPT}~\footnote{\hyperlink{https://openai.com/blog/chatgpt}{https://openai.com/blog/chatgpt}}: Based on GPT-3.5 and GPT-4, it is optimized for dialogue tasks and widely used for conversational AI.
    \item \textbf{GPT-4}~\autocite{achiam2023gpt}: A multimodal model that accepts both text and image inputs, demonstrating exceptional performance in professional exams.
\end{itemize}

\paragraph*{LLaMA Family}

Released by Meta, LLaMA models are open-source and widely used for research and FT.
\begin{itemize}
    \item \textbf{LLaMA}~\autocite{touvron2023llama}: Ranges from 7B to 65B parameters and features architectural modifications like SwiGLU activation.
    \item \textbf{LLaMA-2}~\autocite{touvron2023llama}: Includes chat-specific models fine-tuned for dialogue tasks.
\end{itemize}

\paragraph*{PaLM Family}

Developed by Google, PaLM models excel in large-scale learning tasks.
\begin{itemize}
    \item \textbf{PaLM}~\autocite{chowdhery2023palm}: A 540B-parameter model achieving state-of-the-art results in FSL.
    \item \textbf{PaLM-2}~\autocite{anil2023palm}: An improved version of PaLM with better multilingual and reasoning capabilities.
\end{itemize}

\paragraph*{Other LLMs}
\begin{itemize}
\item \textbf{Mistral-7B}~\autocite{jiang2023mistral}: A smaller, high-performing model that outperforms LLaMA-2-13B in several benchmarks.
\item \textbf{DeepSeek-R1}~\footnote{\hyperlink{https://www.deepseek.com}{https://www.deepseek.com}}: General-purpose LLM optimized for technical reasoning and coding tasks, featuring 7B-128B parameter variants with Mixture of Experts architectures.
\end{itemize}

There are also a large number of other LLMs. We refer the reader to the Appendix for other LLMs. All the above-mentioned LLMs contribute to the advancement of this field.
%Additionally, frameworks and techniques like InnerMonologue~\autocite{huang2022inner}, Megatron-Turing NLG~\autocite{smith2022using}, LongFormer~\autocite{beltagy2020longformer}, OPT-IML~\autocite{iyer2022opt}, MeTaLM~\autocite{hao2022language}, Dromedary~\autocite{sun2024principle}, Palmyra~\footnote{\hyperlink{https://dev.writer.com}{https://dev.writer.com}}, Camel~\footnote{\hyperlink{https://dev.writer.com}{https://dev.writer.com}}, Yalm~\footnote{\hyperlink{https://github.com/yandex/YaLM-100B}{https://github.com/yandex/YaLM-100B}}, MPT~\autocite{mosaicml2023introducing}, ORCA 2~\autocite{mitra2023orca}, Gorilla~\autocite{patil2023gorilla}, PAL~\autocite{gao2023pal}, Claude~\footnote{\hyperlink{https://www.anthropic.com/news/introducing-claude}{https://www.anthropic.com/news/introducing-claude}}, CodeGen2~\autocite{nijkamp2023codegen2}, Zephyr~\autocite{tunstall2023zephyr}, Grok~\footnote{\hyperlink{https://grok.x.ai/}{https://grok.x.ai/}}, Qwen~\autocite{bai2023qwen}, Mamba~\autocite{gu2023mamba}, Mixtral-8x7B~\footnote{\hyperlink{https://mistral.ai/news/mixtral-of-experts/}{https://mistral.ai/news/mixtral-of-experts/}}, DocLLM~\autocite{wang2023docllm}, DeepSeek-Coder~\autocite{guo2024deepseek}, FuseLLM-7B~\autocite{wan2024knowledge}, TinyLlama-1.1B~\autocite{zhang2024tinyllama}, and LLaMA-Pro-8B~\autocite{wu2024llama} have also contributed to LLM advancements.

%% file: Application/Applications.tex
\section{Applications in traffic and transportation research}
\label{section:Application}
\quad In this section, we present how state-of-the-art LLM methodologies are applied in traffic and transportation research. While traffic and transportation are closely related, they differ in scope and focus. Traffic research refers to the flow of vehicles and pedestrians and the associated operational aspects within a transportation system, emphasizing how the different entities interact and move. Transportation, on the other hand, encompasses the entire system and infrastructure that enable the movement of people, goods, and services between locations. This section includes all transportation modes (e.g., road, air, and sea), associated facilities (e.g., airports, ports, and stations), policies, and planning strategies.

%We begin this section with the organization of this review, article selection criteria then start to introduce relevant traffic and transportation research with LLMs.

Before delving into the detailed review of the LLM applications in traffic and transportation research, we first present the organization of our review. The way we identify and select the relevant studies is also described.

\subsection{Organization of review of applications}
The review is structured into three main subsections based on the scope of the selected studies: traffic, transportation, and multi-task applications. As the studies in the first two subsections each deal with one specific task application using LLMs, we put multi-task applications as a separate subsection. In the traffic and transportation subsections, the selected studies are classified mainly by their application domains. Methodologies are discussed within these contexts. Characteristics of articles under the same application domain are summarized at the end of each subsection for clarity and coherence.

Subsection \ref{subsec:articleselection} describes our criteria for searching and selecting literature.

Subsection \ref{subsection:traffic} focuses on studies examining the operational aspects of traffic systems. These include topics such as traffic management, travel behavior, traffic safety, and traffic infrastructure.

Subsection \ref{subsection:transportation} explores broader transportation systems and infrastructure, encompassing research on logistics, supply chain management, and autonomous driving. These studies investigate the movement of goods and people across various modes, including road, air, and sea.

Finally, subsection \ref{subsection:Micellaneous} addresses multi-task applications of LLMs that span multiple aspects of traffic and transportation. These works integrate diverse domains, offering insights into complex, multi-faceted challenges.

%\textcolor{blue}{[Zou: This paragraph seems to be already said in the first paragraph of 3.1. Delete?]} The classification framework for the reviewed papers is organized along three dimensions: application domains, levels of application, and methodologies. Articles are first categorized by their domain and then by their level of application (if applicable), with methodologies discussed within these layers. %This structured approach ensures a logical and well-organized review of the literature.

%Characteristics of the reviewed articles under the same application are summarized at the end of the subsection.

\input{Application/ArticleSelection}

\subsection{Traffic Research}
\label{subsection:traffic}
\quad Traffic research refers to the dynamic flow and movement of vehicles, pedestrians, and other entities within a transportation network. It focuses on the operational aspects of mobility, including real-time interactions between road users, traffic flow optimization, and accident prevention. Unlike the broader study of transportation, which encompasses infrastructure development and policy, traffic research emphasizes the immediate and localized behavior of entities within existing systems.

In this section, we explore the role of LLMs in transforming traffic systems. Key areas of focus include autonomous driving, accident analysis, emergency management, traffic safety, signal control and traffic forecasting. These applications demonstrate how LLMs can enhance decision-making, improve system efficiency, and address complex challenges in modern traffic networks.
\input{Application/AutonomousDriving}
\input{Application/TravelBehavior}
\input{Application/TrafficSafety}
\input{Application/Emergency}
\input{Application/TrafficForecasting}
\input{Application/TrafficSignalControl}
\input{Application/TrafficSimulation}

\input{Application/RoadNetwork}

\subsection{Transportation Research}
\label{subsection:transportation}
\quad Transportation encompasses the broad movement of people, goods, and services across land, air, and water. Transportation research involves designing, planning, and management of infrastructure, policies, and systems to ensure efficient and sustainable mobility. Unlike traffic, which focuses on localized operational dynamics, transportation research analyzes large-scale networks, long-term planning, and multimodal integration. In this section, we examine how LLMs help enhance operational efficiency, optimize resource allocation, and provide intelligent solutions to tackle the multifaceted challenges of transportation systems.

\input{Application/Aviation}
\input{Application/Maritime}
\input{Application/LogisticsSCM}
\input{Application/PublicTransport}
\input{Application/Micellaneous}

In conclusion, the surveyed literature highlights the versatility, effectiveness, and remarkable potential of LLMs in transforming various aspects of the transportation domain. As research in this field progresses, it is evident that the integration of LLMs with domain-specific models and frameworks will play a crucial role in making traffic and transportation systems more intelligent, efficient, and adaptable.

%% file: Application/ArticleSelection.tex
\subsection{Criteria for Literature Search and Selection}
\label{subsec:articleselection}
%\quad\;To comprehensively explore the application of Large Language Models (LLMs) in traffic and transportation studies, a systematic approach to article selection was employed. This methodology ensures that the review captures relevant, high-quality research while maintaining transparency and reproducibility.
This section introduces our literature survey synthesis method and search strategy.

\subsubsection{Synthesis Method}

We employ a scoping review approach, which is particularly suited for mapping the existing literature on a broad topic and identifying key concepts, theories, sources of evidence, and research gaps \autocite{pham2014scoping}. Unlike systematic reviews which aim to answer specific research questions through a detailed appraisal of evidence, scoping reviews are designed to provide an overview of the available literature regardless of study quality. 

The selection of a scoping review methodology is motivated by several key considerations of LLM applications in traffic and transportation research. First, the field encompasses a broad spectrum of research objectives and problem definitions, making a comprehensive systematic review challenging. Second, the diverse applications have led to highly specialized and customized implementations of LLMs. Third, the absence of standardized performance metrics and universal benchmarks complicates direct comparisons across studies, with some research utilizing traditional methods as benchmarks while others lack comparative frameworks entirely. Furthermore, the cross-disciplinary nature of the topic and its strong industry applications have resulted in research dissemination across various platforms, including academic journals, preprint archives, conference proceedings, and book chapters. Given these characteristics, a scoping review methodology enables us to effectively categorize this diverse body of literature into coherent subtopics, analyze methodological approaches, compile relevant statistics, and identify critical research gaps in the field.

%This approach is appropriate for the current review due to the evolving nature of LLM applications in traffic and transportation, the diverse range of study methodologies, and the broad spectrum of application areas within the field.

\subsubsection{Search Strategy}

A comprehensive literature search was conducted using the Web of Science database, selected for its extensive coverage of peer-reviewed journals, conference proceedings, and industry reports relevant to traffic and transportation research. The search was performed in November 2024, ensuring the inclusion of the most recent advancements and trends in the application of LLMs.

To identify relevant studies, a systematic search was conducted using the Web of Science database. The selection criteria were defined using the following Boolean string:
TS=(``Prompt Engineering'' OR ``Large Language Model'' OR ``Vision Language Model'' OR ``GPT'') AND (TS=(``Autonomous driving'' OR ``Scenario generation'' OR ``Safety'' OR ``Traffic'' OR ``Traffic Forecast'' OR ``Travel Behavior'' OR ``Traffic and Signal Control'' OR ``Pollution'' OR ``Transportation Emission'' OR ``Sustainable Transport'' OR ``ITS'' OR ``Intelligent Transport System'' OR ``Shared Mobility'' OR ``Emergency Evacuation'' OR ``Emergency Response'' OR ``Traffic Simulation'' OR ``Pedestrian'' OR ``Modular Vehicle'' OR ``Vehicle'' OR ``Modality'' OR ``Traffic Accident'' OR ``Transport'' OR ``Logistics'')). We do not restrict the publication period in our search.

3,122 studies are identified from our initial search using the Web of Science database. We further set the categories to only include ``Transportation,'' ``Industrial Engineering,'' ``Management,'' ``Engineering Multidisciplinary,'' and ``Computer Science.'' This further reduces our paper number to 1,187. Afterward, we manually filter papers according to the following criteria:

\textbf{Relevance}: Studies must focus on the application of LLMs or related language models in traffic and transportation contexts.

\textbf{Publication Type}: Peer-reviewed journal articles, conference papers, good-quality pre-prints, and reputable industry reports.

\textbf{Timeframe}: Studies published up to November 2024 to capture the most recent advancements.

\textbf{Empirical Evidence}: Studies should provide empirical data or case studies demonstrating the application and effectiveness of LLMs in transportation tasks.

% We have set the following exclusion criteria:

% \textbf{Irrelevance}: Articles that do not directly apply LLMs to transportation or traffic-related problems. For example, papers that apply LLMs for customer service of logistics companies.

% \textbf{Lack of Empirical Data}: Studies that are purely theoretical without practical applications or case studies.

% \textbf{Duplicate Publications}: Multiple entries of the same study.

% Finally, we included 109 papers in this literature review work. 

Applying those criteria results in 109 studies for subsequent detailed review. We acknowledge that despite our efforts to provide an exhaustive review through detailed search and selection criteria, by no means we can capture every relevant study in this field. Our methodology faces several limitations. Primarily, non-English publications may elude detection in databases like the Web of Science. Additionally, the use of specific jargon can obscure relevant studies; for example, researchers might use terms like ``zero-shot learning'' rather than ``LLM'' in their titles. Similarly, in the fields of traffic and transportation, terms like ``routing'' might be used instead of more general descriptors. The diversity of terminology across these disciplines can lead to omissions. To mitigate these issues, our approach includes leveraging the domain expertise of authors to identify pertinent papers and exploring tracing studies that are cited by the studies that we have selected to uncover further relevant literature.

%% file: Application/AutonomousDriving.tex
 \subsubsection{Autonomous Driving}
AD is a rapidly evolving field that aims to reduce accidents, enhance road safety, and improve mobility \autocite{kuo2023public}. AD systems used to rely on rule-based and optimization-based methods, which provided reliable and interpretable results but struggled with complex, real-world scenarios \autocite{yuan2024evolutionary, aksjonov2021rule, guanetti2018control,dai2020joint}. The advent of learning-based methods, such as RL, marked a significant improvement in handling complexity \autocite{yan2022reinforcement}. However, these methods face challenges in interpretability and in managing the long tail case, i.e., rare, unpredictable events that traditional models often fail to address effectively. %\textcolor{blue}{[Zou: Which the arrows point to in Figure 5 is not very clear. Also, may want to explain a bit more what is the ``long tail case''.]}

\begin{figure}
    \centering
    \includegraphics[trim=6cm 5.5cm 6cm 5.5cm, clip, width=\linewidth]{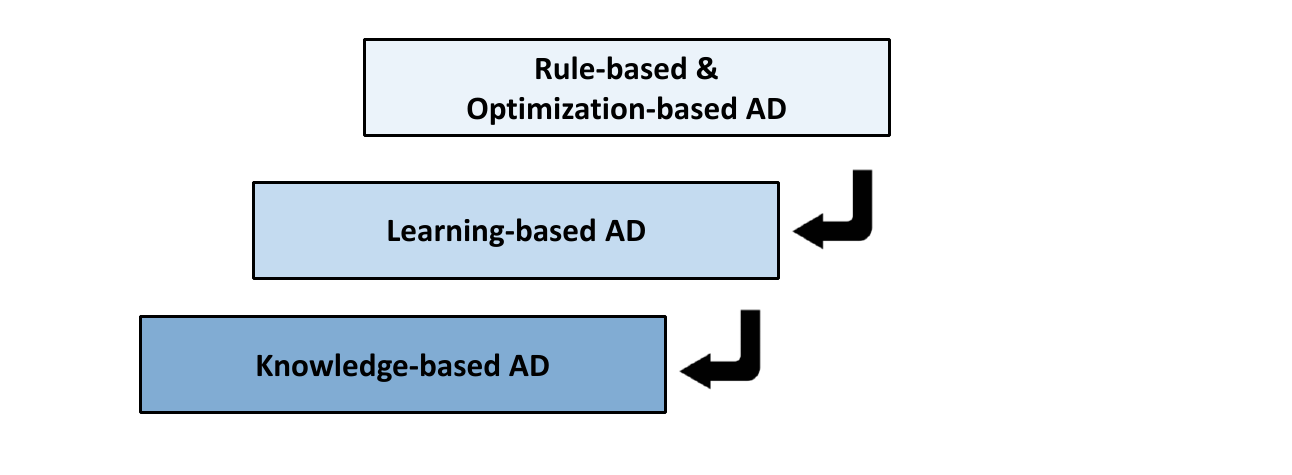}
    \caption{Advancements in Autonomous Driving}
    \label{fig:enter-label2}
\end{figure}

\begin{figure}
    \centering
    \includegraphics[trim=6cm 0cm 6cm 0cm, clip, width=0.8\linewidth]{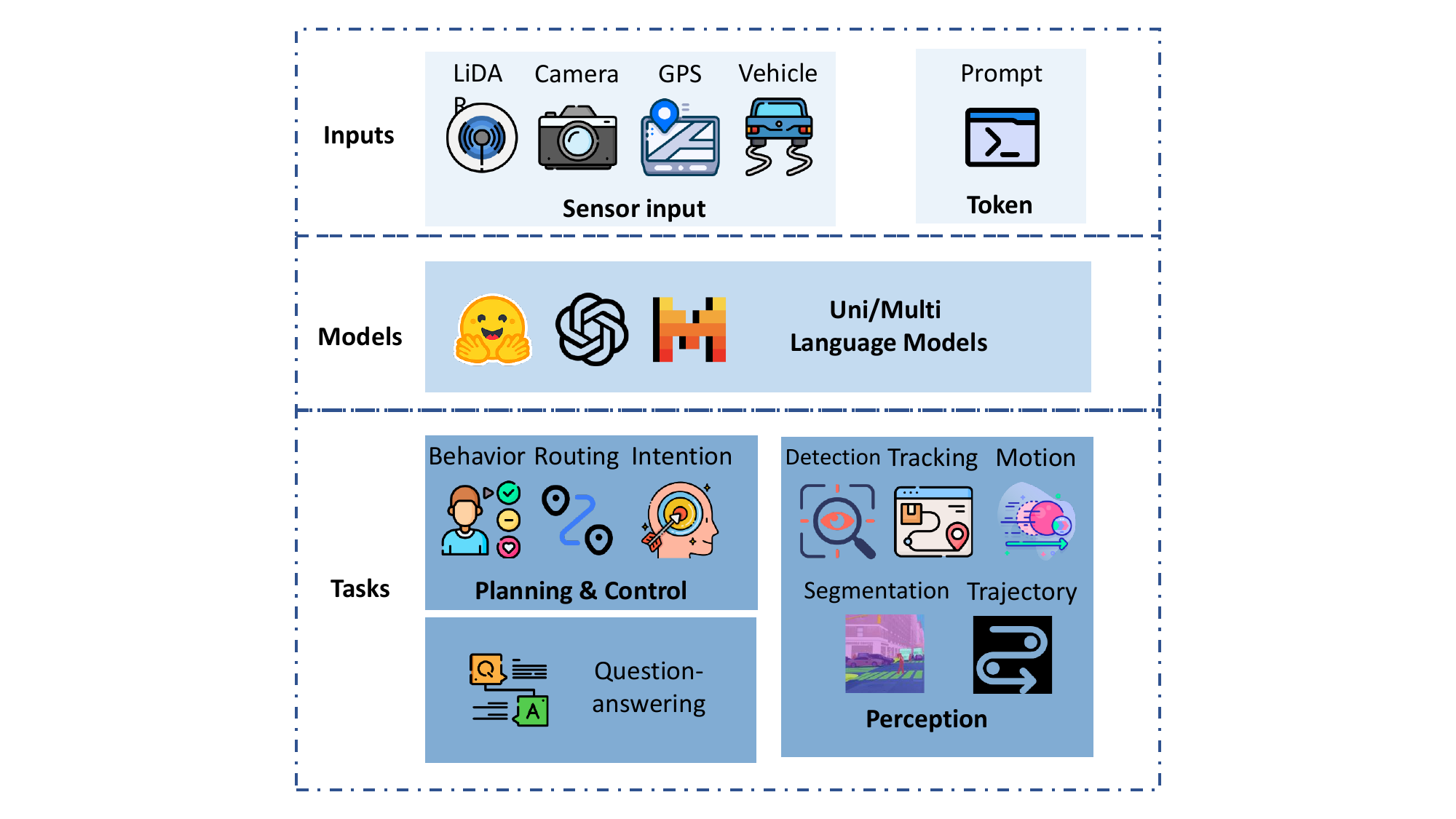}
    \caption{Autonomous Driving Framework \autocite{yang2023llm4drive}}
    \label{fig:AD}
\end{figure}

The introduction of LLMs has enabled better reasoning, knowledge accumulation, and adaptability in AD, and is well-suited for addressing long-tail scenarios. Recent surveys \autocite{yang2023llm4drive, li2024large, cui2024survey, zhou2024vision} highlight their ability to perform a range of tasks, including perception, decision-making, and human-vehicle interaction. The evolution of methods is shown in Figure \ref{fig:enter-label2}. Unlike conventional methods, LLMs excel at integrating multimodal inputs (e.g., visual, sensor, and textual data) and reasoning across diverse scenarios, bridging gaps in traditional AD systems. Key components include inputs, models, and tasks as summarized in Figure \ref{fig:AD}.

State-of-the-art AD frameworks increasingly incorporate LLMs through modular architectures. These systems typically include human instruction, perception, reasoning, reflection, and memory modules to ensure adaptability and safety. For instance, human instruction allows varying levels of intervention, while perception modules gather situational data. Reasoning modules then integrate this information with memory (past experiences) and scenario descriptions to guide decision-making. Reflection evaluates decisions, identifies unsafe outcomes, and updates the memory module for continuous improvement. Frameworks such as \textit{DILU} \autocite{wen2023dilu}, \textit{DRIVELLM} \autocite{cui2023drivellm}, and \textit{DriveGPT4} \autocite{xu2024drivegpt4} exemplify this structure.

Our review focuses on understanding the integration of LLMs into four core components of AD -- perception, decision-making, trajectory planning, and vehicle control. Each component plays a fundamental role in enabling AVs to navigate complex traffic environments safely and efficiently. %In addition, this section further introduces benchmarks and evaluation criteria, which are important for the learning process of autonomous driving. \textcolor{blue}{[Zou: Depending on whether we want to keep the subsection, the last sentence may be deleted.]}

% \begin{figure}
%     \centering
%     \includegraphics[trim=5cm 2cm 5cm 2cm, clip, width=0.7\linewidth]{Application/Figures/longtail.pdf}
%     \caption{Human Driving vs. Autonomous Driving}
%     \label{fig:enter-label1}
% \end{figure}

\paragraph*{Perception}
\quad Perception is the foundation of AD, enabling vehicles to interpret their surroundings and make informed decisions. Perception systems use sensors like LiDAR, cameras, and radar to perform tasks such as object detection, environmental mapping, and situational awareness, ensuring safe operation in dynamic traffic environments.

Traditional rule-based and deep learning (DL)-based methods are effective in structured scenarios but struggle with real-world complexities. They often fail to handle rare or unpredictable events, lack contextual understanding (e.g., why pedestrians gather at a crosswalk), and require extensive retraining to adapt to new situations. LLMs have the ability to process multimodal data, which enables a richer understanding of complex scenarios, better generalization, and adaptation with minimal retraining \autocite{wang2023empowering}. This makes LLMs uniquely suited to enhance perception capabilities in AD.

The first stream of the relevant literature in this field directly applies existing LVLMs with ZSL for recognizing rare and unexpected scenarios. \cite{li2024automated} creates a benchmark named \textit{CODA-LM} for assessing the quality of perception in AD with different large vision-language models. In their benchmark, there are three foundational aspects of perception tasks, including general perception, regional perception, and driving suggestions. Unlike prior works that rely on rigid, natural image-based datasets (e.g., SEED-Bench, MMBench), CODA-LM integrates a hierarchical multimodal framework, combining textual and visual data to assess models' reasoning and decision-making in dynamic driving environments. This proposed benchmark enables a fair comparison between current and future versions of LLMs in AD. The second paper in this area, \cite{li2024UnstrPromptLargeLanguage} introduces the UnstrPrompt model, which leverages CLIP to independently process image and text features. By utilizing zero-shot semantic segmentation, the model accurately identified drivable areas without requiring paired datasets.

Despite these advancements, current LVLMs like GPT-4V excel in analyzing individual images but lack temporal reasoning and struggle with 3D spatial understanding. For instance, GPT-4V achieves only 50\% accuracy in predicting vehicle behaviors, such as moving forward or backward, and performs poorly in identifying speeding vehicles \autocite{sreeram2024probing, wen2024road}. To address these limitations, \cite{wang2024omnidrive} propose OmniDrive, a 3D multimodal architecture which converts visual inputs into 3D representations, to significantly enhance spatial reasoning and decision-making. By incorporating 3D information, OmniDrive enables vehicles to better interpret spatial relationships and navigate complex environments, marking a crucial step toward more robust and intelligent perception systems.

The second stream of the literature focuses on understanding and predicting the motion of surrounding vehicles. Traditional methods often struggle to model the complex interactions between traffic actors. To address this gap, \cite{lan2024traj} introduce Traj-LLM, a model designed to predict the future trajectories of dynamic traffic actors. Traj-LLM leverages observed past trajectories and semantics of surrounding vehicles to predict their future motion without the need for explicit prompt engineering. This approach marks a significant departure from heuristic methods, achieving 7\% to 30\% improvements over several trajectory prediction benchmarks.

\paragraph*{Decision-Making using LLM}
\quad Decision-making is concerned with the selection of appropriate actions based on input from the perception system. This process entails analyzing the environment, predicting potential outcomes, and determining the optimal course of action while considering factors such as safety, legality, efficiency, and passenger comfort \autocite{azarafza2024hybrid}. 

A major advancement brought by LLMs in decision-making is the ability to facilitate natural language interaction. Drivers can issue conversational commands, such as ``merge into the left lane'' or ``adjust speed for the vehicle ahead,'' and the LLM interprets these commands while analyzing perception data such as vehicle speed, localization, and in-cabin monitoring information. Based on this data, the system determines the feasibility of the action and communicates its reasoning back to the driver in natural language. This capability not only makes decision-making more intuitive for passengers but also enhances situational transparency by explaining why a particular action is safe or unsafe \autocite{cui2023drivellm,cui2024drive, xu2024drivegpt4}. By integrating LLMs into the decision-making pipeline, autonomous systems are able to provide passengers with real-time feedback and contextual insights, fostering trust in their operations.

LLM-driven decision-making in autonomous vehicles can address both high-level and low-level actions, which represent two distinct streams of research. High-level actions involve strategic decisions such as left turns, right turns, merging, and lane changes. These actions are often managed through a combination of LLM-based reasoning and traditional rule-based systems, where the LLM provides contextual analysis and generates decisions that adhere to traffic rules and safety protocols \autocite{cui2023drivellm}. On the other hand, low-level actions focus on precise vehicle control, such as determining the turning angle or adjusting speed. These actions are approached through end-to-end systems that leverage LLMs trained on general multimodal datasets (e.g., CC3M, WebVid-2M), enabling real-time predictions of control signals directly from video inputs \autocite{xu2024drivegpt4, cui2024drive, wen2023dilu}. By addressing both streams, LLMs bridge the gap between strategic planning and real-time execution, ensuring that the vehicle can respond robustly to dynamic scenarios.

The Memory Module plays a pivotal role in enhancing LLM-driven decision-making by enabling the system to learn from past experiences and continuously improve. This module stores scene descriptions and reasoning processes, allowing the system to recall and apply solutions to long-tail problems—rare or complex scenarios that traditional systems struggle to address \autocite{wen2023dilu, fu2024drive}. Additionally, the Memory Module records driver preferences, historical actions, maps, and laws, which facilitates personalized decision-making and ensures that the system adapts to individual user behaviors over time. For instance, a driver’s typical preferences for merging speeds or following distances can influence how the system interprets and executes commands. Research shows that incorporating both successful experiences and corrected unsafe decisions into the Memory Module significantly enhances the system’s ability to learn from mistakes, improving safety and adaptability \autocite{wen2023dilu, fu2024drive}.

Another critical contribution of LLMs to decision-making is their ability to handle zero-shot reasoning, which enables them to manage novel or unfamiliar situations without prior exposure. By leveraging their broad training on diverse datasets, LLMs can reason through complex scenarios that involve unusual traffic patterns or unexpected obstacles \autocite{cui2024drive}. For domain-specific challenges, FT the model with specialized datasets further ensures optimal performance in specific contexts, such as urban intersections or highways \autocite{xu2024drivegpt4}. This dual capability allows LLMs to adapt to dynamic environments while maintaining a high level of reliability.

Safety and interpretability are foundational to decision-making in autonomous driving, and LLMs address these aspects effectively. For instance, OpenAI’s GPT-4 has been used as a safety-checking LLM, ensuring that passenger instructions comply with traffic rules and safety protocols \autocite{cui2023drivellm}. Beyond safety, LLMs enhance interpretability by providing clear, natural language explanations for their decisions. They articulate why certain actions -- such as overtaking or merging -- are feasible or not, considering factors like road conditions, traffic density, and vehicle speeds. This transparency is critical in building user trust, especially in safety-critical systems, as passengers can understand the reasoning behind the vehicle’s actions \autocite{xu2024drivegpt4, fu2024drive}.

Despite their strengths, LLMs are not without limitations. A common issue across the literature is hallucination, where the model generates inaccurate or irrelevant outputs. This highlights the importance of robust safety mechanisms and extensive FT to mitigate such errors.

\paragraph*{Trajectory Planning using LLM}
\quad Trajectory planning enables vehicles not only to make decisions but also to execute them safely, efficiently, and comfortably. Traditionally, vehicle motion planners have relied on heuristic methods to plan driving trajectories. While these approaches are effective in familiar, structured environments, they often fail in novel or complex scenarios, where adaptability and generalization are crucial. This limitation has spurred a growing interest in leveraging LLMs to enhance trajectory planning.

One of the primary challenges in trajectory planning is generating accurate and adaptable trajectories for the ego-vehicle, which is the vehicle for which the sensors, decision-making algorithms, and driving actions are being studied or implemented. Existing heuristic methods often lack the flexibility to handle diverse and dynamic driving conditions. To address this, \cite{mao2023gpt} introduces \textit{GPT-Driver}, an innovative approach that uses language descriptions of coordinate positions to generate driving trajectories. By framing trajectory generation as a language modeling task, GPT-Driver eliminates the reliance on handcrafted heuristics, offering a more generalizable solution for ego-vehicle trajectory planning.

However, trajectory planning does not exist in isolation. It is part of a larger closed-loop system where real-world decisions are continuously refined based on feedback. Recognizing this, \cite{fu2024limsim++} propose LimSim++, a closed-loop simulation platform for deploying multimodal large language models (MLLMs) in AD. LimSim++ takes ego-vehicle trajectory planning to the next level by converting decisions from MLLMs into trajectories and assessing their performance through a built-in evaluation module. The platform introduces a reflection and memory module to iteratively improve decision-making accuracy, enabling MLLMs to learn from past mistakes. This refined reasoning is then used as few-shot instances to further enhance future decision-making, creating a feedback loop that traditional heuristic methods cannot achieve.

Building on the idea of modeling interactions, \cite{xia2024language} propose InteractTraj, a method specifically designed to generate interactive traffic trajectories. Unlike Traj-LLM, which focuses on individual actor predictions, InteractTraj emphasizes the interactions between multiple traffic actors, capturing the dynamic interplay that occurs in real-world driving scenarios. This approach leads to a 16\% improvement over baseline methods, showcasing the importance of interaction modeling in trajectory prediction. 

The future of trajectory planning lies in the ability to generalize across diverse scenarios, anticipate interactions with surrounding vehicles, and continuously improve through feedback. The integration of language models, as demonstrated by these works, offers a promising path forward for AD innovation.

\paragraph*{Vehicle Control using LLM}
\quad Vehicle control is the critical step in autonomous driving, translating high-level decisions into physical actions such as steering, acceleration, and braking. These actions must be precise and smooth to ensure safety and user comfort. However, traditional control systems often rely on rigid, rule-based methods, which can struggle with dynamic or complex scenarios, leading to erratic or uncomfortable vehicle behavior. This challenge has become particularly relevant as LLM-controlled vehicles emerge, where the accuracy of decision-making directly impacts the smoothness and safety of vehicle maneuvers. Recent research has begun to explore the potential of LLMs to address these challenges by introducing more adaptable, human-aligned control systems.

To align vehicle control with human driving preferences, researchers have turned to reinforcement learning from human feedback (RLHF) as a key strategy. RLHF enables LLMs to adapt their outputs based on human evaluation, allowing them to fine-tune steering, speed, and other control parameters in a way that mimics human behavior. Building on this, \cite{cui2024drive} leverages GPT-4 to incorporate direct human input into vehicle control. By integrating data from perception, localization, and in-cabin monitoring, their system refines decision-making accuracy in real time, improving adaptability and creating a smoother, more personalized driving experience. 
%This approach prioritizes user comfort and safety, as demonstrated by \cite{stiennon2020learning} and \cite{rafailov2024direct}, marking a shift from static control systems to dynamically learning models. 

In addition to aligning with human preferences, efforts are also made toward enabling LLMs to perform end-to-end driving tasks with minimal or no human intervention. Traditional vehicle control systems use modular pipelines where perception, planning, and control are separate components, which often leads to inefficiencies or errors in decision-making. In contrast, LLMs offer the potential to unify these processes by combining reasoning and perception into a single framework. For example, \cite{chen2024driving} propose a driving question-answering framework that uses LLMs to generate accurate control actions. By framing control tasks as reasoning problems, their system determines appropriate vehicle maneuvers while maintaining transparency, allowing users to understand the rationale behind each action. This marks a significant step toward autonomous and explainable vehicle control systems.

By combining human feedback, real-time adaptability, and reasoning-driven autonomy, these advancements pave the way for safer, more efficient, and explainable autonomous driving systems. A full list of literature surveyed is presented in Table \ref{Table:Autonomous}.

\begin{table}[htbp]
\tiny
\centering
\caption{LLMs in Autonomous Driving}
\label{Table:Autonomous}
\begin{threeparttable}
\begin{tabular}{>{\raggedright}p{2cm} l>{\raggedright}p{2cm} >{\raggedright}p{2.2cm} >{\raggedright}p{2.2cm} >{\raggedright}p{2cm} >{\raggedright\arraybackslash}p{2.2cm}}
\toprule
\rowcolor{gray!20}
\textbf{Literature}  & \textbf{Model Name} & \textbf{Model Backbone} & \textbf{Input} & \textbf{Output} & \textbf{Modality} & \textbf{Task} \\
\midrule
\cite{azarafza2024hybrid} & - & GPT-4 & Road images & Driving action & Image & Driving decision making \\
\cite{chen2024driving} & Driving with LLMs & GPT-3.5, LLaMA-7b & Driving scenario description, object & Driving action & Text & Driving decision making \\
\cite{cui2023drivellm} & DriveLLM & GPT-3.5 & Image, location, vehicle state, weather, and experience & Driving action & Text and image & Perception and decision making \\
\cite{cui2023drivellm} & - & GPT-4 & - & - & - & - \\
\cite{cui2024drive} & - & GPT-4 & Driver command, perception & Trajectory planning & Image and sensor data & Prediction, decision making and motion planning \\
\cite{fu2024limsim++} & LimSim++ & GPT-3.5, GPT-4, GPT-4V & Road image and video & Trajectory & Image and video & Trajectory planning \\
\cite{fu2024drive} &  Drive Like a Human&  GPT-3.5&  Road environment, vehicle state&  Driving decisions & Text&  Driving decision making with interpretation and self-reflection\\
\cite{hu2023enabling} & - & ChatGLM, CLIP, BLIP-2 & Road images, human instruction & Navigation reasoning & Image and text & Vehicle navigation \\
\cite{hussien2025rag} & GPT-4  & Pedestrian and vehicle motion information & Predicted pedestrian actions and reason & Text and data & Behavior understanding of road users & JAAD, PSI and highD \\
\cite{lan2024traj} & Traj-llm & LLM & Historical vehicle trajectories, lane information, surrounding information & Trajectory & Spatial-temporal state data & Trajectory prediction \\
\cite{li2024miningllm} & MiningLLM & GPT-4 & Road images & Mining operation & Image and text & Collaborative operations for mining and safety assessment \\
\cite{li2024automated} &  -&  LLaVA-Llama-3-8B-v1.1&  Vehicle and road information&  Predictions of road entities, descriptions of corner cases, and driving suggestions&  Image and text&  Corner case detection in autonomous driving\\
\cite{li2024UnstrPromptLargeLanguage} & GPT-3.5 with CLIP & Road images and texts describing scene & Segmentation masks with improved accuracy & Image and text & Semantic segmentation and language-guided perception & Talk2Car dataset \\
\cite{mao2023gpt} & GPT-driver & GPT-3.5 & Object images, ego-states, historical trajectories & Predicted trajectories & Text and data & Motion planning, safety compliance, interpreting complex driving scenario and reasoning \\
\cite{pan2024vlp} & VLP & - & Images of other vehicles & Action prediction of other vehicles & Image and text & Perception, prediction and motion planning \\
%\cite{rafailov2024direct} &  double check &  &  &  &  &  \\
%\cite{stiennon2020learning} &  double check &  &  &  &  &  \\
\cite{sreeram2024probing} & - & GPT-4V & Sequence of vehicle images & Predicted action of other vehicles & Image, text & Scene understanding \\
\cite{tanahashi2023evaluation} & - & LLaMA 2, GPT 3.5, GPT 4 & Road images & Driving decisions & Image and text & Spatial recognition and traffic rules compliance in decision making\\
\cite{tian2023vistagpt} & VistaGPT & - & Road images, vehicular states and navigation command & Driving task, waypoints & Image and scalar data & End-to-end autonomous driving system composition and optimization \\
\cite{wang2023chatgpt} & - & GPT-3.5 & Road images and human intention & Driving commands (e.g., steering) and trajectories & Image and text & Co-pilot \\
\cite{wang2023empowering} &  -&  GPT-4&  Scene description and sensor data&  driving task and explanation&  Text&  enhance safety and interpretability in autonomous driving\\
\cite{wang2024bevgpt} & BEVGPT & GPT & Bird eye view image & Driving decisions & Image & Prediction, decision making and motion planning \\
\cite{wang2024drive} & Drive as veteran & LLaMA-7B & Scenario description & Driving tasks & Text & Driving task \\
\cite{wang2024omnidrive} & OmniDrive & GPT-4 & Multiview image & Decision making, planning & Image & Action prediction \\
\cite{wen2023dilu} & DiLu & GPT-3.5 and GPT4.0 & System prompts, textual description, and few-shot experience & Driving action & Text & Decision making \\
\cite{wen2024road} & - & GPT-4V & Image & Driving action & Image & Action prediction \\
\cite{xia2024language} & InteractTraj & GPT-4 & Scenario description & Predicted trajectories & Text and data & Trajectory prediction \\
\cite{xu2024drivegpt4} & Drivegpt4 & GPT-4 & Road videos & Driving decisions & Video & Driving decision making \\
\bottomrule
\end{tabular}
\begin{tablenotes}[flushleft]
    \item EV: Electric Vehicle
    \item JAAD: Joint Attention for Autonomous Driving, PSI: Pedestrian Situated Intent
\end{tablenotes}
\end{threeparttable}
\end{table}

%% file: Application/TravelBehavior.tex
\subsubsection{LLMs in Travel Behavior and Mobility Prediction}
\label{subsubsection}
%Travel behavior is a critical aspect of transportation planning. 
Travel behavior is a broad concept encompassing many attributes, including mode choice, travel purpose, destination selection, departure time, route choice, and travel frequency. Understanding these factors enables agencies to enhance transit services and forecast short-term travel conditions while informing long-term transportation infrastructure investments, policy development, and land use planning. For individuals, mode choice predictions integrated into map applications support informed and efficient travel decisions. 

Traditional methods like decision trees, support vector machines, and regression models rely on predefined features and structured datasets \autocite{yin2024examining, koushik2020machine}. While effective in certain contexts, these methods struggle with the complexity of human travel behavior, especially when dealing with contextual relationships or unstructured data like text and images. They often require extensive feature engineering and lack generalizability in dynamic environments. In contrast, LLMs introduce a shift to travel behavior prediction by integrating both structured and unstructured data. Their ability to process multimodal information and learn contextual relationships enables more precise insights and predictions. LLMs excel in areas such as mode choice prediction, human mobility modeling, and analyzing pedestrian and driver behavior, addressing the limitations of traditional approaches while unlocking new possibilities for intelligent transportation systems.

\paragraph*{Human Mobility Prediction}
\quad Realistic human mobility prediction serves as key components of many research fields, including transportation, disaster management, and urban planning. Traditional methods of human mobility prediction rely primarily on mathematical modeling, statistical analysis, and classical Machine Learning approaches. These methods are rooted in well-established theories of travel behavior and decision-making, often focusing on structured data inputs like socioeconomic characteristics, travel times, and costs. Among them, the most widely studied methods include gravity model \autocite{pappalardo2016human}, discrete choice model \autocite{bierlaire1998discrete}, activity-based models \autocite{bowman2001activity}, agent-based models \autocite{maggi2016understanding} and machine learning (ML) methods \autocite{toch2019analyzing}. These methods for predicting human mobility are based on structured data, extensive feature engineering, and domain-specific assumptions. However, as urban mobility becomes more dynamic and diverse, these methods struggle to handle complex semantic information and unstructured inputs, such as social media updates and sensor-generated textual description, highlighting the need for more adaptable approaches.
% \begin{itemize}
%     \item \textbf{Gravity Model:} Use population and geographic distance to estimate travel flow between regions. It is simple and interpretable \autocite{pappalardo2016human}. 
%     \item \textbf{Discrete Choice Model:} Multinomial Logit (MNL) model, Nested Logit, and Mixed Logit models assume that individuals make rational decisions by maximizing utility. It is simple and interpretable \autocite{bierlaire1998discrete}. 
%     \item \textbf{Activity-based Models:} These models incorporate a detailed representation of human activities, schedules, and constraints. It considers time use and interdependencies between trips but computationally demanding \autocite{bowman2001activity}.
%     \item \textbf{Agent-Based Models:} Agent-based models simulate the behavior of individual agents, incorporating rules for decision-making and interactions. It is highly flexible and detailed, but very computationally demanding \autocite{maggi2016understanding}.
%     \item \textbf{Machine Learning Methods:} Decision trees, random forests, and support vector machines, have been applied to improve mobility prediction accuracy \autocite{toch2019analyzing}.
% \end{itemize}

To address these challenges, researchers have applied NLP techniques to process unstructured data like textual descriptions and contextual information. NLP models convert text into tokens, enabling the integration of numerical and textual data for mobility prediction. For example, the ``SHIFT'' framework \cite{xue2021TranslatingHumanMobility} incorporates contextual information such as location categories, weather, and temporal data into a unified natural language model, outperforming traditional methods like ARIMA and LSTMs in accuracy and flexibility. Similarly, a BERT-based Masked Language Model \autocite{yang2024ApplyingMaskedLanguage} is used to predict transportation modes using two million trip records from major Chinese cities, demonstrating its superiority over the traditional MNL approach.

Despite their successes, NLP methods like SHIFT and BERT-based models rely on predefined templates or rigid frameworks for mobility-to-language transformation, limiting their adaptability to novel scenarios or irregular patterns. To overcome these limitations, researchers have increasingly turned to LLMs, which offer greater flexibility, scalability, and reasoning capabilities. Through our literature search of LLMs applied in this area, two streams of research are identified. The first stream explores the potential of using LLMs to process and generate semantic-rich information on mobility patterns similar to the study of NLP. The second stream applies LLMs to make explainable trajectory modeling at the individual level. In both application streams, semantic reasoning, few-shot learning and explanability capabilities of LLMs are used.

%While NLP methods like SHIFT and BERT-based models excel in integrating diverse data sources, they often rely on predefined templates or rigid frameworks for mobility-to-language transformation. This reliance limits their adaptability to novel contexts, such as unforeseen urban scenarios or irregular mobility patterns. To overcome these constraints, researchers have increasingly turned to LLMs, which offer enhanced reasoning capabilities, flexibility, and scalability.

In the first stream of research, scholars have applied LLMs to make travel mode choice analysis and predictions. Unlike NLP-based methods that rely on manual annotation or fixed models for specific objectives, LLMs allow for automated inference of travel modes, sentiment analysis, and summarization of reasons behind sentiments in a unified framework. A study by \cite{ruan2024twitter} has explored the use of LLMs to analyze social media data from tweets for summarization of reasons behind their travel mode choices and has shown good performance. 

Additional studies have looked into using LLMs for travel mode prediction. By leveraging prompt engineering, LLMs can transform travel attributes—such as time, cost, and individual preferences—into textual inputs, enabling zero-shot predictions without requiring training data. \cite{mo2023large} demonstrate this approach, showing that LLMs could achieve competitive accuracy compared to traditional MNL and random forest models. However, this study also highlights challenges including reasoning errors and hallucinations, emphasizes the need for improved prompt design. Building on this, \cite{liu2024can} introduce a few-shot learning approach, where a small number of training samples were used to guide the model's understanding of bounded rationality in human decision-making. This approach significantly improves the alignment between model predictions and observed behaviors, underscoring the importance of contextual learning in mobility analysis.

 %showcases the application of LLMs in travel choice prediction under metro delays. \textcolor{blue}{[Zou: It is unclear why this sentence is placed in this paragraph?]} 
 Traditional methods like statistical models, ML, and survey-based approaches often struggle with sparse, imbalanced datasets and fail to adequately capture passenger heterogeneity or the complexities of travel behaviors during delays. To address these challenges, DelayPTC-LLM \autocite{chen2024delayptc} leverages advanced reasoning, NSL, and NLP capabilities to analyze delay logs and passenger behavioral data. By integrating prompt engineering and the CoT strategy, DelayPTC-LLM not only predicts travel choices but also explains its reasoning process. Experimental results show that DelayPTC-LLM outperforms traditional models and even standalone LLMs without CoT.

The second stream of research focuses on generating personal mobility trajectories by leveraging historical data and psychological factors, such as attitudes, subjective norms, and perceived behavioral control. For instance, the Chain-of-Planned-Behaviour (CoPB) framework incorporates psychological insights to enable step-by-step reasoning for mobility intention generation, significantly reducing error rates and token consumption compared to pure LLM methods \autocite{shao2024copb}. Similarly, the LLMob framework integrates self-consistency evaluation to align trajectories with historical data and employs retrieval-augmented strategies to infer motivations from temporal and contextual information \autocite{wang2024LargeLanguageModels}. LLMob not only generates semantically rich activity patterns but also outperforms state-of-the-art models, such as DeepMove and DiffTraj, by adapting effectively to external factors like pandemics.

The application of LLMs in mobility prediction marks an evolution in the field. Unlike traditional methods, which rely on structured data and domain-specific assumptions, LLMs excel in processing unstructured inputs, integrating diverse contexts, and generating interpretable outputs. Their semantic reasoning and FSL capabilities enable them to adapt to new contexts, making them highly versatile for tasks ranging from travel mode choice prediction to trajectory generation.

\paragraph*{Pedestrian Motion Prediction}
\quad Pedestrians are the most vulnerable road users. Every day, people walk through urban environments, sharing spaces with motor vehicles at all times and conditions. These interactions expose pedestrians to significant traffic risks. Thus, understanding human behavior is crucial to creating safe and intelligent transportation systems in dynamic urban settings. Traditional methods like rule-based systems and statistical models have been used to study pedestrian and driver behavior, but often fail to capture the complexity and variability of real-world scenarios.  Innovative solutions to accurately model complex pedestrian-vehicle interactions are desired to enhance pedestrian safety.

NLP techniques have been used for structured textual data analysis in pedestrian safety-related studies. For example, models like BERT have successfully classified pedestrian maneuver types \autocite{das2023ClassifyingPedestrianManeuver}, showcasing the effectiveness of pre-trained transformers. While NLP provides a foundation for integrating language models into transportation research, LLMs surpass NLP with broader reasoning and adaptability. In pedestrian behavior modeling, LLMs enable more dynamic and realistic simulations. Traditional trajectory planning models, such as kinematic simulations, have often failed to capture the variability of real human movement. \cite{ramesh2024WalkTalkLLMDriven} demonstrate how the Flan-T5 model could simulate realistic pedestrian movements by leveraging the reasoning capabilities of LLMs to generate human-like movements. This approach provides a more flexible and reliable foundation for safety assessments and autonomous system development, representing a step forward from static, rule-based simulations.

\paragraph*{Driver Behavior Analysis}
% \quad\; Understanding human behavior is essential for designing safe and intelligent transport systems in dynamic urban environments. Traditional methods, such as rule-based systems and statistical models, have long been used to analyze pedestrian and driver behavior. However, these approaches often fall short in capturing the complexity, variability, and multimodal nature of real-world scenarios. Recent advancements in LLMs have opened new possibilities, enabling more robust and data-driven solutions to longstanding challenges in transport research.

% NLP techniques have been applied to analyze structured textual data in transport-related tasks. For instance, classification of pedestrian maneuver types has been achieved using models like BERT \autocite{das2023ClassifyingPedestrianManeuver}. These NLP-based approaches demonstrate the ability of pre-trained transformers to process and analyze textual data effectively. However, while NLP has laid the groundwork for integrating language models into transport research, LLMs surpass these methods by offering broader reasoning and greater adaptability. Despite their potential, our literature review suggests that the overlap between NLP-based approaches and LLM applications in this domain remains minimal, leaving significant room for exploration.

%However, our review indicates limited overlap between NLP-based approaches and LLM applications in this field, highlighting opportunities for further exploration.
\quad Driver behavior is a crucial factor in both driving safety and efficiency. From a safety perspective, modeling elements such as driver fatigue and braking performance is essential for reducing risk. In terms of efficiency, decisions regarding cruise speed, lane-changing, and car-following are critical. A better understanding and accurate modeling of these behaviors not only enhances safety and efficiency for human drivers but also strengthens the capabilities of AD systems. In traditional AD, behavior decision-making is typically based on rule-based methods. However, these methods cannot deal with long-tail cases, such as irregular driving behavior. The advent of LLMs shows the potential to address the challenges with powerful generalization and commonsense reasoning abilities, enabling them to infer information from previously unseen scenarios.  \cite{chen2024GenFollowerEnhancingCarFollowing} address these limitations with the GenFollower model, a prompt-based LLM approach designed to predict car-following behaviors while maintaining interpretability -- an essential feature for practical applications. Similarly, \cite{zhang2024IntegratingVisualLarge} develop a visual large language model that combines visual and textual data through an FSL approach to analyze distracted driving patterns. This multimodal integration showcases the versatility of LLMs in combining contextual and visual information to tackle safety-critical challenges.

In summary, LLMs have demonstrated their potential applications in pedestrian and driver behavior modeling by enabling realistic pedestrian trajectory simulations and integrating diverse data types to analyze driver behaviors. A full list of the literature surveyed is presented in Table \ref{Table:TravelBehaviour}.

\begin{table}[htbp]
\footnotesize
\centering
\caption{Travel Behavior Prediction Using LLMs}
\label{Table:TravelBehaviour}
\begin{threeparttable}
\begin{tabular}{>{\raggedright}p{2cm} >{\raggedright}p{2cm} >{\raggedright}p{2.2cm} >{\raggedright}p{2.2cm} >{\raggedright}p{2cm} >{\raggedright}p{2.2cm} >{\raggedright\arraybackslash}p{2.2cm}}
\toprule
\rowcolor{gray!20}
\textbf{Literature} & \textbf{Model Backbone} & \textbf{Input} & \textbf{Output} & \textbf{Modality} & \textbf{Task} & \textbf{Data Source} \\
\midrule
\cite{chen2024ChatGPTpoweredInquirybasedLearning} & ChatGPT & Student prompts & Explanations, code & Text & Inquiry-based learning & Interaction logs \\
\cite{chen2024GenFollowerEnhancingCarFollowing} & GPT-4 & Natural language prompts & Car-following predictions & Text & Driver behavior analysis & None \\
\cite{chen2024delayptc}& GPT-4 & Event logs and passenger travel choice datasets  & Predicted passenger travel choices during metro delays &  Text & Passenger travel choice prediction & Shenzhen Metro AFC data \\
\cite{liang2024ExploringLargeLanguage} & GPT-4 & Event descriptions and mobility data & Mobility patterns & Text & Human mobility prediction & Public events data \\
\cite{liu2024can} & GPT-3.5/GPT-4 & Designed prompts & Predictions, reasoning & Text & Behavioral prediction & Translated dataset variables \\
\cite{mo2023LargeLanguageModels} & GPT-3.5 & Contextual prompts & Mobility predictions & Text & Human mobility prediction & None \\
\cite{ramesh2024WalkTalkLLMDriven} & Flan-T5-Base & Text descriptions & Pedestrian motion trajectories & Text & Pedestrian behavior simulation & Simulation datasets \\
%\cite{richards2019ModellingModeChoice} & Traditional ML & Feature-based data & Mode choice predictions & Text & Mode choice prediction & None \\
\cite{ruan2024twitter} & GPT-3.5/Llama2 & Social media data & Travel mode predictions & Text & Travel mode prediction & Social media platforms \\
\cite{shao2024copb} & LLaMA3-8B & Intention generation prompts & High-quality predictions & Text & Intention generation & Tencent and China Mobile data \\
\cite{wang2024LargeLanguageModels} & GPT-3.5-turbo & Historical data and contextual prompts & Mobility activities & Text & Human mobility prediction & None \\
\cite{zhang2024IntegratingVisualLarge} & LLaMA & Images and text cues & Driver distraction classification & Image and text & Driver behavior analysis & Driver images \\
\cite{zhao2024DriveLLaVAHumanLevelBehavior} & LLaVA& Image-text pairs & Driving behavior strategies & Image and text & Driver decision-making & nuScenes dataset \\
\bottomrule
\end{tabular}
\end{threeparttable}
\end{table}

%\cite{yang2024ApplyingMaskedLanguage} & BERT-based & Text and numerical data & Mode choice predictions & Text & Mode Choice Prediction & Mixed dataset \\

%\cite{das2023ClassifyingPedestrianManeuver} & BERTBASE & Crash narrative text & Pedestrian maneuver classification & Text & Pedestrian Behavior Analysis & Crash narratives \\
%\cite{oliaee2024AutomatingPedestrianCrash} & BERT, RoBERTa & Crash narratives & Crash typology & Text & Pedestrian Safety Analysis & Crash reports \\

%% file: Application/TrafficSafety.tex
\subsubsection{Traffic Safety}
Traffic accidents remain a global challenge, causing approximately 1.2 million deaths and 20-50 million severe injuries annually. Traditional approaches in traffic safety rely on statistical methods for crash frequency and severity analysis, offering essential insights but with limitations. They depend heavily on structured datasets like crash statistics and vehicle counts, leaving unstructured data sources -- such as police reports, incident logs, and social media -- largely untapped. These methods lack the ability to capture nuanced, context-specific information and often fail to model complex interactions among factors like driver behavior, road conditions, and environmental influences, resulting in incomplete understanding and less effective preventive measures \autocite{xu2025enhancing}.

NLP has emerged as a tool to address these challenges by converting unstructured text into structured formats, enabling more comprehensive analysis. For instance, NLP has been used to transform crash reports into spatial data for more accurate accident location detection \autocite{wang2017exploring}. Social media updates, analyzed using NLP techniques, have also been employed for real-time traffic prediction \autocite{sampath2023traffic}. By bridging structured and unstructured data, NLP enhances ITS, enabling more responsive traffic management \autocite{ali2021traffic, wan2020empowering}. LLMs build on NLP’s foundation, offering unique advantages for accident analysis, multimodal integration, and proactive traffic prediction, as explored in the following subsections.

\paragraph*{Accident Analysis}
\quad  While NLP systems have provided some progress through the extraction of structured insights from textual data, their reliance on predefined rules and limited adaptability have constrained their effectiveness. LLMs change how accident analysis is conducted by automating the extraction and interpretation of unstructured textual data, such as accident reports and police logs. For instance, models like TrafficSafetyGPT have shown the ability to process vast volumes of textual accident data, extracting critical details such as causes, severity, and contributing factors with high accuracy \autocite{zheng2023trafficsafetygpt}. By FT general-purpose LLMs with domain-specific knowledge, such as government guidelines and historical crash data, these models streamline the process of data extraction and summarization, significantly reducing human effort and the potential for error. This capability enables researchers and policymakers to uncover deeper insights into accident causes and mechanisms, which were previously inaccessible with traditional methods or simpler NLP systems.

Beyond data extraction, LLMs enable the generation of synthetic scenarios to address the challenges posed by rare, high-impact accidents. Scenario engineering, a novel application of LLMs, allows researchers to simulate diverse crash scenarios, including edge cases such as vehicle trajectory deviations or lane-change conflicts, which are rarely captured in real-world datasets \autocite{Chang2024LLMScenario}. By augmenting existing datasets with these synthetic scenarios, LLMs enhance the robustness of safety research and enable the testing of preventive measures under a wider range of conditions.

LLMs also excel in multidimensional accident analysis by integrating textual and visual inputs. VLMs, which combine the capabilities of LLMs with computer vision, have been applied to analyze crash scenes in unprecedented detail. For example, recent studies have demonstrated how LLMs, paired with techniques like segment extraction and dynamic prompts, can generate granular descriptions of accident contexts, including pedestrian behavior, vehicle dynamics, and environmental conditions \autocite{Xuan_2024_CVPR}. Similarly, V2X-perception architectures integrate data from panoramic multi-camera views and road devices to construct a comprehensive understanding of traffic environments for accidents \autocite{Wang2023AccidentGPT}. These advancements enable a level of detail and context in accident analysis that far exceeds the capabilities of traditional methods or stand-alone NLP systems.

\paragraph*{Proactive Traffic Prediction and Safety Management}
\quad  LLMs have demonstrated a potential in predicting high-risk traffic conditions and accident hotspots. By analyzing historical accident records, real-time traffic updates, and environmental data, these models can identify patterns and correlations that elude traditional methods. For example, GPT-based models have been applied to predict accident types and traffic flow disruptions with remarkable precision, enabling traffic authorities to implement proactive measures such as adjusting traffic signals or deploying resources to high-risk areas \autocite{baumler2024predicting, wang2024real}. These predictions are not only more accurate than those generated by traditional methods but also provide actionable insights for preemptive traffic management.

Furthermore, LLMs enhance traffic crash response planning by prioritizing high-risk scenarios and improving communication with first responders. Techniques such as CoT reasoning and prompt engineering (PE) enable LLMs to analyze crash scenarios and severity outcomes in real time, providing detailed recommendations for response strategies \autocite{Zhen2024LeveragingLLM}. This capability improves the speed and effectiveness of accident responses, minimizing the impact of accidents and saving lives \autocite{Fan2024LearningCrashes}.

In summary, LLMs can provide deeper insights into traffic accidents, have real-time capabilities, and facilitate proactive interventions. LLMs are paving the way for safer and more efficient road systems. A full list of the literature surveyed is presented in Table \ref{table:TravelSafety}.

\begin{table}[htbp]
\footnotesize
\centering
\caption{Travel Safety Analysis Using LLMs}
\label{table:TravelSafety}
\begin{threeparttable}
\begin{tabular}{>{\raggedright}p{2cm} >{\raggedright}p{2cm} >{\raggedright}p{2.2cm} >{\raggedright}p{2.2cm} >{\raggedright}p{2cm} >{\raggedright}p{2.2cm} >{\raggedright\arraybackslash}p{2.2cm}}
\toprule
\rowcolor{gray!20}
\textbf{Literature} & \textbf{Model Backbone} & \textbf{Input} & \textbf{Output} & \textbf{Modality} & \textbf{Task} & \textbf{Data Source} \\
\midrule
\cite{zheng2023trafficsafetygpt} & GPT (version unknown) & Textual accident reports & Accident-based responses & Text & Multisensory safety analysis & CA DMV, NHTSA, OSM \\
\cite{Chang2024LLMScenario} & GPT-4 & Scenario tokens, instruction tokens, experience tokens, etc. from naturalistic driving datasets & Generated driving scenarios (vehicle trajectories, lane changes, and risk analysis) & Multimodal (vehicle trajectories, semantic understanding, and text) & Scenario generation for risky driving conditions & HighD dataset \\
\cite{Xuan_2024_CVPR} & Qwen-VL & Traffic video segments & Traffic safety descriptions (pedestrians, vehicle behavior, road conditions, contextual information) & Text and video & Traffic safety description and analysis & Woven Traffic Safety (WTS) and BDD100K datasets \\
\cite{Wang2023AccidentGPT} & GPT-4V (in V2X) & Mutli-camera images & 3D object detection, BEV maps, trajecotry detection, and safety assessments & Text and video & Accident analysis and prevention & DeepAccident dataset \\
\cite{baumler2024predicting} & BERT & Textual accident descriptions & Accident type & Text &  Classify traffic accidents & German federal states accident dataset \\
\cite{wang2024real} & GPT-4 & Language queries related to traffic conditions & Traffic advisory reports & Text &  Provide traffic advisories & Inductive loop detectors \\
\cite{Zhen2024LeveragingLLM} & GPT-3.5-turbo & Structured tabular data on various traffic crash attributes & Predictions on crash severity & Text and data & Crash severity inference & CrashStats dataset \\
\cite{Fan2024LearningCrashes} & LLaMA-2 & Crash events dataset (general information, infrastructure details, event descriptions, and unit information) & Injury and severity predictions & Text, data and image & Injury and severity predictions and classification & Highway Safety Information System (HSIS) \\
\bottomrule
\end{tabular}
\end{threeparttable}
\end{table}

\label{subsubsection:TrafficSafety}

%% file: Application/Emergency.tex
\subsubsection{Emergency Management}

Transportation systems heavily rely on precise, real-time information for effective disaster response. However, there's a notable gap in quickly obtaining detailed disaster information, such as the extent and location of an event. Traditional tools like remote sensing lack the needed detail. For social media platforms, while they can provide instant data, they often contain excessive irrelevant information. Moreover, traditional data analysis methods struggle to address the complex, multi-dimensional nature of disasters. 

LLMs offer a solution to address the limitations of some traditional approaches in transportation emergency management. Unlike traditional models, LLMs excel in context-sensitive reasoning, multi-modal data integration, and the real-time extraction of actionable information. By leveraging techniques like CoT reasoning and prompt engineering, LLMs can provide transparent, step-by-step analysis and adapt to specific emergency management tasks. In this section, we introduce three streams of research identified in our literature search, namely real-time information processing and decision support, communication and overcoming language barriers, and risk assessment and prediction.

\paragraph*{Real-Time Information Processing and Decision Support}
\quad Effective disaster response hinges on the ability to process and analyze large amounts of fragmented, real-time information -- something human operators often struggle with. LLMs address this gap by synthesizing unstructured data from diverse sources, such as social media posts, news reports, and emergency calls. These sources may contain critical details like the location and type of emergency, which LLMs can extract to provide actionable instructions for dispatchers \autocite{otal2024LLMAssistedCrisisManagement}.

During emergencies, information is noisy and evolves rapidly. LLMs help filter irrelevant data, prioritize critical updates, and create a comprehensive picture of the crisis. This capability enhances collaboration between human responders and AI systems, enabling faster resource mobilization and reducing response delays.

Despite their strengths, LLMs are prone to hallucination -- an unacceptable limitation in high-stakes scenarios. To address this, frameworks like E-KELL integrate knowledge graphs with LLMs, structuring verified emergency-related data for improved reliability. Using a prompt chain mechanism, E-KELL guides LLMs step-by-step through logical reasoning, ensuring outputs align with regulations and factual data \autocite{chen2023EnhancingEmergencyDecisionmaking}. This structured approach significantly reduces hallucinations, making LLMs more dependable for emergency management.

\paragraph*{Communication and Overcoming Language Barriers}
\quad Language diversity in multicultural and multilingual communities poses a significant challenge during emergencies. Traditional solutions, such as pre-trained translation models or static multilingual databases, lack the flexibility to handle nuanced, real-time interactions. LLMs fill this gap with advanced multilingual support and natural language understanding capabilities.

For example, LLMs can translate emergency calls in real time, generate follow-up questions for dispatchers, and adapt messages to the linguistic and cultural context of the affected population \autocite{jiang2024ApplicationsLargeLanguage, otal2024LLMAssistedCrisisManagement}. As such, LLMs act as ``virtual institutional memories,'' learning from past interactions to improve future responses. Unlike rigid translation models, LLMs dynamically adjust to evolving communication needs, ensuring that critical information reaches all individuals during a crisis.

\paragraph*{Risk Assessment and Prediction}
\quad Proactive risk assessment is essential for mitigating the impact of disasters, yet traditional methods relying on static models or historical data often fail to account for real-time changes or integrate diverse datasets. LLMs excel in identifying patterns, analyzing past emergencies, and predicting emerging risks by processing data from sources such as scientific reports, weather forecasts, and community feedback \autocite{jiang2024ApplicationsLargeLanguage}.

For instance, LLMs can dynamically integrate real-time meteorological data with historical patterns to detect evolving risks during natural disasters. This enables emergency managers to anticipate threats and implement mitigation strategies, thereby reducing the disaster impacts. LLMs outperform traditional systems by effectively synthesizing diverse data streams, which are often siloed in conventional approaches \autocite{jiang2024ApplicationsLargeLanguage}.

In summary, LLMs assist emergency response by offering dynamic and intelligent solutions with their powerful ability of multi-source summarization and analysis. A full list of the selected literature is presented in Table \ref{Table:Emergency}.% However, we remark that, there are a number of other papers we have identified in emergency management, but we find them irrelevant to transportation or evacuation. Readers are directed to relevant papers for more information \autocite{gupta2024UtilizingLargeLanguage, Yin2024CrisisSenseLLM}.

\begin{table}[htbp]
\footnotesize
\centering
\caption{Emergency Management Using LLMs}
\label{Table:Emergency}
\begin{threeparttable}
\begin{tabular}{>{\raggedright}p{2cm} >{\raggedright}p{2cm} >{\raggedright}p{2.2cm} >{\raggedright}p{2.2cm} >{\raggedright}p{2cm} >{\raggedright}p{2.2cm} >{\raggedright\arraybackslash}p{2.2cm}}
\toprule
\rowcolor{gray!20}
\textbf{Literature} & \textbf{Model Backbone} & \textbf{Input} & \textbf{Output} & \textbf{Modality} & \textbf{Task} & \textbf{Data Source} \\
\midrule
\cite{chen2023EnhancingEmergencyDecisionmaking} & ChatGLM-6b and GPT-3.5 & Structured queries related to emergency management scenarios & Precise, actionable guidance for decision making & Text & Emergency decision support & Official data in China \\
%\cite{gupta2024UtilizingLargeLanguage} & GPT-3.5 & Emergency management related prompts & Textual response to user's questions & Text & Answering emergency-related questions & Participants questions and feedbacks \\
\cite{jiang2024ApplicationsLargeLanguage} & GPT-3.5 & Textual data from diverse sources (historic, scientific reports, etc.) & Processed insights, predictions, etc. & Text & Real-time information processing & Historic emergency data, scientific literature, real-time situational updates, etc. \\
\cite{otal2024LLMAssistedCrisisManagement} & LLaMA2 & Emergency call data, social media messages, etc. & Classification and interpretation of emergency data & Text and voice & Classification of emergency & Turkey's Emergency-Disaster Messages Dataset \\
\cite{Yin2024CrisisSenseLLM} & LLaMA-2 & Disaster-related social media posts & Multi-label classifications & Text & Extract critical information from disaster-related posts & CrisisBench dataset \\
\bottomrule
\end{tabular}
\end{threeparttable}
\end{table}

%% file: Application/TrafficForecasting.tex
\subsubsection{Traffic Forecasting}
\label{subsubsection:TrafficForecasting}
Traffic forecasting is essential for improving transportation systems by enabling better traffic management, reducing congestion, and optimizing infrastructure efficiency \autocite{liu2024end, ma2015long}. Accurate predictions of traffic support informed decision-making for both transportation agencies and users, whether for route planning or emergency response \autocite{liu2024heterogeneous}. Key metrics such as traffic volume, speed, travel time, and congestion levels can all benefit from the insights provided by traffic data. The application of NLP and LLMs in traffic flow forecasting helps to incorporate textual data and predict spatial-temporal traffic flow series.

Recent advances have seen the integration of NLP into traffic flow prediction by extracting insights from textual data, such as social media and news reports, to complement traditional traffic and weather datasets. For instance, \cite{jin2021trafficbert} integrates large-scale textural road information and weather information with BERT to predict long-term traffic flow. Furthermore, \cite{essien2021deep} and \cite{yan2025multimodal} demonstrate how combining tweet-based information with traffic and weather data improves prediction accuracy. \cite{tsai2022traffic} incorporate social media features into long-term traffic forecasting models. These approaches highlight the potential of NLP to enhance traffic predictions by providing richer, context-aware datasets that address real-world complexities.

While LLMs are generally not designed as a tool for forecasting purposes, the existing research has seen applications in traffic flow prediction as well as reasoning of traffic flow, as described below.

%prediction, recent studies have applied this technique in various applications in this field. Specifically, the application of LLMs in traffic flow forecasting can be categorized into two key areas: traffic flow prediction and traffic flow reasoning.

\begin{table}[htbp]
\footnotesize
\centering
\caption{Traffic Forecasting}
\begin{threeparttable}
\label{tab:trafficForecasting}
% \begin{tabular}{>{\raggedright}p{2cm} >{\raggedright}p{2cm} >{\raggedright}p{2.2cm} >{\raggedright}p{2.2cm} >{\raggedright}p{2cm} >{\raggedright}p{2.2cm} >{\raggedright\arraybackslash}p{2.2cm}}
\begin{tabular}{>{\raggedright}p{56pt} >{\raggedright}p{56pt} >{\raggedright}p{62pt} >{\raggedright}p{62pt} >{\raggedright}p{56pt} >{\raggedright}p{62pt} >{\raggedright\arraybackslash}p{62pt}}
\toprule
\rowcolor{gray!20}
\textbf{Literature} & \textbf{Model Backbone} & \textbf{Input} & \textbf{Output} & \textbf{Modality} & \textbf{Task} & \textbf{Data Source} \\
\midrule
\cite{guo2024explainable} & LLaMA2 & Traffic flow series + graph & Explanation of link traffic flow & Text and data & Traffic flow reasoning & LargeST dataset, OpenStreet, NOAA\\
 \cite{gebre2024ai} & GPT-4 & Traffic flow density + prompt & Explanation of link traffic density & Text and data & Traffic flow reasoning & NGSIM dataset \\ 
 \cite{huang2024enhancing} & GPT (version not mentioned)  & Traffic flow series + semantic information & Future traffic flow & Text and data & Traffic flow prediction & NYC bike share data \\
 \cite{liu2024spatial}& GPT-2, LLaMA2 & Traffic flow series + graph & Future traffic flow & Text and data & Traffic flow prediction & NYCTaxi, CHBike\\
 \cite{ren2024tpllm}& GPT-2 & Traffic flow series + Graph & Future traffic flow & Text and data & Traffic flow prediction & PeMS dataset \\
 \cite{wang2024traffic} & GPT-4 & Network-wide mobility info + prompt & SQL queries + interpretations & Text and data & Traffic flow reasoning & Network-wide mobility database\\
 \cite{ying2024WordsEvaluatingLarge} & 	GPT-4 \& Phi-3-mini & Prompts for writing codes & Python code& Texts & Generate codes for congestion pricing computation& MATSim-generated synthetic travel data\\
  \cite{zhang2024semantic} & GPT-3.5 & Missing traffic data query & Imputed traffic data & Text and data & Traffic data imputation query & PeMS dataset\\
\bottomrule
\end{tabular}
\end{threeparttable}
\end{table}

\paragraph*{Traffic Flow Prediction}
\quad Traffic flow prediction models aim to forecast traffic volumes and speeds across spatial networks and varying time horizons. Compared to NLP models, LLMs provide a unique advantage as they inherently encode real-world knowledge (e.g., ``heavy rain increases taxi demand''), enabling them to generalize to rare or unseen events (e.g., concerts and extreme weather) that lack historical data. LLMs are fed with text information, and their final hidden layer outputs (or pooled outputs) are used as embeddings. These embeddings are fused with historical traffic data and fed into traditional spatiotemporal models (e.g., STGCN and diffusion convolutional recurrent neural network (DCRNN)). Using this method, \cite{huang2024enhancing} demonstrate how LLMs could represent text-based regional information as nodes in a traffic network, enriching the spatial-temporal embeddings and improving the prediction accuracy. 

%The second innovation we identified in this field involves using LoRA fine-tuning techniques

Low-rank adaptation (LoRA) fine-tuning techniques are also used to make LLMs more efficient for traffic flow prediction, even with limited labeled data. \cite{ren2024tpllm} demonstrate that LoRA FT preserves pre-trained knowledge while enhancing the model's ability to extract temporal and spatial patterns, ensuring high prediction accuracy with minimal computational overhead.

%For example, LLMs can embed spatial-temporal relationships by redefining time steps and locations as tokens, allowing the model to learn global patterns across large networks. This approach has been shown to improve regional- and node-level traffic predictions by combining textual information (e.g., event descriptions) with historical traffic flow data. 

%\paragraph*{Traffic data imputation}
%\quad\: Incomplete traffic data is a persistent challenge in Intelligent Transportation Systems (ITS), often caused by sensor malfunctions, communication issues, or data loss. Traditional imputation methods, such as interpolation or matrix completion, fail to capture the rich semantic and spatial-temporal dynamics present in traffic networks. LLMs, with their ability to understand contextual relationships and semantic meaning, offer a more nuanced approach to filling these data gaps.

%By integrating semantic descriptions with spatial-temporal data, LLMs can impute missing traffic information more effectively. \cite{zhang2024semantic} demonstrates how combining Sentence Transformers with GNNs enhances imputation accuracy. This method leverages sensor connectivity and semantic descriptions to infer missing values, outperforming traditional methods by providing a deeper understanding of traffic behavior.

\paragraph*{Traffic Flow Reasoning}
\quad  Beyond prediction and imputation, LLMs excel in reasoning and providing actionable insights for traffic management. Traditional models often focus solely on forecasting traffic conditions, leaving the interpretation of results and decision-making to human experts. LLMs, however, go a step further by enabling contextual understanding, root cause analysis, and actionable recommendations for alleviating traffic issues.

For instance, large-scale traffic flow reasoning can benefit from the interpretive capabilities of LLMs. These models can process human prompts to generate SQL queries, interpret traffic flow predictions, and explain the underlying causes of congestion or bottlenecks. Using FSL and CoT prompts, \cite{guo2024explainable} and \cite{wang2024traffic} demonstrate how LLMs can align traffic flow predictions with natural language explanations, providing interpretability and enhancing trust in model outputs. 

%Similarly, GPT-4 has been used for traffic analysis by generating Chain-of-Thought prompts for SQL queries and natural language interpretations. This enables detailed, step-by-step reasoning that identifies bottlenecks and suggests management strategies to optimize traffic flow \autocite{wang2024traffic}.

Similar approaches can also be applied for human mobility analysis, addressing challenges like integrating unstructured textual data (e.g., concert details or artist popularity) with historical human mobility patterns. By employing CoT prompting, these models generate step-by-step reasoning for human mobility predictions, with improved transparency and interpretability in the case study of public events \autocite{liang2023exploring}.

Another advancement involves interacting physics-informed models with LLMs, allowing users to query specific traffic conditions and receive human-like explanations. \cite{gebre2024ai} use GPT-4 to combine traffic flow reasoning with physical models, enabling more accurate and explainable analyses of traffic density and flow patterns.

%In this way, LLMs not only predict traffic conditions but also provide the reasoning and actionable insights needed for effective traffic management, offering a significant leap beyond traditional methods.

In summary, LLMs advance real-time traffic reasoning and management, by playing three key roles: (i) interpreting human prompts and SQL queries or code to extract relevant data; (ii) extracting temporal and semantic information to capture the correlation between traffic flows; and (iii) interpreting the traffic flow prediction results and suggesting ways to alleviate bottlenecks. Apart from the aforementioned literature, two additional studies are worth noting. \cite{zhang2024semantic} study the use of an LLM for querying the traffic speed imputation system. \cite{ying2024WordsEvaluatingLarge} and \cite{zhang2024semantic} study the decision support aspect of congestion pricing with LLMs. They are not included in the above review as we consider their applications of LLMs indirectly related to the research topic. The key studies in this field are summarized in Table \ref{tab:trafficForecasting}.

%% file: Application/TrafficSignalControl.tex
\subsubsection{Traffic Signal Control}
\label{subsubsection:TrafficSignalControl}
%LLMs are increasingly being integrated into traffic signal control (TSC), especially adaptive traffic control, due to their advanced reasoning capabilities, adaptability, and ability to handle complex urban traffic environments.  

Traditional adaptive traffic signal control (TSC) struggles with flexibility and generalization, particularly in unfamiliar or dynamic scenarios, and thus fails to adapt effectively. For example, rule-based systems, while effective under stable conditions, are not capable of adequately handling sudden changes, such as emergency vehicle arrivals or sensor failures \autocite{pang2024illm_TS4, wang2024llm_TS3}. RL-based methods, although capable of learning from real-time data, can suffer from overfitting to specific conditions or a lack of flexibility when faced with unforeseen circumstances \autocite{greguric2020application, chu2021traffic}.

To overcome these limitations, integrating LLMs like GPT-4 into TSC frameworks is becoming increasingly prevalent. These models enhance decision-making processes by leveraging their extensive knowledge and reasoning capabilities \autocite{pang2024illm_TS4,wang2024llm_TS3}. For instance, the iLLM-TSC (Integration of RL and LLM for TSC) framework combines RL’s capacity for learning traffic control policies from real-time data and making decisions with the LLM’s ability to evaluate these decisions to verify their reasonableness \autocite{pang2024illm_TS4}. This integration not only helps in filling the gaps left by RL models -- such as missing state information or unconsidered events -- but also increases the robustness of TSC systems under diverse conditions like communication degradation \autocite{pang2024illm_TS4,villarreal2023can_TS9}.

Another focus in the field is the adaptation and fine-tuning of LLMs like GPT-4 specifically for TSC tasks. Models like LA-Light \autocite{wang2024llm_TS2} and LightGPT \autocite{lai2023large_TS5} demonstrate frameworks that incorporate LLMs as central agents in their decision-making processes, allowing the traffic signal systems to modify their strategies in real-time based on evolving traffic conditions. These models go beyond generic language processing to directly interact with real-time traffic environments, providing context-aware control strategies \autocite{masri2024leveraging_TS8}. Such specialization is critical because generalist LLMs, while powerful, may not always accurately interpret traffic-specific inputs unless they are trained on highly relevant data.

In conclusion, integrating LLMs into traffic signal control systems addresses the limitations of traditional methods by enhancing flexibility, generalization, and decision-making under dynamic conditions. By combining the strengths of RL and LLMs, frameworks like iLLM-TSC improve robustness and adaptability, even in complex or unforeseen scenarios. The full list of the literature is provided in Table \ref{Table:Signal}.

%\textcolor{blue}{[YM: need a conclusion and cite the table]}

% \ym{[YM: where is the table]}
% % pang2024illm_TS4
% wang2024llm_TS3
% % villarreal2023can_TS9
% lai2023large_TS5
% % masri2024leveraging_TS8

\begin{table}[htbp]
\label{Table:Signal}
\footnotesize
\centering
\caption{Travel Signal Control Using LLMs}
\label{Table:Signal}
\begin{threeparttable}
\begin{tabular}{>{\raggedright}p{2cm} >{\raggedright}p{2cm} >{\raggedright}p{2.2cm} >{\raggedright}p{2.2cm} >{\raggedright}p{2cm} >{\raggedright}p{2.2cm} >{\raggedright\arraybackslash}p{2.2cm}}
\toprule
\rowcolor{gray!20}
\textbf{Literature} & \textbf{Model Backbone} & \textbf{Input} & \textbf{Output} & \textbf{Modality} & \textbf{Task} & \textbf{Data Source} \\
\midrule
\cite{pang2024illm_TS4} & GPT-4 & Real-time traffic data and scenario descriptions& Refined signal control decisions &  Text and data & Optimize traffic signal control & Traffic simulation data\\
\cite{wang2024llm_TS3} & GPT-4 & Static and dynamic traffic data & Optimized traffic signal phase &  Text and data & Optimize traffic signal control & Shanghai traffic data and traffic simulation data\\
\cite{villarreal2023can_TS9} & GPT-4 & Natural language queries& Suggested state spaces and reward functions for RL tasks &  Text & Assist in defining RL components for mixed traffic control & Traffic simulation data\\
\cite{lai2023large_TS5} & LightGPT & Real-time traffic data and task descriptions & Traffic signal control decisions & Text and data & Optimize traffic signal control & Jinan and Hangzhou traffic data \\
\cite{masri2024leveraging_TS8} & GPT-4o-mini &  Real-time traffic data and scenario descriptions & Traffic signal control recommendations &  Text and data & Optimize traffic signal control and improve mixed traffic safety & Generated data from GPT-4o-mini\\
\bottomrule
\end{tabular}
\end{threeparttable}
\end{table}

%% file: Application/TrafficSimulation.tex
\subsubsection{Traffic Simulation}
\label{subsubsection:TrafficSimulation}
Traffic simulation is the process of using computational models to represent and analyze the movement of vehicles and pedestrians within a transportation network. This powerful tool enables researchers and engineer to study traffic behavior under diverse conditions and scenarios. By simulating traffic dynamics, it helps explain traffic patterns (e.g., \cite{xu2023non}) and evaluate the potential impacts of traffic control and management strategies (e.g., \cite{yu2020incorporating}) prior to real-world implementation.

To perform a traffic simulation, a traffic scenario must be constructed first. The key elements of a traffic scenario include the network (e.g., geometry, road type, and lane configuration), traffic demand (e.g., origin-destination trips), and traffic infrastructure (e.g., traffic signals) \autocite{SUMO2018}. Defining these components accurately is crucial for creating a realistic traffic scenario. However, this process is often complex, as these elements originate from diverse data sources, making data integration and post-processing both challenging and time-intensive.

In recent years, a few studies have emerged that utilize LLMs for intuitive and efficient creation of complex scenarios for traffic simulation. For example, \cite{tan2023language} introduce a language-conditioned model, LCTGen, which utilizes an LLM to transform natural language descriptions of driving scenarios into simulation scenarios that include both the initial states and motions of traffic actors (e.g., vehicles and pedestrians). Similarly, \cite{zhong2023language} combine an LLM with a scene-level diffusion model to bridge user queries and traffic generation. By translating linguistic commands into differentiable loss functions, the proposed model guides the simulation to produce realistic and query-compliant multi-agent traffic interactions such as car following and lane changing.

\cite{guzay2023generative} extends the application of LLMs beyond simple scenario generation. In this study, LLMs are employed to convert natural language prompts describing traffic scenarios -- such as road layouts, intersections, vehicle types, and traffic conditions -- into simulation parameters in ``XML'' format, directly compatible with the traffic simulator ``Simulation Of Urban Mobility" \autocite{SUMO2018} and ready to be simulated.

In summary, LLMs aid traffic simulation by simplifying the creation of realistic and complex scenarios through NLP. LLM-driven approaches enhance the efficiency and adaptability of traffic simulation, supporting more effective traffic management and planning as summarized in Table \ref{Table:TravelSimulation}.

\begin{table}[htbp]
\footnotesize
\centering
\caption{Travel Simulation Using LLMs}
\label{Table:TravelSimulation}
\begin{threeparttable}
\begin{tabular}{>{\raggedright}p{2cm} >{\raggedright}p{2cm} >{\raggedright}p{2.2cm} >{\raggedright}p{2.2cm} >{\raggedright}p{2cm} >{\raggedright}p{2.2cm} >{\raggedright\arraybackslash}p{2.2cm}}
\toprule
\rowcolor{gray!20}
\textbf{Literature} & \textbf{Model Backbone} & \textbf{Input} & \textbf{Output} & \textbf{Modality} & \textbf{Task} & \textbf{Data Source} \\
\midrule
\cite{tan2023language} & GPT-4 & Traffic scenario descriptions& Traffic scenarios with vehicle parameters &  Text & Generate realistic traffic scenarios & Traffic simulation data\\
\cite{zhong2023language} & GPT-4 & Traffic scenario descriptions and traffic data & Traffic scenarios with vehicle trajectories &  Text and data & Generate realistic traffic scenarios & Vehicle trajectory data\\
\cite{guzay2023generative} & GPT-4 & Traffic scenario descriptions& Traffic simulation files in XML format &  Text & Generate realistic traffic scenarios and simulation files & Waymo Open Dataset\\
\bottomrule
\end{tabular}
\end{threeparttable}
\end{table}

%% file: Application/RoadNetwork.tex
\subsubsection{Road Network Generation}
\label{subsubsection:RoadDesign}
The generation of road networks is critical for a wide range of applications, including traffic simulation, autonomous navigation systems, and urban planning. Accurate and efficient road network models are essential for tasks such as optimizing traffic flow, planning transportation infrastructure, and supporting real-time navigation. Traditional methods for generating road networks, such as segmentation-based techniques and manual modeling with traffic simulation software, have been widely used but suffer from significant limitations.

Segmentation-based techniques rely on satellite or aerial imagery and involve generating binary segmentation masks to distinguish between roads and non-road areas. These techniques demand extensive labeled datasets, require time-intensive pre-processing, and are sensitive to image quality. On the other hand, manual modeling with traffic simulation software requires significant human intervention to input road layouts, design intersections, and configure traffic parameters. This process is labor-intensive, prone to errors, and requires domain expertise.

LLMs can generate road networks without these deficiencies, offering advantages such as reduced dependence on labeled datasets, automation of manual tasks, and the ability to process multimodal inputs like text, images, and hand-drawn maps. Their reasoning and recognition capabilities enable faster, more accurate, and cost-effective road network generation.

Through our literature search, we have identified two relevant works in this field: NavGPT introduced by \cite{rasal2024beyond} and network generation AI (NGAI) proposed by \cite{chen2024multimodal}. These studies highlight the potential of LLMs in generating road networks and demonstrate innovative approaches that address the shortcomings of traditional methods.

NavGPT is a multi-modal LLM designed to generate navigable road networks directly from aerial images \autocite{rasal2024beyond}. It takes raw aerial imagery as input and outputs detailed road networks in a recognized format. Unlike traditional segmentation-based methods, NavGPT bypasses the need for binary segmentation masks by aligning visual features from aerial images with language-based outputs. NavGPT builds upon the MiniGPT-4 architecture, leveraging frozen pre-trained components for both vision and language. The model is fine-tuned using a novel training methodology to identify road geometries. It requires only one A100 GPU and approximately 26 hours for retraining, making it lightweight and cost-effective.

NGAI takes a step further to support multimodal inputs, including text, satellite images, and hand-drawn maps, and outputs road networks \autocite{chen2024multimodal}. NGAI employs a U-Net model for satellite image segmentation and advanced corner detection algorithms to extract road geometries. Unlike NavGPT, NGAI relies on pre-trained models for image processing but integrates them with LLMs through the LangChain framework. This combination allows NGAI to autonomously select and invoke plugins for modeling tasks, ensuring flexibility and adaptability for various user queries.

The use of multimodal LLMs for road network generation simplifies workflows, reduces dependence on labeled datasets, and enables efficient processing of diverse input formats. This area of research remains limited as shown in Table \ref{Table:RoadNetwork}. Future research is expected to focus on improving the accuracy of image-based recognition, enhancing the adaptability of LLMs to varied traffic scenarios, and expanding their capabilities to handle dynamic and large-scale transportation systems.

\begin{table}[htbp]
\footnotesize
\centering
\caption{Road Network Generation}
\label{Table:RoadNetwork}
\begin{threeparttable}
\begin{tabular}{>{\raggedright}p{2.2cm} >{\raggedright}p{2.2cm}>{\raggedright}p{2.2cm}  >{\raggedright}p{2.2cm}>{\raggedright}p{2.2cm}>{\raggedright}p{2cm} >{\raggedright\arraybackslash}p{2.2cm}}
\toprule
\rowcolor{gray!20}
\textbf{Literature} & \textbf{Model Name} & \textbf{Model Backbone} & \textbf{Inputs} & \textbf{Outputs}& \textbf{Modality}& \textbf{Data Source} \\
\midrule
 \cite{rasal2024beyond} & NavGPT  & Mini-GPT4 & Satellite images and texts & Detailed road networks & Text and image & Satellite images \\
 \cite{chen2024multimodal} & NGAI & GPT (version not specified) & Texts and hand-drawn images & Detailed road networks & Text and image & Open Street Map  \\
\bottomrule
\end{tabular}
\end{threeparttable}
\end{table}

%please write a section according to the following train of thought
%Introduction: For what purpose we generate these road networks. Traditional methods of generating road networks: Segmentation-Based Techniques, Manual Modeling with Traffic Simulation Software. Introduce problems with traditional techniques. LLM methods can generate road networks without these deficiencies and advantages of LLMs.

%main content: through our literature search, we have found two relevant literature in this field. respectively introduce \autocite{rasal2024beyond} and \autocite{chen2024multimodal} 1.  input and output of the proposed LLM, briefly introduce is it based on other's work or is it trained independently, or is it fine tuned? 2. strength of this method (bypasses the need for binary segmentation masks). 3. training methods and lightweightedness.

%Conclusion: multimodal LLM can assist road network generation based on ..., it has advantages of ...,  but this area remains unexplored.

%NavGPT excels in precision and lightweight training, while NGAI showcases versatility through its support for multimodal inputs and plugin-based automation. Despite these promising developments, the field of LLM-based road network generation remains relatively unexplored. 

%% file: Application/Aviation.tex
\subsubsection{Aviation and Air Traffic Control}
\label{subsubsection: Aviation and Air Traffic Control and UAVs}
Aviation and air traffic control (ATC) require precision, real-time decision-making, and smooth human-machine collaboration. These fields grapple with numerous challenges, such as managing dense airspace, directing uncrewed aerial vehicles (UAVs), accommodating noisy environments, and dealing with diverse accents while adhering to stringent regulations.

While traditional methods have advanced speech recognition, predictive modeling, and navigation systems, they sometimes fail to handle the complexity and variability of these areas, particularly with unstructured data or when dynamic responses are crucial. LLMs have proven valuable assets in overcoming these limitations. In this section, we introduce a few research areas, including strategic decision-making in ATC, situational awareness and regulatory compliance in ATC, situational awareness of UAVs, as well as autonomous flight control.

\paragraph*{Strategic Decision-Making in ATC}
\quad  Strategic decision-making in ATC focuses on proactive, long-term planning to manage traffic flows, mitigate demand-capacity imbalances, and address recurring constraints in the National Airspace System. Such decisions are critical in ensuring air traffic operations' overall efficiency and reliability. Strategic decisions include implementing ground delay programs (GDPs) to manage flight arrival demand at capacity-constrained airports, rerouting flights to avoid restricted airspace or severe weather, and recalling similar past events to inform current decision-making. These tasks require traffic managers to analyze complex data, such as weather forecasts, operational constraints, and historical traffic patterns, to optimize traffic flow while minimizing delays.

Traditional methods for supporting strategic decisions in ATC rely heavily on manual processes, requiring traffic managers to sift through vast, unstructured datasets or rely on experience and intuition. For example, identifying historical GDP patterns or recalling similar traffic scenarios is often a time-intensive and cognitively demanding task, compounded by the need to filter and extract relevant information from loosely organized data sources. While existing tools, such as the flight schedule monitor (FSM), offer some degree of automation, they are limited in their ability to integrate diverse data types, generate summaries, or provide actionable insights tailored to specific queries. 

\cite{abdulhak2024chatatc} introduce an LLM-driven conversational agent that is explicitly designed to address traditional methods' limitations in managing strategic air traffic flow. By FT LLMs on a 23-year dataset of over 80,000 GDP issuances, revisions, and cancellations, CHATATC provides a powerful tool for summarizing historical GDP patterns, extracting insights, and answering context-specific queries in natural language. Unlike traditional methods, CHATATC automates labor-intensive data retrieval and summarization, enabling traffic managers to focus on unique challenges rather than repetitive, routine tasks. The tool also reduces the barriers to training new traffic managers by offering accessible insights into historical patterns, enhancing their understanding of strategic traffic flow management. 

%Additionally, these tools often lack the flexibility to adapt to the unique needs of individual traffic managers, making it challenging to address both routine and exceptional traffic management scenarios.

%By integrating a graphical user interface (GUI) and leveraging generative AI capabilities, CHATATC demonstrates how LLMs can transform strategic decision-making in ATC, addressing inefficiencies and improving situational awareness in non-safety-critical applications.

\paragraph*{Situational Awareness in ATC}
\quad  Situational awareness in ATC refers to the ability of controllers to perceive, understand, and predict the status of air traffic and environmental conditions to ensure safe and efficient operations. A significant challenge in training AI for situational awareness is the lack of anomaly datasets. \cite{fox2024leverage} address this using GPT-3.5 to generate synthetic data sets, including regular and anomalous air traffic conversations, using FS. These datasets provide diverse training samples for a variational auto-encoder (VAE), which learns a latent representation of ``normal'' communication patterns. The VAE detects ``off-nominal'' (anomalous) scenarios by measuring reconstruction loss and identifying communications that deviate from typical patterns. This approach effectively processes unstructured natural language data while overcoming dataset scarcity, offering a robust framework to enhance anomaly detection and situational awareness in civil aviation.

%By leveraging historical data and real-time inputs, LLMs can predict potential disruptions and suggest optimal flow management strategies, thus enhancing situational awareness \autocite{abdulhak2024chatatc}.

%In civil aviation, situational awareness is crucial, yet air traffic controllers may struggle to promptly detect anomalies in unusual communications with aircraft, necessitating specialized detection methods as a "wingman." 

%Stringent regulations govern RATC operations; understanding and implementing them can be time-consuming and error-prone. LLMs simplify the processing of regulatory texts, as demonstrated by GPT-3's ability to classify and summarize ATFM regulations. This application not only aids comprehension but also reduces the cognitive load on controllers, fostering more focused and safer air traffic management.

\paragraph*{Regulatory Compliance in ATC}
\quad  Regulatory compliance involves adhering to established aviation rules, procedures, and standards to maintain safety, prevent conflicts, and support the orderly flow of air traffic within controlled airspace. Regulatory compliance is very text-intensive and has a strong cognitive burden for traffic managers. LLMs have been applied to automate the classification and simplification of regulatory texts, making them more accessible to air traffic controllers. For example, GPT-3 has been used to classify and summarize air traffic flow management (ATFM) regulations, improving comprehension and accountability \autocite{jarry2024effectiveness}. These tools reduce the cognitive burden, allowing controllers to focus on operational tasks and ensuring safer, more efficient air traffic management.

\paragraph*{Situational Awareness of UAVs}
\quad UAVs increasingly rely on semantic scene understanding to navigate complex environments and communicate effectively with human operators. Traditional UAV navigation systems, while robust in structured settings, often struggle in dynamic or unstructured environments, such as disaster zones or crowded urban areas. LVLMs enable UAVs to process visual and textual data, giving them a contextual understanding of their surroundings. For example, LVLM-driven frameworks can generate semantically rich real-time descriptions of environmental characteristics by combining object detection output (for instance, from YOLOv7) with advanced natural language generation capabilities (for example, GPT-3) \autocite{de2023semantic}. By providing detailed, human-readable descriptions, LLM-enhanced UAVs improve operators' situational awareness and reduce cognitive load, enabling faster and more informed decision-making.

\paragraph*{Autonomous Flight Control}
\quad  In air combat scenarios, high precision, adaptability, and responsiveness are crucial. Traditional flight control methods fall into two categories: model-based approaches and model-free approaches, such as DRL. Model-based methods rely on accurate mathematical modeling which is often limited by uncertainties. DRL offers a model-free framework capable of adaptive control through agent-environment interactions, making it a promising technique to tackle complex and dynamic flight control tasks. However, standalone DRL has limitations such as sparse rewards, low sample efficiency, and slow convergence, which hinder its effectiveness in high-dimensional, six-degree-of-freedom (6-DOF) flight control tasks.

%Although model-based methods rely on accurate mathematical modeling to achieve stability and effective control, real-world uncertainties and incomplete system understanding often limit them. 

LLMs provide contextual knowledge and logical reasoning capabilities that complement DRL’s trial-and-error learning paradigm. For example, in 6-DOF flight control, LLMs serve as a guide mechanism during the training process, improving the quality of agent-environment interactions by injecting domain knowledge and timely feedback into the learning loop \autocite{yang2024large}. This integration allows the intelligent flight controller (IFC) to address sparse reward issues by evaluating agent actions against predefined criteria derived from domain-specific knowledge bases, such as flight operation manuals. LLMs can accelerate the learning process and improve sample efficiency by rejecting suboptimal actions and guiding the agent toward more promising strategies.

Another notable contribution of LLM-guided DRL is its ability to handle the feasibility of complex tactical maneuvers, such as looping, immelmann turn, and split S-flight actions that are often required in air combat. LLMs enhance the training process of DRL by providing a structured evaluation of action feasibility. For example, during early training episodes, LLMs evaluate the agent’s control actions (e.g., roll, pitch, yaw, and throttle commands) concerning the aircraft’s current state and target goals, rejecting actions that deviate from the desired trajectory. This structured feedback reduces exploration inefficiencies and accelerates convergence, enabling the IFC to achieve precise control over flight attitudes and perform highly adaptive maneuvers. By integrating logical reasoning and domain knowledge into the learning process, LLMs address the key limitations of DRL, resulting in reduced reliance on exhaustive sample exploration and improved training efficiency.

The intersection of LLMs, RL, and other AI technologies with aviation and ATC heralds a new era of intelligence, automation, and efficiency. These technologies support strategic decision-making, enhance autonomous flight control, and advance UAV situational awareness capabilities. Despite the aforementioned research works, this area remains largely unexplored. A summary of the existing studies is shown in Table \ref{tab:aviationAirTrafficControl}. 

\begin{table}[htbp]
\footnotesize
\centering
\caption{Aviation and Air Traffic Control}
\label{tab:aviationAirTrafficControl}
\begin{threeparttable}
\begin{tabular}{>{\raggedright}p{2cm} >{\raggedright}p{2cm} >{\raggedright}p{2.2cm} >{\raggedright}p{2.2cm} >{\raggedright}p{2cm} >{\raggedright}p{2.2cm} >{\raggedright\arraybackslash}p{2.2cm}}
\toprule
\rowcolor{gray!20}
\textbf{Literature} & \textbf{Model Backbone} & \textbf{Input} & \textbf{Output} & \textbf{Modality} & \textbf{Task} & \textbf{Data Source} \\
\midrule
\cite{abdulhak2024chatatc} & GPT-4 & Historical air traffic data & ATFM suggestions & Text & Conversational ATFM support & Historical ATC data \\
\cite{carranza2023need} & Custom LLM & Text prompts & UAV responses & Text & UAV communication & Custom dataset \\
\cite{de2023semantic} & GPT-4 & Scene descriptions & Semantic scene understanding & Image and text & Scene analysis & UAV scene dataset \\
\cite{fox2024leverage} & GPT-3.5 & Air traffic conversations &  Reduces the controller's cognitive load & Text and data & Measuring communications &  VAE detects "off-nominal" (anomalous) scenarios \\
\cite{jarry2024effectiveness} & GPT-3 & ATFM regulations & Categorized regulations & Text & Regulation analysis & ATFM regulatory data \\
\cite{yang2024large} & ChatGLM-6B & State variables + commands & Flight control actions & Text and data & RL for 6-DOF control & Flight simulations \\
\bottomrule
\end{tabular}
\begin{tablenotes}[flushleft]
    \item ATFM: Air Traffic Flow Management
    \item ATC: Air Traffic Control
    \item 6-DOF: Six Degrees of Freedom
\end{tablenotes}
\end{threeparttable}
\end{table}

%\cite{guo2021context} & Encoder-Decoder & Speech spectrograms + context & Transcribed speech & Audio + Text & Speech Recognition & ATCSpeech Corpus \\
%\cite{zuluaga2022atco2} & Hybrid ASR & ATC voice communications & Text transcripts & Audio & ASR + NLU & ATCO2 Corpus \\
%\cite{ma2024text} & Transformer Encoder & Flight callsigns & Predicted flight counts & Text + Sequence & Traffic Flow Prediction & Flight Record Data \\
%\cite{pueyo2024clipswarm} & CLIP Vision-Language & Text prompts & Drone formations & Visual + Text & Drone Show Generation & Simulation Data \\
%\cite{liu2023aerialvln} & Vision-Language Navigation & Visual scenes + commands & UAV navigation paths & Visual + Text & Vision-Language UAV Navigation & UAV Visual Dataset \\
%\cite{alves2022dress} & Domain-Specific Language & Scenario specs & Adaptive drone behaviors & Text & Self-Adaptive UAV Scenarios & Scenario-Based Data \\

%% file: Application/Maritime.tex
\subsubsection{Maritime Transportation}
\label{subsubsection:Maritime}
Maritime transportation plays a vital role in global trade and economy, connecting nations and facilitating the movement of goods across the world. As the industry continues to evolve and face new challenges, researchers are exploring innovative solutions to enhance efficiency, safety, and sustainability in various aspects of maritime operations. Through our literature search, we have identified three key areas of research that applied LLMs: satellite image detection of ships, navigation and path finding, and unmanned ships.

\paragraph*{Satellite Image Detection}
\quad  Monitoring the status and locations of vessels is crucial for efficient port operations. Traditionally, this has been achieved through remote sensing technologies that provide operators with visualizations of vessel placements. This process has the potential to be automated thanks to the advancements in language models, which permit integrating and interpreting data from multiple sources into human-readable text. The Popeye model proposed by \cite{zhang2024popeye} integrate visual perception with the generalization capabilities of LLMs. By aligning visual features with language features and employing a cross-domain joint training strategy, the Popeye model achieves superior zero-shot performance on ship interpretation tasks.

\paragraph*{Navigation and Path Finding}
\quad  Enhanced vessel location monitoring serves as a stepping stone to improve maritime navigation, overcoming limitations of traditional methods that rely on human expertise, fragmented data, and manual processing of variables like weather, tides, and traffic patterns. These traditional approaches face challenges in handling dynamic maritime complexities -- such as growing vessel sizes, fluctuating conditions, and data silos -- leading to inefficient routes, higher fuel consumption, and safety risks. 

To address these challenges, researchers have turned to NLP and LLMs as powerful tools for enhancing vessel navigation and path planning \autocite{wang2024kunpeng}. KUNPENG is a comprehensive model in this field that combines NLP, LLMs, and other AI techniques. It enables the real-time processing and analysis of vast amounts of heterogeneous data, including weather conditions, vessel performance, and traffic patterns. This allows for the generation of optimal route recommendations, the early detection of potential hazards, and the coordination of multi-vessel operations, all of which contribute to enhanced safety, efficiency, and sustainability in the maritime industry.

\paragraph*{Unmanned Ships}
\quad  The emerging field of unmanned ship systems (USSs) promises enhanced safety and efficiency in maritime operations by eliminating the need for human crews, thus reducing the risks associated with human error. However, the absence of onboard personnel poses challenges in detecting operational anomalies. To address this, recent research by \cite{lirobust} has developed a sophisticated anomaly detection framework that integrates LLMs with a Bi-LSTM model. This approach preprocesses operational data to generate vectorized representations, which are then used to detect anomalies in real time. It identifies deviations from established patterns, enabling prompt responses to potential issues. This method has shown superior performance over traditional machine learning techniques, significantly boosting the survivability and reliability of autonomous maritime systems.

The application of LLMs and other advanced AI techniques in maritime transportation has the potential to improve various aspects of the industry. From satellite image detection of ships to navigation and path finding, and to the development of unmanned ships, LLMs are enabling more efficient, accurate, and data-driven decision-making processes. Nonetheless, the existing research remains scant, as shown in Table \ref{Table:Maritime}, suggesting that the area of maritime transportation remains fertile for future exploration of LLM applications.

\begin{table}[htbp]
\footnotesize
\centering
\caption{Maritime Transportation}
\label{Table:Maritime}
\begin{threeparttable}
\begin{tabular}{>{\raggedright}p{2cm} >{\raggedright}p{2cm} >{\raggedright}p{2.2cm} >{\raggedright}p{2.2cm} >{\raggedright}p{2cm} >{\raggedright}p{2.2cm} >{\raggedright\arraybackslash}p{2.2cm}}
\toprule
\rowcolor{gray!20}
\textbf{Literature} & \textbf{Model Backbone} & \textbf{Input} & \textbf{Output} & \textbf{Modality} & \textbf{Task} & \textbf{Data Source} \\
\midrule
\cite{lirobust}& BERT\& GPT& Vectorised log data  &Real-time anomaly &Text & Anomaly detection& Simulated data  \\
 %\cite{han2023semantics}& BERT BiLSTM & Pretrained Chinese BERT& VHF sentences & Entity relations & Extract entity relations &  Dalian Maritime VTS \\
%\cite{li2022semantic}& Latent Dirichlet Allocation & Detailed ship movements& Semantics describing movement behaviours & Text and quantity & Movement categorisation & AIS data \\
%\cite{bin2021design}& N-gram & Specailised corpus for the ship maneuvering field & 1, 2, 3 grams & Text & Speech recognition & Maritime English conversation data \\
\cite{wang2024kunpeng}& Not mentioned& sensor data, satellite data, navigation data, meterological data &Autonomous navigation instructions & Text and image & Intelligent maritime navigation & Maritime knowledge base \\
\cite{zhang2024popeye}& LLaMA& Remote sensing imaginery & Horizontal bounding boxes& Text and image & Ship detection& Satellite images  \\
%\cite{huang2023deep}&BERT & Tag name and description from ship data I/O lists & Local thing ID and local property generation type & Text & Mapping ship data tags to onshore platform data & Ship I/O lists \\
%\cite{mei2020port}& Word2vec & AIS data  & Alternative ports & Text & Location recommendation &  Berth records of container ships \\
\bottomrule
\end{tabular}
\begin{tablenotes}[flushleft]
    \item VHF: Very High Frequency
    \item VTS: Vessel Traffic Service
    \item AIS: Automatic Identification System
    \item Ship I/O: Ship input/output list
\end{tablenotes}
\end{threeparttable}
\end{table}

%% file: Application/LogisticsSCM.tex
\subsubsection{Supply Chain Management}
\label{subsubsection:LogisticsSCM}
There has been a broad spectrum of research in the field of supply chain management. While supply chain management encompasses a wide range of topics, this subsection focuses specifically on research related to transportation and logistics. This area of research involves handling complex networks, managing dynamic decision-making processes, and extracting actionable insights from unstructured data.

%While the primary focus of this review paper is not on SCM, it is important to acknowledge the research conducted in this area \autocite{zhou2021main, schopper2021using, aguero2024potential, DharaDelgado2024}. SCM shares many challenges with traffic and transport systems, such as handling complex networks, managing dynamic decision-making processes, and extracting actionable insights from unstructured data. As such, advancements in LLMs for SCM offer valuable methodologies that could inspire future research in traffic and transport studies. This section reviews key applications of LLMs in SCM and explores their potential relevance to transport-related challenges.

According to the research agenda provided by \cite{DharaDelgado2024} and \cite{aguero2024potential}, LLMs have the potential to be applied in several aspects of supply chain management, including knowledge management, demand forecasting, customer service, contract analysis, and quality control and maintenance. Among these aspects, knowledge management and demand forecasting are particularly relevant to research in the field of transportation logistics. Knowledge management involves automating data extraction, information summarization, and providing access to vast amounts of structured and unstructured data for risk analysis and decision-making \autocite{almahri2024enhancing}. Demand forecasting utilizes real-time market data, consumer behavior, and historical trends to predict future demands and provide optimization \autocite{quan2024invagent}.

\paragraph*{Knowledge Management and Risk Identification}
\quad  Modern manufacturers often source raw materials from suppliers across multiple tiers, making supply chain visibility crucial for risk management and decision-making. However, obtaining and managing knowledge beyond tier 1 suppliers can be challenging. \cite{almahri2024enhancing} proposes a framework that leverages LLMs to facilitate smoother knowledge extraction and management from unstructured sources. The framework collects data and information about suppliers from unstructured web resources and applies ZSL for named entity recognition (NER). NER identifies and classifies entities such as company names, locations, and product types within large volumes of text. Subsequently, relation extraction (RE) is employed to delineate the relationships between these entities, forming the edges of the knowledge graph that illustrate the flow of materials, information, and finances across the supply chain.

%To ensure the accuracy and consistency of entity representation, Entity Disambiguation is performed. This process resolves ambiguities where multiple identifiers or descriptions may refer to the same entity or distinguishes between entities with similar names but different identities. By clarifying these ambiguities, the framework maintains the integrity and reliability of the KG, ensuring that each node and link accurately reflects the real-world supply chain structures.

The purpose of knowledge extraction is to serve risk identification in the supply chain. \cite{shahsavari2024empowering} applt LLM+Bayesian Networks approach for the probabilistic reasoning of risky events. This approach extracts information from news feeds of events and datasets, feeding it into Bayesian Networks to model the probabilities of events that lead to significant supply chain risks based on causal relationships. This method enables merchants to better identify potential events that may cause significant disruptions to the supply chain. \cite{shahsavari2024empowering} successfully tests this method in the transportation sector of Victoria, Australia in 2021, identifying four contributing events to supply chain disruptions: rising COVID cases, vaccination mandates, construction worker strikes, and a bridge blockage. The method provides detailed probabilities of each contributing event causing disruptions.

While historical data offers valuable insights into past risks, historical data may not account for novel or unprecedented scenarios. To address this limitation, simulation modeling has emerged as a complementary approach for risk identification. According to \cite{jackson2024natural}, LLMs can generate complex Python code for logistics simulations using appropriately designed prompts. These models enable researchers to simulate hypothetical scenarios and assess their potential impacts on supply chains. 

%This research highlights the ability of LLMs to enhance coordination and scalability in dynamic environments, suggesting similar applications for flexible and adaptive supply chain systems.

\paragraph*{Demand Forecasting and Inventory Decisions}
\quad  With the simulation model in place, inventory decisions can be made. \cite{quan2024invagent} address inventory decisions using LLMs. The purpose of the research is to replace traditional regression models with LLMs. The authors feed the LLM agent with historical demand information, inventory levels, and other essential state features, requesting the LLM to output an order quantity. It is found that this method works fairly well under simple demand scenarios, offering better explainability than deep models. However, this method underperforms regression models when it comes to more complicated scenarios.

\cite{li2023large} propose a more powerful application of LLMs in modeling what-if scenarios in supply chains. Instead of directly asking LLMs to output results, they design an agent named `OptiGuide' that automatically interacts with optimization models from the backend to derive results. The interaction with the LLM within OptiGuide involves inputting queries in natural language, which the LLM then translates into optimization code. This code is executed by traditional optimization solvers that interact with databases to fetch necessary data and compute outcomes. The results are then converted back into natural language by the LLM, providing users with actionable insights and detailed explanations of the optimization results. OptiGuide represents a significant step forward in making complex supply chain optimizations accessible to a broader range of business users without requiring them to have specialized knowledge in optimization algorithms or machine learning.

Despite the aforementioned advantages, most industry interviewees indicate that LLMs should serve an assistive role rather than taking full autonomy \autocite{DharaDelgado2024}. Users must also be cautious about data privacy, the accuracy of responses, and the operational dependence on LLMs. The future of supply chain management with LLMs needs to take an integrated approach where technology complements human skills rather than replacing them.

LLMs offer unique advantages to improve supply chain management practices by enhancing efficiency, reducing costs, minimizing risks, and improving service delivery. However, they must also be used carefully to mitigate the associated risks. The full list of the literature is provided in Table \ref{Table:SCM}. 

%While the primary focus of this review paper is not on supply chain management, the methodologies and approaches discussed in this section may inspire future research in traffic and transport studies. The ability to extract knowledge from knowledge graphs, model complex relationships, perform probabilistic reasoning and interacting with optimization solvers could be valuable in addressing challenges related to traffic congestion, transportation network optimization, and risk assessment in transportation systems. As the field of LLMs continues to evolve, it is essential for researchers in traffic and transport studies to stay informed about the advancements and potential applications of these technologies in order to drive innovation and improve the efficiency and resilience of transportation networks.

\begin{table}[htbp]
\footnotesize
\centering
\caption{Supply Chain Management}
\label{Table:SCM}
\begin{threeparttable}
\begin{tabular}{>{\raggedright}p{2cm} >{\raggedright}p{2cm} >{\raggedright}p{2.2cm} >{\raggedright}p{2.2cm} >{\raggedright}p{2cm} >{\raggedright}p{2.2cm} >{\raggedright\arraybackslash}p{2.2cm}}
\toprule
\rowcolor{gray!20}
\textbf{Literature} & \textbf{Model Backbone} & \textbf{Input} & \textbf{Output} & \textbf{Modality} & \textbf{Task} & \textbf{Data Source} \\
\midrule
\cite{almahri2024enhancing} & GPT-4 & EV manufacturers info & Knowledge graph & Text \& quantity & Find relationships of suppliers & Wikipedia \\
\cite{chen2024scalable} & GPT-3.5 \& 4 & Agent states & Movement & Text and quantity & Collaborative boxes & None \\
\cite{jackson2024natural} & GPT-3 & Inventory simulation prompt & Python code & Text & Generating simulation code & None \\
\cite{li2023large} & GPT-3.5 \& 4 & If-then questions & New solutions to changes & Text & Implement optimisation and interpret solutions & Open source benchmarks \\
\cite{quan2024invagent} & GPT-4 & Stage, state information of supplier & Order quantity & Text and quantity & Inventory decisions & Generated data \\
\cite{shahsavari2024empowering} & GPT-3.5 & News & Probabilities of contributing events & Text & Identify contributing events and risks & Google news \\
\cite{zhao2024optimizing} & GPT-3.5 \& open-source pre-trained models & News, merchant and supplier information & Risk information & Text & Label risks given news & Tier 1 suppliers, Google news \\
\cite{aguero2024potential} & \multicolumn{6}{c}{Research agenda paper} \\
\cite{schopper2021using} & \multicolumn{6}{c}{Review paper} \\
\cite{DharaDelgado2024} & \multicolumn{6}{c}{Research agenda paper} \\
\cite{wang2024large} & \multicolumn{6}{c}{Research agenda paper} \\
\cite{zhou2021main} & \multicolumn{6}{c}{Research agenda paper} \\
\bottomrule
\end{tabular}
\begin{tablenotes}[flushleft]
    \item EV: Electric Vehicle
\end{tablenotes}
\end{threeparttable}
\end{table}

%% file: Application/PublicTransport.tex
\subsubsection{Public Transportation}
\label{subsection:PublicTransport}
\quad  %Public transportation is a system designed to transport a group of people, and therefore subject to severe consequences if malfunctioning occurs. 
When LLMs are applied to public transportation systems, three research areas are identified: passenger demand forecasting, bus holding control, and infrastructure analysis.

\paragraph*{Passenger Demand Forecasting}
\quad  Accurate demand prediction is vital for optimizing resource allocation and service reliability, especially under irregular conditions like delays or disruptions. Conventional methods, such as ARIMA and LSTM-based models, rely on structured numerical data (e.g., historical ridership) but struggle to adapt to dynamic, unstructured variables like real-time delays or weather events. For instance, while GC-LSTM incorporates spatial dependencies, it fails to contextualize how sudden delays propagate through a network, leading to inaccurate forecasts. \cite{huang2024enhancing} tackles this by reformulating numerical and spatial data into natural language prompts (e.g., “A 15-minute delay occurred at Station X; predict passenger flow at Station Y”). The proposed LLM-based framework employs CoT reasoning, enabling step-by-step analysis of delay impacts and passenger behavior. This approach significantly outperforms traditional models in accuracy during irregular scenarios, offering interpretable predictions that aid transit agencies in proactive scheduling.

\paragraph*{Bus Holding Control}
\quad  Bus holding, i.e., delaying departures to maintain schedule adherence, is a key strategy for reducing congestion and improving service regularity. Traditional model-based methods use rigid algorithms to predict bus states and demand, but they lack flexibility in dynamic environments. RL emerged as a promising alternative by framing holding decisions as a reward-maximization problem. However, RL’s reliance on manually designed reward functions (e.g., penalizing deviations from schedules) limits its adaptability and requires labor-intensive tuning. \cite{yu2024large} integrate LLMs into RL to automate reward function design. Here, LLMs interpret real-time operational data (e.g., passenger load and traffic) and generate context-aware rewards (e.g., prioritizing crowded buses for priority dispatch). This hybrid paradigm reduces human intervention, accelerates policy convergence by 30\%, and improves on-time performance by 15\% in simulations. The LLM’s ability to parse unstructured data (e.g., driver reports) further enhances decision-making in edge cases, such as accidents or road closures.

\paragraph*{Infrastructure Analysis}
\quad  Bus stop accessibility and quality directly impact user experience and equity. Traditional evaluations rely on computer vision tools like YOLO for detecting specific features (e.g., wheelchair ramps). While effective for isolated tasks, these methods lack contextual reasoning -- for example, recognizing that a ramp obscured by debris is non-functional. \cite{oliveira2024towards} proposes an LLM-driven framework that synthesizes multimodal data (street-level images, maintenance logs) to generate holistic assessments. The model identifies physical defects (e.g., cracked pavements) and correlates them with accessibility metrics (e.g., ADA compliance), providing actionable insights like, ``Tactile paving at Stop 12 is 80\% worn, posing risks to visually impaired passengers.'' This approach scales evaluations across entire networks, reducing manual inspection costs by 40\% while prioritizing high-impact repairs.

In summary, LLMs demonstrate great potential in enhancing public transportation systems across three areas: passenger demand forecasting, bus holding control, and infrastructure analysis. The advancement of LLMs helps improve prediction accuracy, streamline decision-making, and optimize resource allocation, aiding transit agencies in providing more reliable and equitable services. A summary of the reviewed literature is presented in Table \ref{Table:PubicTransport}.

\begin{table}[htbp]
\footnotesize
\centering
\caption{Public Transportation}
\label{Table:PubicTransport}
\begin{threeparttable}
\begin{tabular}{>{\raggedright}p{2.2cm} >{\raggedright}p{2.2cm} >{\raggedright}p{2.4cm} >{\raggedright}p{2.4cm} >{\raggedright\arraybackslash}p{2.4cm}>{\raggedright\arraybackslash}p{2.4cm}}
\toprule
\rowcolor{gray!20}
\textbf{Literature} & \textbf{Model Backbone} & \textbf{Inputs} & \textbf{Outputs} & \textbf{Modality} & \textbf{Data Source} \\
\midrule
\cite{huang2024prompt} & GPT-4 & Delay, historical passenger flow data, adjacency matrix & Passenger flow predictions & Text and data & AFC data from Shenzhen Metro \\
\cite{yu2024large} & GPT-4 & State-action data & Optimized reward function & Text and data & Bus data from Beijing \\
\cite{oliveira2024towards} & LLaVA & Street-level images, evaluation criteria & Qualitative and quantitative assessments & Text and image & INACITY platform \\
\bottomrule
\end{tabular}
\end{threeparttable}
\end{table}

%% file: Application/Micellaneous.tex
\subsection{Multi-task Applications}
\label{subsection:Micellaneous}
\begin{figure}[h]
\centerline{\includegraphics[trim=4cm 2cm 4cm 2cm, clip, width=0.9\textwidth]{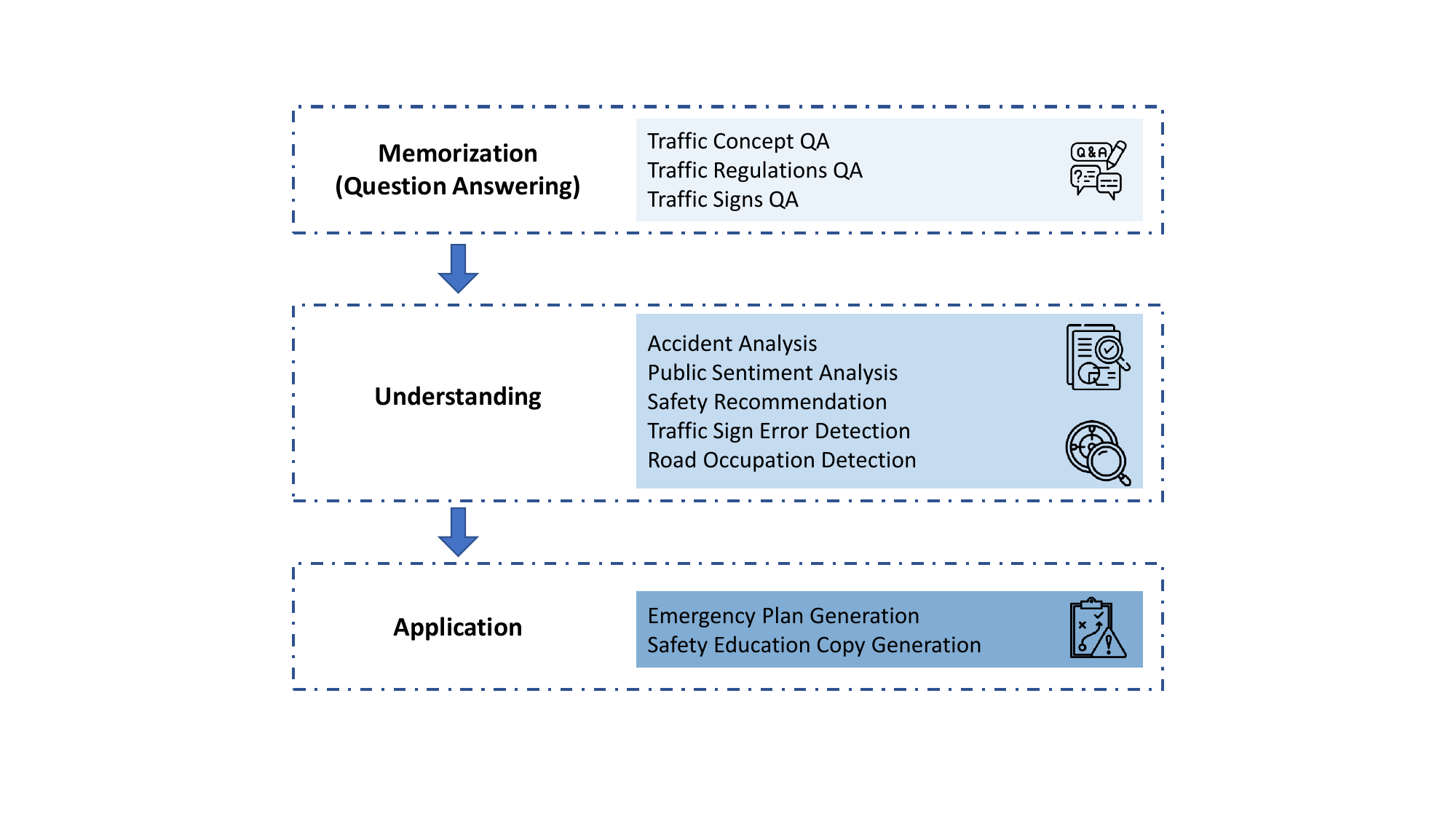}}
\caption{Capabilities of LLMs in the field of traffic and transportation research}
\label{GPTinTransport}
\end{figure}

\quad Many LLMs are versatile and can cover a range of traffic and transportation tasks. This section focuses on reviewing LLM works that span multiple aspects of traffic and transportation research. According to \cite{zhang2024transportationgames}, the various tasks that LLMs cope with can be categorized into three main areas: memorization, understanding, and application as shown in Figure \ref{GPTinTransport}.

Memorization refers to an LLM's ability to recall answers to specific questions. \cite{syed2024benchmarking} evaluate this capability using datasets from TransportBench, which consists of undergraduate-level exam questions covering various traffic- and transportation-related topics, such as transportation economics, driver characteristics, vehicle motion, road geometry design, traffic flow/control, transportation planning, utility/modal split, transportation networks, and public transit systems. The authors employed a zero-shot prompting strategy, directly inputting the original problem descriptions into web-version LLMs. The accuracy rates range from 40\% to 70\%, demonstrating the LLMs' ability to memorize and recall transportation-related information. Besides zero-shot learning, LLMs also show the ability to retrieve information from databases as needed. The ReBoostSQL framework, introduced by \cite{sui2023reboost}, exemplifies the model's prowess in converting natural language queries into executable SQL commands. This capability is particularly beneficial in complex business environments where traffic data is stored across large, intricate databases. By enhancing query accuracy and linking these queries to specific database schemas, ReBoostSQL improves the efficiency of data retrieval and ensures the relevance and precision of the results.

Moving beyond memorization, understanding involves an LLM's capacity to analyze text and data and make recommendations based on its findings. TransGPT, a model fine-tuned with a vast amount of textual data from various transportation-related sources, exemplifies this capability. The model's training data includes books, reports, documents, websites, driving tests, traffic signs, landmarks, and corpora. \cite{wang2024transgpt} showcases the potential applications of TransGPT in traffic analysis and modeling, such as generating synthetic traffic scenarios, explaining traffic phenomena, answering traffic-related questions, providing traffic recommendations, and generating traffic reports. These applications highlight the model's ability to understand and interpret transportation-related information.

The third category, application, refers to an LLM's ability to generate traffic operational plans and strategies. The innovative Open-TI model, introduced by \cite{da2024open}, aims to create a Turing Indistinguishable level of traffic intelligence. This model integrates the theoretical strengths of LLMs with practical traffic management needs, marking a shift towards operational applications that focus on real-time traffic simulation and management. Open-TI processes natural language inputs and generates detailed traffic analyses and simulations, representing a significant step towards creating intelligent transportation systems that can interact seamlessly with human operators and adapt dynamically to changing traffic conditions.

The most comprehensive model is TrafficGPT developed by \cite{zhang2024trafficgpt}, which provides a framework combining the foundational strengths of LLMs with Traffic Foundation Models to enhance traffic management. TrafficGPT facilitates a deeper understanding of traffic data, enabling more nuanced and effective traffic control strategies. The framework uses natural language inputs from users to guide the execution of various traffic-related tasks through a sequence of steps involving understanding, planning, and execution using traffic foundation Models. The full list of the literature surveyed in this subsection is provided in Table \ref{Table:composite}.

\begin{table}[htbp]
\footnotesize
\centering
\caption{Multi-task Applications}
\label{Table:composite}
\begin{threeparttable}
\begin{tabular}{>{\raggedright}p{2.2cm} >{\raggedright}p{2.8cm}>{\raggedright}p{2.2cm}  >{\raggedright}p{2.2cm} >{\raggedright}p{2.4cm} >{\raggedright\arraybackslash}p{2.4cm}}
\toprule
\rowcolor{gray!20}
\textbf{Literature} & \textbf{Model Name} & \textbf{Model Backbone} & \textbf{Modality} & \textbf{Task} & \textbf{Data Source} \\
\midrule
 \cite{da2024open} & Open-ti &Pre-trained traffic LLMs & Text and data & Traffic data analysis, simulation, optimization and control & Traffic datasets, maps, real-time traffic data\\
 \cite{sui2023reboost} & N/A & GPT-4 & Text and SQL & Text-to-SQL, schema linking SQL generation, SQL boosting & CTtraffic and CompanyZ\\
 \cite{syed2024benchmarking}& None & GPT-4V &Text and images & Traffic event recognition, analysis and reporting &Open-source datasets\\
 \cite{tian2023vistagpt} & None &GPT-4, Claude 3.5, Gemini 1.5, Llama &Text & Solving transportation engineering questions & TransportBench dataset\\
 \cite{wang2024transgpt} & TransGPT & ChatGLM, VisualGLM & Text and images & Answering transportation-related questions & Traffic engineering documents, examinations papers \\
 \cite{zhang2024transportationgames} & TransportationGame & 16 LLMs & Text and images & Memorization, understanding, application& Examination papers, news, gov websites, images\\
  \cite{zhang2024trafficgpt}& TrafficGPT & GPT-4 & Text and images& Traffic management and decision support & Various traffic data sources\\
\bottomrule
\end{tabular}
\end{threeparttable}
\end{table}

%% file: FutureDirections/FutureDirections.tex
\section{Current Trends and Future Directions}
\quad Our literature review has uncovered a variety of LLM methodologies and applications in traffic and transportation research. Yet significant challenges remain in addressing the associated problems. In this section, we provide a statistical analysis of the studies examined, to highlight the prevailing trends and explore the potential avenues for future research.

\label{section:FD}

\input{FutureDirections/Statistics}
\input{FutureDirections/Methods}

\input{FutureDirections/Applications}
\input{FutureDirections/ProsCons}

%% file: FutureDirections/Statistics.tex
\subsection{Statistical Analysis}
\label{subsubsection:Statistics}

\begin{figure}
    \centering
    \includegraphics[width=1\linewidth]{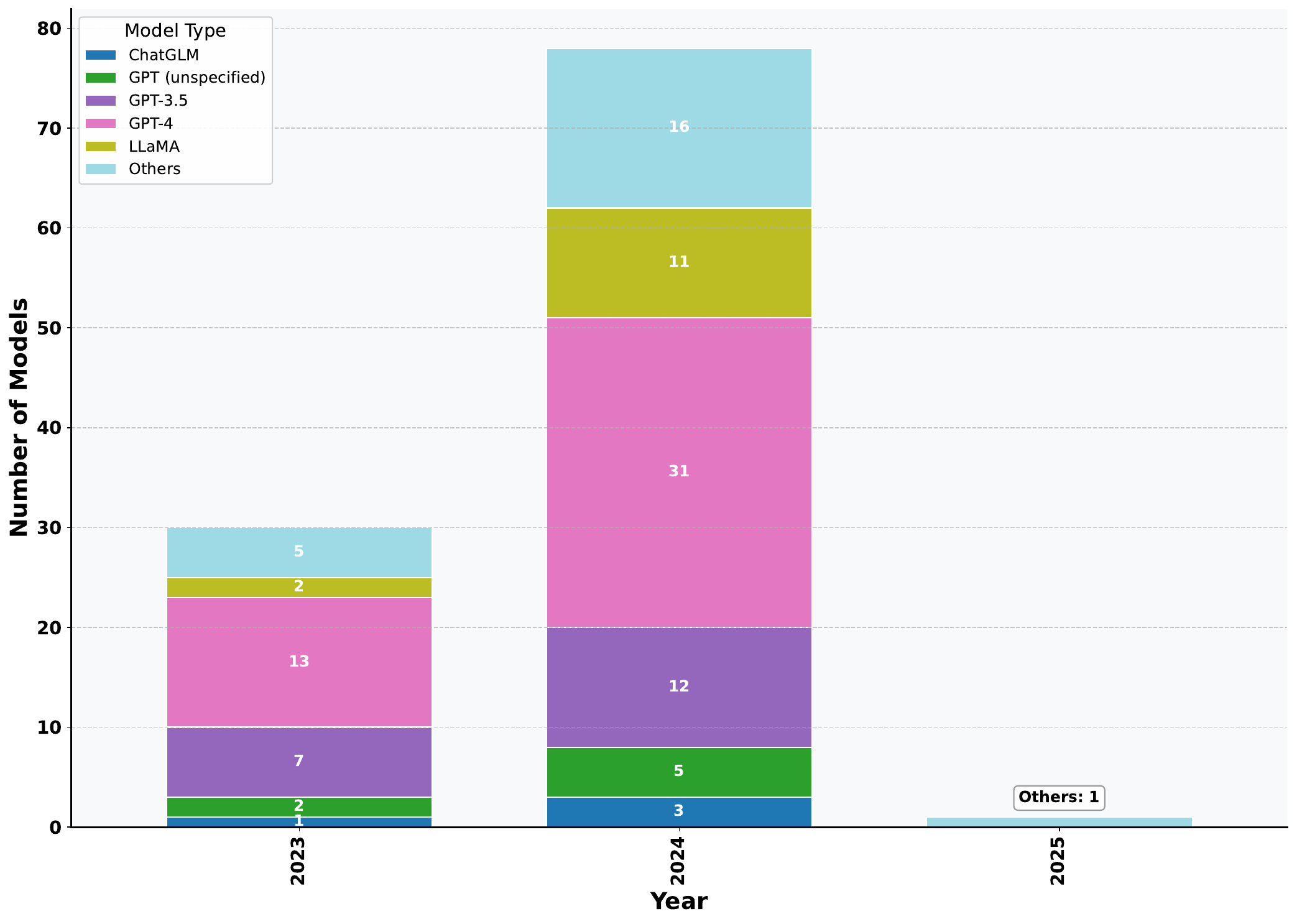}
    \caption{Number of Publications by Year}
    \label{fig:publication_year}
\end{figure}

\quad Figure~\ref{fig:publication_year} presents a temporal distribution of LLM usage in transportation research from 2023 to 2025. We observe a moderate level of adoption in 2023, where GPT‑4 accounts for the largest share among the models used, followed by GPT‑3.5, ChatGLM, and unspecified version of GPT. This initial wave likely reflects early experimentation as researchers and practitioners assessed the feasibility of integrating LLMs into diverse transportation problems. In 2024, there is an increase in overall LLM usage, with GPT‑4 as the most significant increase—surpassing all other models in both absolute and relative terms. LLaMA and ``Other models'' also show substantial growth, indicating a rapidly diversifying research ecosystem. The jump in 2024 could be attributed to the broader release and improved accessibility of newer models, as well as increased recognition of the potential benefits of language-based approaches for tasks such as policy formulation, demand forecasting, and real-time system management. As this paper is finished at the beginning of 2025, papers in this year are mostly not included. We believe that there will be an increasing trend in the number of papers in 2025.

\begin{figure}
    \centering
    \includegraphics[width=1\linewidth]{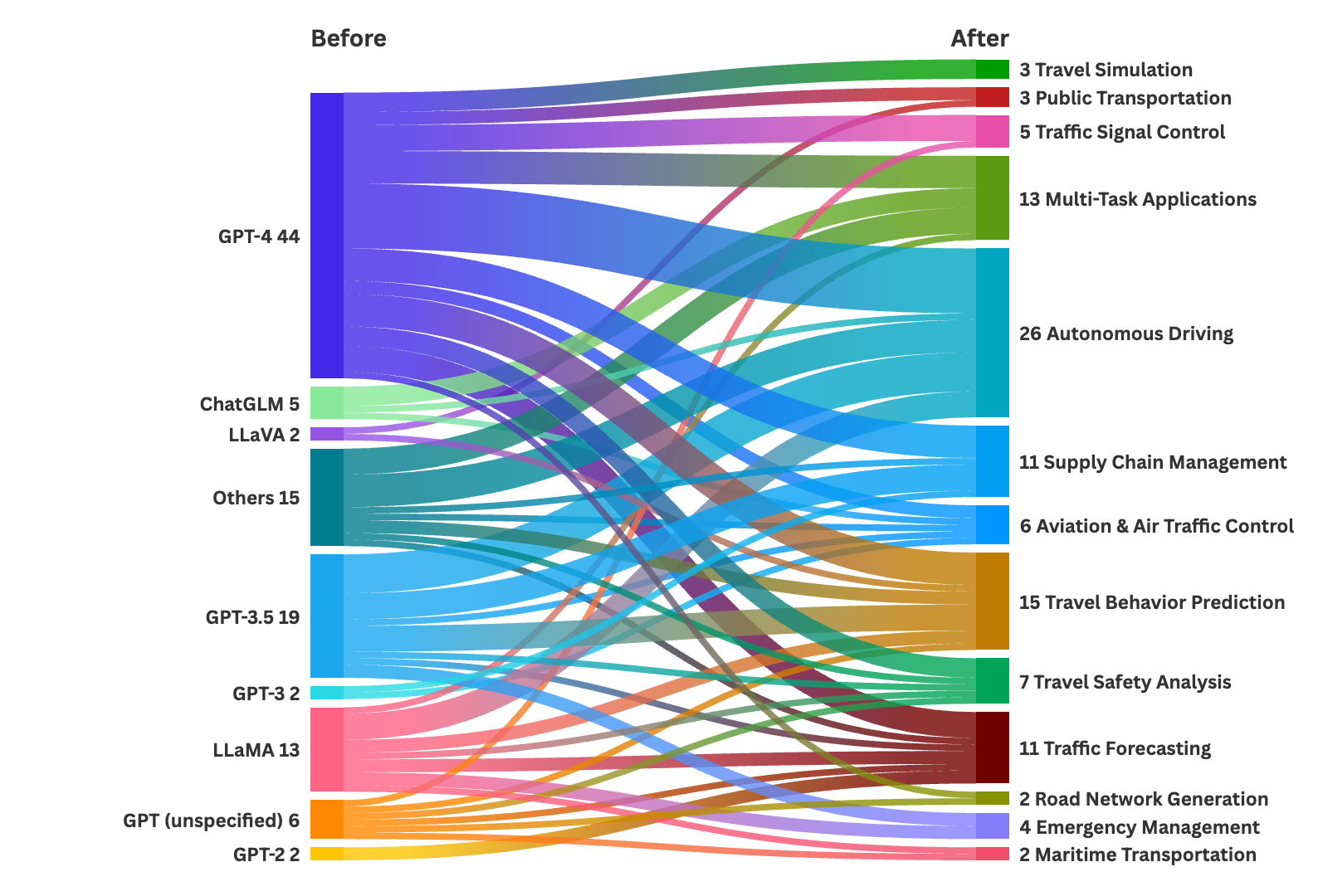}
    \caption{Mapping of LLM Models to Applications}
    \label{fig:model_app}
\end{figure}

Figure~\ref{fig:model_app} illustrates the mapping of LLMs to various transportation applications, capturing the breadth of how different models have been applied in this domain. The applications include traffic signal control, traffic forecasting, autonomous driving, aviation and air traffic control, supply chain management, travel behavior prediction, and several others. We observe that GPT‑4 is prominently featured across a wide range of these applications, likely due to its enhanced reasoning capabilities and broader availability after its introduction. GPT‑3.5 is also widely adopted, reflecting its early presence and established user base. Meanwhile, ChatGLM, LLaMA, and other models each find more specialized applications. Again, if a particular model is used only once, it is categorized under “Others” to maintain clarity in the mapping. This mapping diagram shows that certain application areas, such as autonomous driving and travel safety analysis, attract the widest variety of LLMs, suggesting that these tasks may require flexible language reasoning, scenario analysis, or large-scale data handling. By contrast, more specialized tasks such as travel simulation or emergency management appear to cluster around a smaller set of preferred models.

\begin{figure}
    \centering
    \includegraphics[width=1\linewidth]{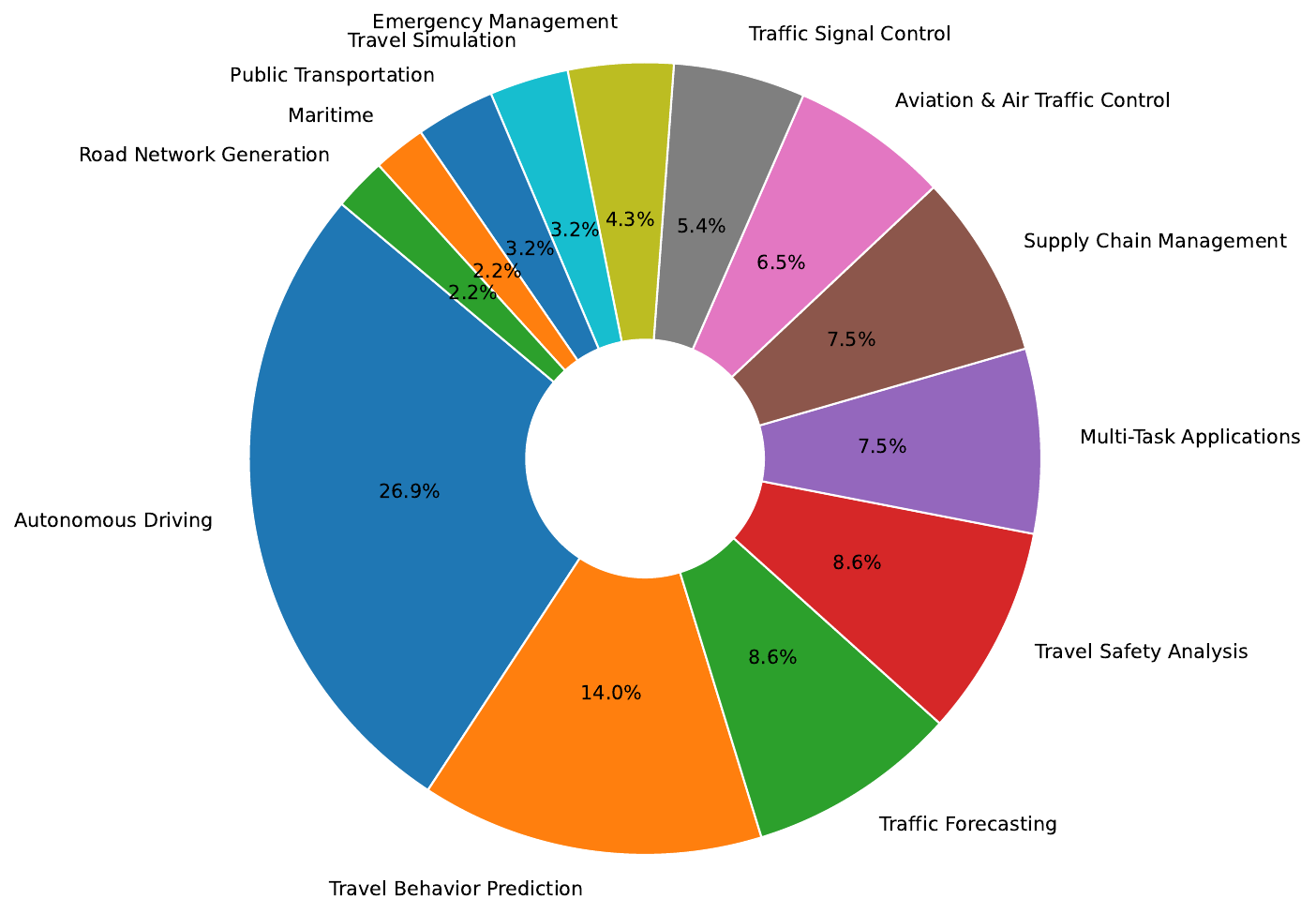}
    \caption{Applications Percentage}
    \label{fig:app_percent}
\end{figure}

Figure~\ref{fig:app_percent} provides another perspective by illustrating the proportion of LLM usage across major transportation application categories. Here, AD (26.9\%) emerges as the most represented application, reflecting the strong interest in leveraging LLMs for tasks such as decision-making in autonomous vehicles and real-time navigation support. Travel Behavior Prediction (14.0\%) ranks second, followed by Traffic Forecasting (8.6\%) and Travel Safety Analysis (8.6\%). Multi-task Applications and Supply Chain Management each account for 7.5\%, suggesting growing diversity in how LLMs are deployed. Meanwhile, Aviation \& ATC (6.5\%) and TSC (5.4\%) occupy the mid-range, while Emergency Management (4.3\%), Travel Simulation (3.2\%), Public Transportation (2.2\%), Maritime Transportation (2.2\%), and Road Network Generation (2.2\%) form smaller but noteworthy slices. Taken together, these distributions indicate that certain areas—particularly those involving real-time or highly dynamic tasks—are more likely to incorporate LLMs, whereas emerging domains may still be experimenting with smaller-scale or pilot deployments.

Overall, these figures reveal two primary trends. First, GPT‑based models (GPT‑3.5 and GPT‑4) maintain a dominant presence across a wide spectrum of transportation research topics. Second, while other LLMs such as ChatGLM and LLaMA are not as universally deployed, their increase in specialized domains suggests that researchers are exploring more diverse solutions to address emerging traffic and transportation challenges. Future studies may benefit from a closer examination of how specific model features -- such as FT protocols, context-window lengths, or multimodal capabilities -- translate into performance gains in particular transportation applications.

%% file: FutureDirections/Methods.tex
\subsection{Future Directions for LLM Methodologies}
\label{subsubsection:FDMethods}

\paragraph*{LLMs for Problem Formulation and Solver Integration} 
\quad LLMs can process natural language descriptions of transportation and traffic problems, translate them into formal mathematical models (e.g., linear programming, integer programming), and interface with backend solvers to find solutions. This integration significantly lowers the barrier for non-experts to utilize OR techniques, by shifting the burden of model construction to the LLM. Through our literature search, we have only identified a relevant paper in supply chain management \autocite{li2023large}. A general LLM-integrated solver named ORLM has been proposed, as demonstrated in the methodology section of this paper, but a field-specific model remains lacking. 

Relevant methods have the potential to be applied to a variety of traffic and transportation challenges. For example, in traffic signal optimization, natural language queries such as ``Optimize green light timings to reduce congestion during peak hours'' can be translated into formal linear programming models, which can then be solved to minimize delays or maximize traffic flow efficiency. Similarly, in transit scheduling, LLMs can process inputs like ``Create a bus schedule to maximize coverage while minimizing idle time for the fleet'' and construct integer programming models to generate optimized schedules. Another promising application is freight route optimization, where LLMs can handle descriptions such as ``Find the cheapest and fastest way to deliver goods from warehouse A to multiple destinations'' by building transportation models, such as network flow or mixed-integer programming models, and interfacing with solvers to identify the most efficient routes. These examples highlight how LLMs can lower the barrier for non-experts to use advanced OR techniques, enabling more accessible and effective decision-making.

%It has the potential to be transferred to our field of studies. For example, a transportation operator might describe an optimization problem in natural language, such as minimizing travel times across a network while respecting capacity constraints. The LLM can parse this description, identify the relevant variables and constraints, and construct the mathematical equations representing the problem. These equations can then be passed to a backend solver, such as Gurobi or CPLEX, to compute a solution. Once the solver generates a result, the LLM can translate the output into user-friendly feedback, offering actionable insights to decision-makers.

%This workflow not only democratizes access to OR methods but also improves efficiency and accuracy in problem-solving. However, challenges such as ensuring the correctness of the generated equations, managing large-scale problems, and maintaining interpretability in the process are key areas for further research.

%\paragraph*{LLMs for Database Retrieval}: LLMs can be used for generating SQL codes for the retrieval of data.

\paragraph*{Automatic Design of Heuristics}
\quad Traditional heuristics (and meta-heuristic) methods are applied to solve a variety of traffic and transportation problems, such as signal control, delivery and scheduling. Traditional manual heuristic design requires significant domain expertise and time, while automatic heuristic design methods such as genetic programming cannot produce new heuristics from scratch \autocite{liu2024evolution}. LLMs have the potential to address these limitations by simultaneously adjusting heuristic parameters and exploring new heuristics.

\cite{liu2024evolution} introduce evolution of heuristics (EoH), a novel framework that combines LLMs and evolutionary computation (EC) to automate the design of optimization heuristics. Unlike traditional methods that manually craft heuristics or rely on NNs, EoH iteratively evolves both the thoughts (natural language descriptions) and codes (executable implementations) of heuristics. It uses LLMs to generate and refine heuristics based on prompt strategies. The proposed framework not only outperforms hand-crafted heuristics but also surpasses other automatic heuristic design methods, while requiring significantly fewer computational resources.

This method can potentially be applied to traffic and transportation systems by designing heuristics for complex optimization problems such as traffic signal control, vehicle routing, dynamic ride-sharing, and public transit scheduling. For example, in vehicle routing, EoH could evolve heuristics to dynamically allocate vehicles to customers in real time, considering factors like traffic congestion, fuel minimization, and service time. Similarly, for traffic signal optimization, EoH can design adaptive heuristics to adjust signal timings dynamically to reduce congestion and improve traffic flow. The ability of EoH to combine linguistic reasoning (thoughts) with code generation enables the generation of domain-specific heuristics for traffic systems that are adaptive and efficient, particularly in scenarios where real-time decisions are crucial.

%EoH offers several advantages over traditional optimization methods in traffic and transportation applications. Unlike manual heuristic design, which requires significant domain expertise and time, EoH is fully automated, reducing human effort. Compared to neural solvers for optimization, it does not require large-scale training data or computationally expensive model training. This area of research remains void through our literature search.

\paragraph*{Prompt Optimization with Operational Research Techniques}
\quad As shown in the survey in the previous section, efficient prompt techniques such as CoT greatly enhance the accuracy of outputs from LLMs. Prompts can be optimized not only with human experience, but also with mathematical optimization methods. Unlike continuous domains where gradient descent is applicable, prompt optimization is inherently discrete.

Pioneering works in this field have used this method to address adversarial prompts—inputs deliberately crafted to bypass a language model’s safety mechanisms and generate harmful outputs \autocite{thompson2024flrt}. The authors introduce a framework called fluent student-teacher redteaming (FLRT), which leverages fine-tuned LLMs. In this framework, objectives are defined to both maximize attack success rates (ASR) and minimize token repetition, thereby preserving natural fluency as shown in Equation (\ref{eq:obj}). At the same time, constraints such as maintaining a reasonable prompt length and meeting fluency criteria are imposed to shape the overall optimization as shown in Equation (\ref{eq:constraint}). 
\begin{equation}
\min_{\text{prompt}} \ \text{Discrepancy}\big(\text{LLM output(prompt)}, \text{target}\big)
\label{eq:obj}
\end{equation}

subject to:
\begin{equation}
\text{Fluency(prompt)} \geq \text{threshold}.
\label{eq:constraint}
\end{equation}

To solve this discrete problem, heuristic methods, including greedy coordinate gradient (GCG) and the BEAST algorithm, are employed. These methods iteratively adjust tokens, propose new sequences, and assess their performance, ultimately generating more effective prompts.
%Operational research techniques provide a robust framework to address the challenges of prompt optimization, such as the discrete nature of tokens and the black-box/white-box dynamics of LMs. Algorithms like GCG and BEAST, as discussed in \textit{FLRT} \autocite{thompson2024flrt}, leverage token-level adjustments (e.g., insertions, deletions, swaps) to iteratively improve the prompt. These methods are analogous to local search heuristics in operational research, where each iteration explores a neighborhood of potential solutions. Additionally, regularization techniques like multi-model perplexity penalties and repetition penalties ensure that the optimized prompts avoid nonsensical or repetitive patterns, aligning with multi-objective optimization principles.

Prompt optimization techniques are relatively novel and have not been applied in any literature in our field throughout our literature search. They have broader applications in tasks like scheduling, logistics, and decision-making, where fluency corresponds to interpretability in operational contexts. For instance, in traffic management, optimized prompts could guide AI systems to generate adaptive routing suggestions during congestion. Similarly, in logistics, multi-task optimization of delivery routes could be achieved by leveraging the same principles described in FLRT. By integrating operations research methodologies, prompt optimization becomes a powerful tool to balance competing objectives like efficiency, safety, and fluency while navigating the constraints of discrete, complex systems. This area remains limited through our literature search.

\paragraph*{Retrieval-Augmented Generation}
\quad RAG enhances LLMs by integrating a retrieval mechanism that fetches external information dynamically during the generation process. Instead of relying solely on the LLM’s pre-trained knowledge, which is static and limited to its training data, RAG dynamically retrieves relevant data from external sources such as text corpora, databases, or APIs. This dynamic retrieval capability enables LLMs to provide up-to-date and accurate information, reducing the likelihood of hallucinations (incorrect or fabricated outputs). Additionally, RAG allows LLMs to handle domain-specific tasks without requiring extensive FT, making LLMs more versatile and cost-effective.

The RAG framework consists of two key components. The first is the retriever, which fetches relevant external data based on the input query. The second is the generator, which uses the retrieved data to synthesize coherent, contextually informed responses. Together, these components create a system that combines the strengths of retrieval-based systems with the generative capabilities of LLMs.

Relevant techniques have been used in travel behavior studies \autocite{wang2024LargeLanguageModels} and for AD \autocite{hussien2025rag}. Application opportunities also abound in other areas. For example, in transportation, an RAG model can retrieve real-time metro delay logs and combine them with its reasoning capabilities to generate actionable insights for passenger flow management. This integration of external, real-time data with the model's reasoning ability makes RAG particularly valuable for dynamic, real-world applications.

\paragraph*{Knowledge Graphs} 
\quad KGs, which represent entities and their relationships in a structured format, complement LLMs by providing a source of grounded and interconnected knowledge. This integration is achieved through methods such as embedding KGs into vector spaces using graph neural networks, enabling LLMs to query KGs during inference via RAG, or using KGs as external reasoning modules alongside LLMs. By combining unstructured generative capabilities with structured relational data, LLMs are capable of generating more accurate, interpretable, and contextually relevant outputs. Through our literature search, only a few works that apply KGs for regulatory compliance can be found, in the areas of emergency traffic management \autocite{chen2023EnhancingEmergencyDecisionmaking}, supply chain management \autocite{almahri2024enhancing}, and autonomous driving \autocite{hussien2025rag}. 

The combination of KGs and LLMs has the potential to open up significant possibilities for optimizing operations and enhancing decision-making. For emergency traffic management, KGs can encode relationships between road networks, traffic incidents, weather conditions, and historical congestion data. LLMs can leverage the information to provide dynamic traffic insights, such as predicting congestion hotspots or suggesting alternate routes. For supply chain management, KGs capture the relationships between warehouses, delivery hubs, and road networks, to allow LLMs to optimize delivery routes, manage inventory, and mitigate risks associated with disruptions.

Autonomous driving may further benefit from KG-LLM integration by enabling context-aware decision-making. KGs can represent the relationships between road entities, traffic conditions, and pedestrian behavior, while LLMs reason over this data to generate safer navigation strategies. Moreover, public transportation systems may use KGs to integrate real-time transit schedules, passenger feedback, and sensor data, allowing LLMs to generate personalized travel recommendations and demand forecasts. These capabilities contribute to improving the system operational efficiency, reducing traffic delays, and enhancing user experiences for passengers.

\paragraph*{Large Vision-Language Models} 
\quad LVLMs can perform complex multimodal tasks such as image captioning, visual question answering (VQA), text-to-image generation, and scene understanding. These models are trained on large datasets of paired visual and textual information, allowing them to understand and align visual features with linguistic concepts. One of the key advantages of LVLMs is their ability to work in few-shot or zero-shot settings, where they can handle new tasks with minimal or no additional training. This adaptability, combined with their ability to retrieve and interpret cross-modal information, makes LVLMs suitable for a wide range of applications. They also excel at improving accessibility by generating descriptive text for visual data, enabling visually impaired individuals to better understand visual content. Furthermore, LVLMs facilitate natural human-computer interactions, where users can interact with AI systems using both text and images.

Through our literature search, a few existing studies are identified that apply this technique in the perception of autonomous driving \autocite{wang2023empowering, li2024automated}, maritime transportation \autocite{zhang2024popeye}, travel behavior analysis \autocite{zhao2024DriveLLaVAHumanLevelBehavior, zhang2024IntegratingVisualLarge}, and scenario engineering of accident analysis \autocite{Xuan_2024_CVPR}.

Beyond these uses, LVLMs has potential for other applications as well. For example, in traffic management, LVLMs could process live camera feeds to deliver text updates on road conditions, congestion, or crashes, while addressing queries like ``Why is Main Street blocked?'' by merging real-time visuals with language skills. In public transit, they could analyze video of passenger flow at stations, suggesting actions like ``Station X is overcrowded; add more trains.'' Likewise, in logistics and supply chain management, LVLMs could streamline warehousing and freight by monitoring goods, overseeing loading processes, and verifying safety compliance via video, linking visual insights to actionable language outputs. These possibilities highlight LVLMs’ game-changing impact across varied domains.

\paragraph*{Light and Specialized LLMs} 
\quad We need lightweight and specialized LLMs to overcome the significant challenges of deploying traditional, resource-intensive models in real-world applications. Standard LLMs, with their massive parameter counts and high computational demands, are typically confined to high-performance cloud servers, making them unsuitable for mobile devices, offline environments, and applications requiring low energy consumption and high privacy. Lightweight models address these limitations by optimizing for lower latency, reduced memory usage, and energy efficiency, enabling their deployment on resource-constrained devices like smartphones, smartwatches, and compact medical tools. These advancements not only expand accessibility but also reduce operational costs and environmental impact, paving the way for broader adoption across various industries \autocite{misra2024era}.

Additionally, specialized LLMs are crucial for delivering high performance within specific use cases while maintaining efficiency. By employing innovative techniques such as quantization, ternary value representation (e.g., BitNet b1.58), and shared compression layers, these models achieve impressive results with limited computational resources. For instance, models like MiniCPM-V and MobileLLM demonstrate the ability to perform complex tasks, such as image and video processing, on smaller devices \autocite{balani2024future}. This specialization enables the integration of advanced AI capabilities into vehicles, airplanes or any other means of mobility.

In our literature search, a few specialized LLMs are identified for traffic data analysis, simulation, optimization, and control \autocite{da2024open, zhang2024trafficgpt} as well as for answering transport-related questions \autocite{wang2024transgpt}. However, the number of specialized models remains small, highlighting the need for more targeted research and development in this area.

\paragraph*{Mixture of Experts}
\quad Mixture of experts (MoE) is an ML technique that combines the strengths of multiple specialized models, called experts, to solve complex tasks. In MOE, each expert is trained to specialize in a specific aspect of the problem, while a gating mechanism (typically another model) determines which expert(s) to activate for a given input. A classic example of success is DeepSeek-V3, which applied MoE method to vastly improve computational efficiency, where only a small subset of experts (specialized sub-networks) is activated for each input \autocite{liu2024deepseek}. 

At the time of conducting our literature review, we find that MoE is rarely used in traffic and transportation applications. On the other hand, MoE can be applied, for instance, with one expert specializing in urban traffic flow prediction while another expert handling freight logistics optimization. This expert specialization reduces computational costs by ensuring that only the most relevant parts of an LLM are used, making MoE ideal for resource-constrained environments. Furthermore, the gating mechanism in MoE dynamically selects the most appropriate experts based on input, enabling an LLM to handle diverse transportation tasks with precision and efficiency. 

%Combined with FP8 mixed precision training, as seen in DeepSeek-V3, MoE models can further reduce memory usage and energy demands during training and inference without sacrificing performance.

For further specialization, ensemble methods can be employed with multiple lightweight models combined and each optimizing a specific task. For example, one model could focus on detecting traffic incidents, while a second model predicts their impacts and a third optimizes route suggestions. These models can work collaboratively to provide comprehensive solutions, ensuring that the system remains versatile and capable of addressing complex and interrelated transportation challenges.

\paragraph*{Hybrid Online and Onboard LLMs}
% Online LLMs: Continuous learning from new scenarios.
% Onboard LLMs (e.g., LLaMa): Reduced reliance on network connectivity, lower computational needs, and faster decision-making.
% Hybrid approaches: Combining the strengths of both online and onboard LLMs to enhance robustness and efficiency.
\quad Hybrid online and onboard LLMs combine the adaptability of cloud-based continuous learning with the reliability of real-time, localized decision-making. While most LLM applications in autonomous driving and traffic systems focus on either centralized, online models or isolated, onboard systems, hybrid approaches are often overlooked due to the technical complexity and the challenges of synchronizing online and onboard functionalities. However, hybrid models are uniquely positioned to address the demands of transportation systems, where both long-term adaptability and immediate responsiveness are essential.

Throughout our literature search, we have not found any applications of a hybrid approach in traffic and transportation research despite multiple potentials. For example, in traffic management, online LLMs analyze global trends and predict long-term patterns, while onboard LLMs optimize local traffic flows in real time. For AVs, onboard systems handle immediate sensor data (e.g., obstacle detection or navigation), while online systems refine driving algorithms by learning from global edge cases. In personalized navigation, online LLMs provide route recommendations based on user preferences, while onboard systems adapt to real-time inputs and hazards. Hybrid systems also enhance traffic safety by combining predictive online insights with onboard hazard detection and improve urban planning by integrating high-level design insights with real-world monitoring through embedded onboard models.

%Hybrid LLMs require specific methodologies to bridge online and onboard systems. Continuous learning and transfer learning enable online LLMs to adapt to new scenarios, while edge computing optimization ensures onboard systems process real-time data efficiently. Federated learning can synchronize updates between online and onboard models, enhancing global intelligence without compromising privacy. Additionally, multi-modal learning allows hybrid LLMs to process diverse inputs like sensor data, maps, and voice commands. These methodologies ensure hybrid LLMs are scalable, responsive, and robust, paving the way for intelligent and adaptive transportation systems.

\paragraph*{Hybrid Reinforcement Learning and LLM}
\quad RL is a very popular technique nowadays for real-time optimization. However, as mentioned earlier the integration of RL and LLMs is still in its infancy in traffic and transportation research. In our literature search, we have only identified papers in autonomous flight control \autocite{yang2024large} and bus holding control \autocite{yu2024large}.

One approach of hybridizing RL and LLM is to use an LLM as the task planner and RL as the task executor: an LLM breaks down high-level goals into actionable sub-tasks, while RL optimizes their execution. For instance, in traffic management, an LLM might propose rerouting during congestion, and an RL agent adjusts signal timings accordingly. In AVs, LLMs could translate instructions like ``avoid construction zones'' into actions refined by RL.

Another approach of hybridization is to use LLMs to enhance RL state representation in multimodal settings, such as traffic systems with textual incident reports, sensor data, and camera feeds. By creating a unified, semantically rich environment model, LLMs enable RL agents to make better-informed decisions, e.g., interpreting ``accident on Interstate Highway 101'' to guide resource allocation. In logistics, LLMs could simulate constraints like delivery delays, aiding RL in real-time routing adjustments. Similarly, in human-AI systems like ride-sharing, LLMs manage user preferences while DRL optimizes fleet distribution. This synergy of LLMs and RL holds significant promise across diverse application areas.

\paragraph*{Integrating 3D Reconstruction with LLMs} 
\quad AD systems often struggle with scene understanding when relying solely on static images \autocite{sreeram2024probing, wen2024road}. Significant challenges exist in interpreting complex, dynamic environments. To address the challenges, the method of integrating 3D reconstruction with LLMs involves using 3D reconstruction techniques to generate detailed spatial data from various sensors, including cameras, LiDAR, and radar \autocite{wang2024omnidrive}. The spatial data is then transformed into a format that LLMs can process, such as text or structured data, enabling LLMs to interpret and reason about the driving environment in a more nuanced way.

Looking ahead, several key areas require further exploration to fully realize the potential of integrating 3D reconstruction with LLMs in autonomous driving. One direction could involve training LLMs to predict and narrate future 3D scene changes, for example anticipating a pedestrian’s path from reconstructed trajectories to enhance proactive navigation. Another idea is integrating LLMs with real-time 3D semantic mapping to enable vehicles to query and reason about unseen areas (e.g., “What’s around the corner?”) using reconstructed spatial context. A third possibility is developing LLMs that optimize 3D reconstruction itself by prioritizing key environmental features (e.g., obstacles over scenery), to reduce the computational load while maintaining accuracy for driving tasks.

%% file: FutureDirections/Applications.tex
\subsection{Future Directions for LLM Applications}
\label{subsubsection:Applications}
\paragraph*{Ethical Considerations} 
\quad Ethical considerations involve evaluating the moral implications and societal impacts of deploying LLMs in traffic and transportation systems, focusing on issues like safety, fairness, privacy, and equity. These considerations address risks such as LLMs generating inaccurate outputs (e.g., misreading traffic signs), making life-or-death decisions in crash scenarios, or mishandling sensitive user data, ensuring that technology aligns with human values. Studying these ethical aspects offers significant advantages: it fosters the creation of safer, more reliable systems by identifying and mitigating risks like algorithmic bias, enhances public trust through transparency, and ensures compliance with legal standards, reducing liability while promoting equitable access for diverse populations, including those with disabilities or in underserved regions.

Throughout our literature search, we have not identified any paper that focuses on ethical considerations of applying LLMs in traffic and transportation research. Future research should prioritize developing ethical decision-making frameworks for LLMs in autonomous vehicles, particularly to address fairness and accountability in unavoidable crash scenarios, drawing inspiration from debates like the ``trolley problem'' in philosophy. Another critical direction is designing privacy-preserving techniques for managing sensitive transportation data, ensuring a balance between system utility and user rights, such as anonymizing location data while maintaining functionality. Additionally, studies should focus on reducing biases in LLMs to promote equity across diverse communities and investigate human-AI interaction to prevent over-reliance, ensuring drivers retain situational awareness and intervention capabilities in emergencies. These efforts will be essential for responsibly integrating LLMs into transportation systems.

\paragraph*{Cross-cultural Adaptability} 
\quad Many current LLMs are developed and trained using datasets that are biased toward specific languages, cultures, or regions, leading to limited effectiveness in global or multicultural contexts. For example, traffic optimization systems may fail to account for regional driving behaviors, informal transportation networks, or cultural preferences for specific modes of travel, thereby reducing their applicability in diverse environments. Neglecting these differences can result in inequitable access to transportation benefits, diminished trust in AI systems, and a lack of global scalability for solutions. Addressing cross-cultural adaptability is vital to ensure inclusivity, fairness, and the success of AI-driven transportation systems in a globalized world.

Existing methods have applied LLMs directly for translation in emergency dispatch scenarios, which are considered indirectly related to traffic and transportation \autocite{jiang2024ApplicationsLargeLanguage, otal2024LLMAssistedCrisisManagement}. Applications that merit further research include real-time assistance systems for public transportation in multilingual or low-literacy contexts, culturally sensitive hazard alert systems, and multimodal transportation integration in regions with informal systems like minibuses or tuk-tuks. Additionally, AV interfaces and virtual assistants must be adapted to accommodate diverse languages, accents, and communication norms. Research on sustainability applications, such as promoting eco-friendly transportation modes, is also critical to understanding how cultural values and attitudes toward the environment influence user behavior. These areas represent opportunities to design more inclusive and context-aware transportation systems.

%A notable example is Tesla’s Full Self-Driving (FSD) system in China, which faces adaptability challenges. The FSD system is designed to mimic human driving behavior in the United States, where practices such as crossing a double line in the middle of the road may be common. However, such behaviors can violate local regulations in other markets, leading to compliance and safety issues. Beyond regulatory differences, traffic patterns and the presence of vulnerable road users (VRUs) vary significantly across regions. For instance, densely populated urban areas may have higher pedestrian and cyclist activity compared to suburban environments. An LLM-based autonomous driving agent should incorporate this transferability to ensure that it meets the demands of safety, comfort, and regulatory compliance across diverse driving environments.

To conduct relevant research, existing LLM methodologies can be leveraged by focusing on FT models with culturally diverse datasets that represent a range of languages, dialects, and regional contexts. Researchers can also employ transfer learning to adapt pre-trained models to specific cultural settings, ensuring the inclusion of underrepresented populations. Comparative studies can be conducted by deploying LLM-powered applications in different cultural contexts and analyzing variations in user interactions, preferences, and outcomes. Moreover, participatory approaches such as co-designing applications with local stakeholders can help identify cultural nuances and tailor solutions accordingly. 

\paragraph*{Resilience and Adaptability} 
\quad Resilience ensures that systems can withstand disruptions like data outages, unexpected traffic incidents, or evolving user needs, while adaptability enables systems to adjust to new conditions, such as emerging transportation technologies or policy changes. These challenges are frequently neglected because much research focuses on the immediate functionality or performance of LLMs without considering their robustness in dynamic real-world environments. However, transportation systems are inherently complex, involving unpredictable events, diverse user behaviors, and rapidly changing external factors. Failing to address resilience and adaptability can lead to unreliable systems, decreased trust among users, and limited scalability across different contexts or regions.

Existing research has studied resilience in the fields of traffic safety \autocite{baumler2024predicting, wang2024real}, supply chain \autocite{shahsavari2024empowering} and ATC \autocite{abdulhak2024chatatc}. Future research can explore applications such as adaptive traffic management systems, which dynamically respond to unforeseen disruptions, or personalized navigation tools that adjust to changing user preferences and real-time data. Another promising area is emergency response planning, where resilient LLM-powered systems can process live data during disasters to facilitate evacuations. Additionally, public transportation could benefit from research on operation adaptability to predicted demand during unusual events, such as festivals or extreme weather. These applications can address critical needs in real-world transportation systems while improving their overall reliability and performance.

To conduct relevant research on these challenges, existing LLM methodologies can be adapted by incorporating real-time feedback loops, multi-modal training (e.g., integrating text, sensor, and geospatial data), and domain-specific fine-tuning. Researchers could simulate real-world disruptions, such as data outages or sudden traffic pattern changes, to test and improve system resilience. For adaptability, continual learning techniques can be employed, allowing LLMs to update their parameters as new data becomes available or as transportation systems evolve. Collaboration with transportation experts to curate high-quality datasets and define evaluation metrics for resilience and adaptability will ensure that the research aligns with practical needs.

\paragraph*{Standard Metrics}
\quad Quantitatively assessing the impact of LLMs requires the use of well-defined methodologies and key performance indicators. Current research employs a variety of datasets and evaluation metrics, often tailored to specific studies, making it difficult to compare the performance of different LLM-based models. For example, traffic prediction models can be assessed using error metrics like mean absolute error or root mean squared Error \autocite{huang2024enhancing, ren2024tpllm}, while incident detection systems can be measured based on classification accuracy, precision, and recall \autocite{zheng2023trafficsafetygpt}. Each task requires scholars to carefully consider and assess evaluation criteria.

In addition to task-specific metrics, operational efficiency and economic impact are important considerations. LLMs can be evaluated based on their ability to reduce costs, such as operational expenses for traffic management systems, or to generate savings in fuel consumption and travel time through better route planning. For applications involving human interactions, usability metrics like user satisfaction scores and task completion rates provide insights into how effectively LLMs assist users in making decisions or accessing information.

Additionally, the dynamic nature of transportation systems complicates the establishment of consistent benchmarks, as real-world conditions often vary significantly. Long-term impacts, such as reductions in accidents, emissions, or system-wide costs, are harder to measure and require continuous monitoring to fully assess the effectiveness of LLM-powered solutions. Therefore, the quantitative evaluation of LLMs in transportation and traffic systems must consider a broad spectrum of metrics.

%Furthermore, issues related to bias, fairness, and energy efficiency must also be considered to ensure that these systems are equitable and sustainable.

%In summary, the quantitative evaluation of LLMs in transportation and traffic systems relies on a combination of task-specific metrics, operational KPIs, and broader economic and environmental considerations.

\paragraph*{Urban Delivery}
\quad In the realm of urban delivery and logistics, several pressing challenges persist, such as optimizing dynamic routing, managing customer interactions, and enhancing last-mile delivery efficiency. These challenges are compounded by the complexity of urban environments, making traditional solutions less effective and highlighting the need for innovative approaches.

To date, our literature search indicates that no studies have explored the application of LLMs to urban delivery, which has a few promising research directions. First, integrating LLMs with real-time spatial-temporal data offers a promising avenue for creating dynamic, context-aware routing systems, potentially cutting delivery times and environmental impact. Second, LLMs can process real-time traffic data, including traffic patterns, weather conditions, and local events, to assist in dynamic route optimization. Third, research could investigate how LLMs might generate augmented instructions for last-mile delivery, aiding couriers in navigating urban complexities—such as apartment layouts or alternative drop-off points—while aligning with customer preferences, thus streamlining operations and enhancing delivery performance.

% \paragraph*{Warehousing}
% \quad As the scale of modern warehouses becomes very large, warehousing operations encounter challenges that affect productivity and profitability, such as optimizing inventory management, ensuring accurate order fulfillment, and coordinating workforce efforts. These difficulties are intensified by the demands of modern supply chains, which require rapid adaptability to fluctuating demand, diverse product portfolios, and real-time operational adjustments.

% A review of existing literature indicates that the application of LLMs to warehousing remains unexplored, indicating significant research gaps and future application potential.

% \begin{itemize}
%     \item \textbf{Integration with Robotics and Autonomous Systems} The integration of LLMs with warehouse robotics and autonomous systems is another promising area. By leveraging LLMs' ability to interpret complex instructions and contextual nuances, robots can dynamically adapt their operations to changing warehouse environments, such as re-routing in congested aisles or handling fragile items with precision.
    
%     \item \textbf{Real-Time Decision-Making} Developing efficient inference techniques for deploying LLMs in resource-constrained environments like warehouses is a critical challenge. Future research could focus on lightweight LLMs or model distillation approaches to enable real-time decision-making for tasks such as order prioritization, shipping label generation, or responding to supply chain anomalies.
% \end{itemize}

\paragraph*{Road Design}
\quad  While we have identified papers that use multi-modal LLMs for generating road networks in recognized computer formats based on aerial images or even hand-drawn graphs \autocite{rasal2024beyond,chen2024multimodal}, the design of roads remains an open area for exploration. Future research could extend the applications of LLMs beyond network generation to address more complex and context-sensitive road design challenges. Below, we outline several potential research directions:

\begin{itemize}
    \item \textbf{Economic and Demographic Factors-Based Road Design:} LLMs could be developed to design roads based on a combination of economic and demographic factors. By integrating the relevant multimodal data, an LLM could propose road layouts optimized for cost-effectiveness and equitable access. For instance, an LLM could identify underserved areas and suggest new road designs that connect isolated populations to urban hubs, markets, or essential services. Integrating geospatial and economic data with the LLM reasoning capabilities could lead to more inclusive and economically sustainable road planning.
    \item \textbf{Road Bottleneck Identification and Mitigation:} Current research does not address the use of LLMs for identifying and mitigating road bottlenecks. Future work could explore how LLMs can analyze traffic flow data, accident reports, and congestion patterns to pinpoint bottlenecks in road networks. By combining this analysis with generative capabilities, LLMs could propose redesigns or expansions of problematic road segments to improve traffic efficiency. This could be particularly valuable for dynamic traffic management systems aiming to adapt road designs in real-time based on evolving traffic conditions.
    \item \textbf{Integration of Environmental and Climate Factors:} Road design often neglects the critical impact of environmental and climate factors. LLMs could integrate environmental data, such as flood zones, heat maps, and greenhouse gas emission levels, to propose climate-resilient road designs. For example, they could suggest elevated roads in flood-prone areas or optimize layouts to reduce vehicle emissions. By incorporating sustainability considerations, LLMs could contribute to the development of environmentally conscious transportation systems.
    \item \textbf{Integration with Real-Time Traffic Systems:} Road design is often treated as a static problem, but real-world traffic systems are highly dynamic. Future LLMs could integrate with real-time traffic monitoring systems to propose adaptive road designs and layouts. For example, they could suggest temporary road expansions, one-way systems, or dedicated bus lanes during peak hours, providing a flexible approach to urban traffic management.
    \item \textbf{Exploration of Lightweight and Cost-Effective Models:} Building on the lightweight training methods demonstrated in \cite{rasal2024beyond}, future research could focus on developing low-resource LLMs for road design. This would make advanced road planning tools accessible to low-income regions or smaller municipalities that lack the computational resources to deploy large-scale AI systems.
    \item \textbf{Multimodal Road Design Optimization:} Current LLM applications focus on generating networks. Future research could explore optimizing road designs for multimodal transportation systems. This could include integrating data from public transit networks, freight logistics, and pedestrian pathways to propose road layouts that balance the needs of different transportation modes. LLMs could also consider the trade-offs between infrastructure costs, travel times, and environmental impacts.
\end{itemize}

%\paragraph*{Explaining navigation decision}

% \paragraph*{Scalability and Real-World Complexity} 
% \quad  As real-world applications of LLMs encounter increasingly complex scenarios, particularly edge cases (e.g., rare or unforeseen traffic events), the scalability of current SOTA methods is called into question. SOTA methods often lack sufficient discussion on their ability to handle these corner cases while maintaining timely responses. The reliance on few-shot learning techniques, while promising, may introduce significant delays and demand higher computational resources, potentially impacting real-time decision-making in transportation systems.

%\paragraph*{Rapid Model Advancements} 
%\quad  LLM backbones, such as GPT, are evolving rapidly, leading to frequent updates in architecture, reasoning capabilities, and performance metrics. It is unclear how these advancements directly affect the requirements for transportation applications, such as reasoning accuracy, response time, and knowledge convergence. Researchers must continuously adapt and re-evaluate their methods to integrate the latest developments while maintaining backward compatibility.

\paragraph*{End-to-end Autonomous Vehicle} 
\quad The integration of LLMs into end-to-end autonomous driving frameworks has attracted significant attention due to their ability to unify perception, prediction, and planning into a single model. Inspired by the research of \cite{zhu2024surveylargelanguagemodelempowered}, we propose two research directions where LLMs can play a role in end-to-end autonomous driving.

First, end-to-end AD frameworks, such as Tesla's full self-driving (FSD), rely heavily on diverse and high-quality datasets to train models capable of directly generating motion plans. However, the collection and annotation of such data are resource-intensive, with rare and critical events—like extreme weather conditions or sudden pedestrian crossings—often underrepresented in datasets. To address this, strategies such as simulation-based data augmentation and active learning can improve dataset efficiency while reducing the dependency on costly real-world data collection. Nonetheless, balancing cost, scenario diversity, and realism remains an ongoing challenge that needs further exploration.

Second, interpretability is a crucial factor for establishing trust in end-to-end AD systems, as their black-box nature limits transparency in decision-making. Techniques such as attention mechanisms and explainable AI (XAI) tools offer ways to shed light on the reasoning behind model predictions, while intermediate representations can enhance transparency by breaking down the decision-making process into interpretable stages. Recent advancements, such as vision-language planning (VLP), leverage LLMs to generate natural language explanations for motion planning decisions, providing an additional layer of interpretability. However, striking a balance between interpretability and system performance continues to be a significant challenge \autocite{pan2024vlp}.

%As highlighted by \autocite{zhu2024surveylargelanguagemodelempowered}, several critical challenges remain in this field: how to construct better-quality datasets for training end-to-end algorithms, improve the efficiency of training processes, and enhance the interpretability of these systems. Addressing these challenges will require breakthroughs in data collection methodologies, model architecture designs, and innovative explainability techniques, paving the way for more reliable and transparent autonomous driving systems.

\paragraph*{Interaction with Other Road Users} 
\quad In autonomous driving, effectively managing interactions with road users like pedestrians, cyclists, and other vehicles is a pivotal challenge that directly influences safety and traffic flow, such as interpreting a pedestrian’s wave to cross, understanding a car’s flashing lights signaling a turn, or reacting to a cyclist’s sudden swerve. These issues are critical because AVs must quickly and accurately decipher these often non-verbal, dynamic cues to avoid collisions and ensure smooth navigation. 

Although there are a number of research outputs that analyze trajectories of other road users \autocite{lan2024traj,wang2024omnidrive}, research on finer granularity remains absent. Future research could significantly advance this field by integrating LLMs with multimodal systems, such as combining them with computer vision to simultaneously analyze verbal warnings and visual gestures, thereby building a richer understanding of road user intentions. Another promising direction involves training LLMs on contextual and historical data to predict behaviors -- like anticipating a pedestrian’s crossing based on subtle cues or past patterns -- enhancing proactive decision-making. Additionally, equipping LLMs to enable vehicles to communicate outwardly, such as broadcasting a message to yield or acknowledging a cyclist's signal via text or sound, could foster clearer, safer interactions in shared traffic spaces, ultimately elevating the adaptability and trustworthiness of AD systems.

%\paragraph*{Integrating LLM with Driver Behavior Database}
%\textcolor{blue}{[YM: you may integrate this in future direction for methodology: RAG]}
%Driver behavior is a complex phenomenon influenced by multiple factors. To fully understand the relationship between decision-making and external elements—such as surrounding vehicles, environmental conditions, and driving preferences—an integrated database, encompassing both structured and unstructured data, is essential. This database should combine information from cameras, LiDAR, radar, and other relevant sources to provide a comprehensive view of travel behavior. Such domain-specific database serves as valuable resources for training specialized LLM.

\paragraph*{Travel Behavior Modeling}
\quad In the research area of travel behavior, LLMs complement traditional methods by excelling at processing vast, unstructured datasets—think social media posts or open-ended survey responses—where conventional models falter. Their strength lies in uncovering nuanced patterns and adapting to multi-modal data, offering a richer understanding of travel behavior. For instance, studies like \cite{ruan2024twitter} showcase LLMs predicting travel mode choices from social media, while \cite{chen2024delayptc} and \cite{shao2024copb} leverage them for public transit and historical data analysis using CoT prompting.

Despite these advancements, several research gaps remain that future studies should address to enhance LLMs' utility in travel behavior modeling. One critical gap is the limited integration of real-time, multi-modal data. Current models excel with historical or static datasets but struggle to adapt to dynamic urban conditions, such as sudden traffic incidents or weather shifts. Future research should focus on developing LLMs that fuse live traffic feeds, weather data, and visual inputs (e.g., from traffic cameras) using techniques like continual learning and transfer learning. For example, in metro delay scenarios, an LLM could combine real-time delay logs, passenger sentiment from social media, and station camera feeds to predict adaptive travel choices, improving upon DelayPTC-LLM's \autocite{chen2024delayptc} static approach. Similarly, for pedestrian safety, integrating live sensor data from smart crossings with LLM reasoning could enhance real-time collision risk predictions, addressing limitations in the simulations of \cite{radford2021CLIP}. Another gap is explainability, particularly in safety-critical applications. Enhancing methods like attention visualization or counterfactual reasoning could make LLM predictions—such as a driver’s lane-changing decision or a pedestrian’s crossing intent—more transparent, aiding AV decision-making and urban planning.

%% file: FutureDirections/ProsCons.tex
\subsection{Strengths and weaknesses of LLMs}
\label{subsubsection:ProsCons}
\quad  By leveraging pre-trained capability and adaptability, LLMs bring significant advantages to traffic and transportation research. The following advantages across different applications to traffic and transportation applications.

\textbf{Versatility with Pre-Trained Models}: LLMs offer significant versatility in traffic and transportation applications due to their pre-trained capabilities. These models can be employed in various scenarios with ZSL or FSL, meaning they can perform tasks without needing extensive retraining on new data. For instance, an LLM can be used to analyze traffic reports, understand road conditions, or even generate summaries of transportation policies without specific training for each task. This flexibility is particularly beneficial in the dynamic field of transportation, where new challenges and data types, such as real-time traffic updates or incident reports, emerge frequently. %Research has shown that pre-trained models like BERT can classify traffic incidents directly, demonstrating their utility without additional training.

\textbf{Adaptability through Fine-Tuning}: Fine-tuning LLMs allows them to be tailored to specific transportation contexts, enhancing their performance in those areas. For example, a transportation agency could fine-tune an LLM on historical traffic data to better predict congestion patterns or optimize public transit schedules. This adaptability ensures that the model is not only general but also highly effective in specific, localized scenarios, such as managing traffic in a particular city or region. Studies have highlighted the effectiveness of fine-tuning in tasks like route optimization, where LLMs can learn from local traffic patterns to improve decision-making, aligning with the need for precision in urban transportation systems.

\textbf{Multimodal Integration}: LLMs can integrate with visual data, providing a more comprehensive understanding of transportation scenarios that involve both text and images. In transportation, this could mean analyzing images from traffic cameras alongside textual reports to better understand and respond to traffic incidents or road conditions. Visual language models, which combine the capabilities of LLMs with image processing, can help in tasks like automatic incident detection or monitoring road maintenance needs. Recent research, such as the use of VAD-LLaMA for traffic anomaly detection, demonstrates how LLMs with visual data can enhance real-time responsiveness in autonomous driving systems, improving safety and efficiency.

\textbf{Handling Unforeseen and Long-Tail Scenarios}: The broad knowledge base of LLMs enables them to understand and respond to unusual or rare situations in transportation. Whether it's dealing with extreme weather conditions, unusual traffic patterns due to events, or new types of vehicles, LLMs can provide insights or predictions based on their general understanding of language and world knowledge. This is crucial for transportation systems to be resilient and adaptive to unpredictable events, such as sudden road closures or emergency evacuations. Their ability to generate unseen scenarios also supports scenario planning, as seen in studies using LLMs for traffic management at urban intersections, where they handle mixed traffic conditions effectively.

\textbf{Generating Synthetic Datasets}: LLMs can create artificial data for training other models or for simulation purposes, which is particularly useful in transportation where real data might be limited or sensitive. For example, synthetic data can be used to train models for rare events like accidents or to simulate different traffic scenarios for testing new control algorithms. This capability helps improve the robustness and safety of transportation systems without relying solely on real-world data, which might be scarce or biased. Research has explored LLMs generating synthetic datasets for traffic flow forecasting, enhancing model training in data-scarce environments.

\textbf{Language Translation and Adaptation}: Transportation is a global issue, and different regions have different languages. LLMs can translate information or adapt to different language contexts, which is helpful for international transportation systems or for providing multilingual information to users. This ensures that transportation information and services are accessible to a diverse range of users, enhancing inclusivity and efficiency in global transportation networks. %For instance, a case study in San Antonio demonstrated LLMs providing personalized travel assistance in multiple languages, improving passenger experience. \textcolor{blue}{[Zou: suggest either adding the ref for the last sentence or deleting it (since here is about general discussions of the pros/cons.]}

\textbf{Rich Embeddings}: The embeddings from LLMs contain nuanced information that traditional NLP methods might not capture, providing deeper insights into transportation data. For instance, these embeddings can help understand the sentiment of public feedback on transportation services or in identifying patterns in traffic-related text data that are not immediately obvious. This can lead to better decision-making and more effective communication strategies in transportation management, such as analyzing social media for real-time traffic sentiment, which traditional NLP might overlook.

Apart from the above strengths, LLMs do have weaknesses. Addressing the weaknesses is essential for ethical and effective use of LLMs. However, this will require further advancements in LLM development and training processes. The major weaknesses include: 

\textbf{High Computational Requirements}: Training and running LLMs require significant computational power, which can be a barrier for some transportation agencies, especially those with limited budgets or in regions with poor Internet connectivity. Real-time applications in transportation, such as traffic signal control or dynamic routing, demand low latency and high processing speeds, which LLMs may not always meet efficiently. This can limit their practical implementation in time-critical transportation systems, particularly in developing regions where resources are constrained.

\textbf{Privacy}: Privacy is another critical concern when using LLMs in transportation, as detailed in \cite{das2025security}. These models are trained on vast amounts of data, which may include sensitive information about individuals' travel habits, such as commuting patterns or frequent destinations, raising risks of personally identifiable information (PII) leakage. There is a risk that LLMs could inadvertently reveal or misuse this data, leading to privacy breaches, especially in cloud-based systems where data might be shared with third parties. Transportation agencies must ensure that any data used in training or fine-tuning LLMs is handled in accordance with privacy regulations, such as GDPR, and steps are taken to prevent data leakage or unauthorized access, which is crucial for maintaining public trust and compliance.

\textbf{AI Hallucinations}: LLMs can produce outputs that are factually incorrect or entirely made up, leading to misinformation or wrong decisions, such as suggesting non-existent road closures or incorrect detour routes. In transportation, such errors could have serious consequences, from misdirecting traffic to providing inaccurate safety information, potentially endangering lives.

\textbf{Up-to-date Data}: LLMs are usually trained on historical data, which may not reflect current traffic conditions, new infrastructure like recently built highways, or regulatory changes such as updated speed limits. LLMs risk providing outdated or irrelevant recommendations, which could compromise transportation efficiency and safety, as seen in \cite{zhang2024large}. 

\textbf{Lack of Uniform Benchmark}: The lack of a uniform benchmark for evaluating LLMs in transportation, especially in generating unseen scenarios and synthetic datasets, is a significant hurdle, as highlighted in \cite{syed2024benchmarking}. Without standardized evaluation methods, it is difficult to compare the performance of different LLMs or to assess how they stack up against traditional approaches based on human expert-defined rules, such as traffic simulation models. Developing such benchmarks is crucial to drive progress in this area and ensure that LLMs are effectively enhancing transportation systems, particularly for tasks like predicting rare events or creating training data for other AI models.

\textbf{Numerical Ability}: While these models excel in processing and generating text, their performance with numerical data, such as traffic counts, travel times, or fuel consumption estimates, may not be as robust as that of traditional numerical models or specialized AI models designed for such tasks. For instance, predicting exact traffic volumes during peak hours or predicting demand for supply chain applications requires precise numerical reasoning, which LLMs might struggle with compared to statistical models. 

\textbf{Trustworthiness and Safety}: LLMs, with their probabilistic outputs and potential for errors or biases, may not always meet the stringent safety requirements of such systems, as they can hallucinate or provide inconsistent results under varying conditions. Besides, LLMs are vulnerable to prompt injection, jailbreak attacks, and data poisoning, which can manipulate their responses or make them reveal sensitive information~\autocite{yao2024survey}.

%% file: Conclusion.tex
\section{Conclusion}
\label{section:Conclusion}
\quad In this paper, we offer a comprehensive overview of LLM methodologies and their applications within traffic and transportation research. We categorize these studies by their specific use cases -- ranging from urban logistics to AD interactions -- while critically examining the associated challenges. LLMs emerge as potent tools for overcoming the shortcomings of conventional approaches, excelling in managing long-tail scenarios, integrating text with diverse data types, and enhancing system adaptability. Despite these advancements, substantial research opportunities persist, both in refining the necessary methodologies and expanding their practical applications, promising further innovation in this rapidly evolving field.

%% file: Appendix.tex
%\appendix
%\begin{landscape}
\newpage
\section{Appendix}
\label{section:appendix}
%\addbibresource{sample.bib}
%\usepackage[utf8]{inputenc}
%\usepackage{natbib}
%\usepackage{CJKutf8}
%\setcitestyle{numbers}
\subsection{Supplementary Details on Transformer Architectures}

\paragraph*{Attention Mechanism Details}
\quad The self-attention mechanism assigns an attention score to each token based on its relevance to others. These scores are computed using query (Q), key (K), and value (V) matrices, as shown in Eq.~\ref{eq:attention}:

\begin{equation}
\label{eq:attention}
    \text{Attention} (Q, K, V) = \text{softmax} \left(\frac{QK^T}{\sqrt{d_k}}\right) V
\end{equation}

Multi-head attention extends self-attention by using multiple attention heads to focus on diverse aspects of input. The combined outputs are linearly transformed as shown in Eq.~\ref{eq:MultiHead_attention}:

\begin{align}
\label{eq:MultiHead_attention}
    \text{MultiHead}(Q, K, V) = \text{Concat}(\text{head}_1, \text{head}_2, ..., \text{head}_h)W^O
    ,\; W^O \in \mathbb{R}^{(h \cdot d_{\text{head}}) \times d_{\text{model}}}
\end{align}

To improve scalability, variants like multi-query attention~\autocite{shazeer2019fast} and grouped-query attention~\autocite{ainslie2023gqa} reduce computational costs while maintaining performance.

\paragraph*{Positional Encoding Details}
\quad Absolute positional encoding, as introduced by Vaswani et al.~\autocite{vaswani2017attention}, is defined as:

\begin{align}
\label{eq:pos_en}
    PE_{pos, 2i} &= \sin \left(\frac{pos}{10000^{2i/d_{model}}}\right) \\
    PE_{pos, 2i+1} &= \cos \left(\frac{pos}{10000^{2i/d_{model}}}\right)
\end{align}

This ensures that nearby positions have similar encodings. Relative positional encoding, on the other hand, encodes distances between tokens, making it effective for tasks requiring contextual understanding.

\paragraph*{Decoder-Only Models} 
\quad Operate unidirectionally, processing text left-to-right, and are trained on next-token prediction loss. Examples include GPT~\autocite{radford2018improving,brown2020language}.

\paragraph*{Encoder-Only Models} 
\quad Use bidirectional attention for masked language modeling (MLM). Examples include BERT~\autocite{devlin2018bert} and RoBERTa~\autocite{liu2019roberta}.

\paragraph*{Encoder-Decoder Models}
\quad Combine bidirectional encoding and autoregressive decoding for sequence-to-sequence tasks like translation (e.g., T5~\autocite{raffel2020exploring}, BART~\autocite{lewis2019bart}). For more mathematical derivations and examples, readers are directed to ~\autocite{hadi2023survey, kalyan2021ammus, huang2023advancing}.

\subsection{Training Details}
\subsubsection{Pre-training Phase Details}
\paragraph*{Next-token prediction}
\quad The objective of next-token prediction is to minimize the autoregressive loss:
\begin{align}
\label{eq:next_token_pred_appendix}
     \mathcal{L}_{\text{AR}}(\theta) = -\mathbb{E}_{(x_1, \dots, x_T) \sim \mathcal{D}} \left[ \sum_{t=1}^{T} \log P_\theta(x_t \mid x_{1:t-1}) \right]
\end{align}

\paragraph*{Masked Language Modeling (MLM)}
\quad In MLM, the model predicts masked tokens by minimizing the following loss:
\begin{align}
\label{eq:mlm_loss_appendix}
         \mathcal{L}_{\text{MLM}}(\theta) = -\mathbb{E}_{(x_1, \dots, x_T) \sim \mathcal{D}} \left[ \sum_{t \in M} \log P_\theta(x_t \mid x_{\backslash t}) \right]
\end{align}

\noindent\subsubsection{Post-training Phase Details}
\paragraph*{Alignment with human preferences}
\quad Alignment with human preferences involves training a reward model \(R_\phi(x)\) using human feedback. The reward model's objective is formulated as:
\begin{align}
\label{eq:reward_loss_appendix}
      \mathcal{L}_{\text{reward}}(\phi) = -\mathbb{E}_{(x_1, x_2) \sim \mathcal{D}_{\text{human}}} \left[ \log \frac{e^{R_\phi(x_1)}}{e^{R_\phi(x_1)} + e^{R_\phi(x_2)}} \right]
\end{align}

The fine-tuning objective for RLHF is:
\begin{align}
\label{eq:rlhf_loss_appendix}
      \mathcal{L}_{\text{RLHF}}(\theta) = -\mathbb{E}_{x \sim \pi_\theta(x \mid z)} \left[ R_\phi(x) \right]
\end{align}

The Proximal Policy Optimization (PPO) objective is used for reinforcement learning:
\begin{align}
\label{eq:ppo_appendix}
      \mathcal{L}_{\text{PPO}}(\theta) = \mathbb{E}_{x \sim \pi_\theta} \left[ \min\left( r(\theta), \text{clip}(r(\theta), 1 - \epsilon, 1 + \epsilon) \right) \cdot R_\phi(x) \right]
\end{align}
where \(r(\theta)\) is the probability ratio:
\begin{align}
\label{eq:probability_ratio_appendix}
  r(\theta) = \frac{\pi_\theta(x \mid z)}{\pi_{\theta_{\text{old}}}(x \mid z)}
\end{align}

\paragraph*{Fine-tuning for downstream tasks}
\quad Fine-tuning loss for code generation tasks is:
\begin{align}
\label{eq:fine_tune_code_appendix}
         \mathcal{L}_{\text{code}}(\theta) = -\mathbb{E}_{(c_1, \dots, c_T) \sim \mathcal{D}_{\text{code}}} \left[ \sum_{t=1}^{T} \log P_\theta(c_t \mid c_{1:t-1}) \right]
\end{align}

For domain-specific datasets, the fine-tuning objective is:
\begin{align}
\label{eq:domain_specific_fine_tune_appendix}
    \mathcal{L}_{\text{domain}}(\theta) = -\mathbb{E}_{(x_1, \dots, x_T) \sim \mathcal{D}_{\text{domain}}} \left[ \sum_{t=1}^{T} \log P_\theta(x_t \mid x_{1:t-1}) \right]
\end{align}

\subsection{Other Prompt Engineering Techniques}
\subsubsection*{CoT Prompting with Self-Consistency}
Also known as the multiple chain-of-thought prompting technique,~\autocite{wang2022self} introduces a ``sample-and-marginalize'' procedure to identify the most consistent answer for complex tasks requiring deliberate reasoning. LLMs generate diverse reasoning paths using chain-of-thought prompting and then reach a consensus through majority voting. This method is effective for tasks with fixed sets of correct answers, where self-consistency can be leveraged. However, for open-ended tasks, careful design of consistency metrics is necessary to ensure reliability.

\subsubsection*{Tree-of-Thought Prompting}
The tree-of-thought~\autocite{yao2024tree} technique generalizes chain-of-thought prompting by exploring multiple reasoning paths that branch out like a tree, enabling self-evaluation, backtracking, and looking ahead. Unpromising branches are pruned, and each reasoning step acts as an intermediate stop to help decompose complex tasks. For instance, in solving a multi-step problem like the game of 24, tree-of-thought breaks the task into intermediate steps and expands reasoning paths from each node. Heuristics evaluate branches, and search algorithms like breadth-first or depth-first search guide the process, improving performance in multi-step reasoning tasks.

\subsubsection*{Graph-of-Thought Prompting}
The graph-of-thought~\autocite{besta2024graph} technique extends tree-of-thought by connecting intermediate reasoning steps into a network, forming a large graph of thoughts. The key innovation lies in the aggregation process, where edges from separate reasoning paths converge at shared vertices. Sorting and ranking are used to organize branches selectively. This method excels in tasks like sorting, keyword counting, and sequential planning, where multiple goals must be achieved in a specific order.

\subsection{Additional LLMs}
\quad In this section, we provide a brief overview of additional LLMs that were not included in the main content but are noteworthy for their unique contributions and applications.

\subsubsection*{Vicuna}
~\footnote{\hyperlink{https://lmsys.org/blog/2023-03-30-vicuna/}{https://lmsys.org/blog/2023-03-30-vicuna/}} is an open-source chatbot fine-tuned from LLaMA on approximately 70K user-shared conversations from ShareGPT. It achieves 90\% of ChatGPT’s quality in preliminary evaluations using GPT-4 as the judge. Vicuna is widely recognized for its ability to produce detailed, well-structured responses, and it serves as a strong alternative to proprietary models in open-source research.

\subsubsection*{BLOOM}
\textbf{BLOOM}~\autocite{le2023bloom} is a multilingual, open-access model with 176 billion parameters, trained on 59 languages. It emphasizes inclusivity and accessibility for diverse linguistic communities. BLOOM is particularly notable for its large-scale collaborative development by the BigScience project, making it a pioneering effort in democratizing LLM research.

\subsubsection*{FLAN}
\textbf{FLAN}~\autocite{wei2021finetuned} is Google's instruction-tuned model designed to improve zero-shot and few-shot performance. By fine-tuning on chain-of-thought and instruction data, FLAN achieves significant gains in reasoning and comprehension tasks, making it a strong contender in instruction-following applications.

\subsubsection*{Other Notable LLMs}
The following models, while not included in the main content, are significant for specific tasks or domains:
\begin{itemize}
    \item \textbf{Alpaca}~\autocite{taori2023alpaca}: Fine-tuned from LLaMA using self-instruct techniques, Alpaca is designed for cost-effective fine-tuning and task-specific applications.
    \item \textbf{Mistral-7B}~\autocite{jiang2023mistral}: A smaller, high-performing model that outperforms larger open-source models like LLaMA-2-13B in several benchmarks.
    \item \textbf{Med-PaLM and Med-PaLM 2}~\autocite{singhal2023large,singhal2023towards}: Domain-specific versions of PaLM optimized for medical question answering.
    \item \textbf{CodeGen}~\autocite{nijkamp2022codegen}: An open-source model tailored for programming tasks, capable of generating code in multiple programming languages.
    \item \textbf{T0}~\autocite{sanh2021multitask}: A model designed for natural language tasks by mapping them into a prompted form, showcasing strong generalization in zero-shot settings.
\end{itemize}

\subsection{Prompt Optimization Relevant Concepts (mentioned in Future Research Directions)}

\subsubsection*{Attack Success Rate}
Attack Success Rate (ASR) is a metric used to evaluate how vulnerable a machine learning model, including LLMs, is to adversarial attacks. These attacks are deliberate attempts to manipulate the model into producing incorrect, biased, or harmful outputs. ASR quantifies the effectiveness of such attacks.

\subsubsection*{Perplexity}
Perplexity is a key metric used to assess the performance of language models in NLP. It measures how well a model predicts a sequence of words, essentially capturing the model's uncertainty about the next word in a sequence. A lower perplexity score indicates that the model is better at making accurate predictions.

Perplexity (PPL) is mathematically defined as the exponentiation of the average negative log-likelihood of a word sequence. For a sequence of words $w_1$, $w_2$, $\dots$, $w_n$, it is calculated as:

\[
\text{PPL} = \exp \left( -\frac{1}{n} \sum_{i=1}^{n} \log P(w_i \mid w_{1:i-1}) \right)
\]

where \( P(w_i \mid w_{1:i-1}) \) is the probability of word \( w_i \) given prior words, and \( n \) is the sequence length.
%\end{landscape}